\documentclass[a4paper,preprintnumbers,floatfix,twocolumn,aps,prb,unsortedaddress,superscriptaddress]{revtex4-2}

\usepackage[para]{threeparttable} 
\usepackage{amsmath}
\usepackage{graphicx} 
\usepackage{booktabs}
\usepackage{caption}
\usepackage{subcaption}
\usepackage{nicefrac}
\usepackage{url}
\usepackage{hyperref}
\usepackage{miller}
\usepackage{bm}
\usepackage{enumitem}
\usepackage{color}
\usepackage{multirow}
\usepackage{tabularx}
\newcommand{\etal}{{\it et al. }}
\usepackage[paperwidth=210mm,paperheight=297mm,centering,hmargin=1.7cm,vmargin=2.5cm]{geometry}

\usepackage[utf8]{inputenc}

\begin{document}

\title{Fast and accurate machine-learned interatomic potentials for large-scale simulations of Cu, Al and Ni}

\author{A. Fellman}
\thanks{Corresponding author}
\email{aslak.fellman@helsinki.fi}
\affiliation{Department of Physics, P.O. Box 43, FI-00014 University of Helsinki, Finland}
\author{J. Byggmästar}
\affiliation{Department of Physics, P.O. Box 43, FI-00014 University of Helsinki, Finland}
\author{F. Granberg}
\affiliation{Department of Physics, P.O. Box 43, FI-00014 University of Helsinki, Finland}
\author{K. Nordlund}
\affiliation{Department of Physics, P.O. Box 43, FI-00014 University of Helsinki, Finland}
\author{F. Djurabekova}
\affiliation{Department of Physics, P.O. Box 43, FI-00014 University of Helsinki, Finland}

\date{\today}

\begin{abstract}
Machine learning (ML) has become widely used in the development of interatomic potentials for molecular dynamics simulations. However, most ML potentials are still much slower than classical interatomic potentials and are usually trained with near equilibrium simulations in mind. In this work, we develop ML potentials for Cu, Al and Ni using the Gaussian approximation potential (GAP) method. Specifically, we create the low-dimensional tabulated versions (tabGAP) of the potentials, which allow for two orders of magnitude higher computational efficiency than the GAPs, enabling simulations of large multi-million atomic systems. 
The ML potentials are trained using diverse curated databases of structures and include fixed external repulsive potentials for short-range interactions. The potentials are extensively validated and used to simulate a wide range of fundamental materials properties, such as stacking faults and threshold displacement energies. Furthermore, we use the potentials to simulate single-crystal uniaxial compressive loading in different crystal orientations with both pristine simulation cells and cells containing pre-existing defects.
\end{abstract}

\maketitle

\section{Introduction}
\label{sec:intro}
Machine learning (ML) approaches have become increasingly commonplace in the development of interatomic potentials for atomistic simulations. ML potentials have shown excellent accuracy compared to traditional analytical potentials~\cite{bartok_machine_2018,deringer_machine_2017}. Therefore, ML potentials can bridge the gap between \textit{ab initio} methods, such as density functional theory (DFT), and classical analytical models. A growing number of different ML frameworks exist for the creation of interatomic potentials and the number of available ML potentials is increasing rapidly. ML potentials do not assume fixed functional forms, which allows them to be more flexible than their analytical counterparts, mainly at the cost of computational efficiency. Although numerous different ML potentials for single elements have been developed~\cite{pun2020development,lysogorskiy_performant_2021,zuo_performance_2020}, these potentials are usually trained mainly with near equilibrium simulations in mind. In extreme environments, the simulated systems exhibit far from equilibrium phenomena and care must be taken to avoid the models breaking down due to the poor extrapolation capability of ML~\cite{byggmastar_machine-learning_2019}.

Molecular dynamics (MD) has been widely used to study the fundamental properties of compressive loading~\cite{lund2004tension,spearot2009orientation,tschopp2007atomistic,10.1063/1.2715137,li2020molecular,zhang2018nonlinear,xie2014tension}. Even though direct comparison between MD simulations and experiments is difficult due to time-scale limitations and, therefore, exaggerated strain-rates, MD can still provide valuable insights into the fundamental mechanisms with atomic resolution. In particular, MD can provide understanding of the dynamics of dislocation nucleation caused by straining of the materials. Single crystal compression has been studied extensively in elemental metals with a good portion of the work focusing on Cu. However, most of the simulations are performed on pristine systems without any pre-existing defects, even though real materials are never defect-free. In BCC metals, such as tantalum, the effects of pre-existing defects on compression has been studied in some detail~\cite{zepeda2017probing}. There exist some studies of effects of defects on materials response during uniaxial loading of the FCC materials as well 
~\cite{salehinia2014crystal,zhang2017stacking,hu2019effects,tang2015molecular,WANG2017116}. However, the defects in these simulations under compressive loading were well-defined, 
such as voids or stacking fault tetrahedra (SFT). 

Additionally, it has been shown that the choice of interatomic potential in terms of accuracy is crucial for 
loading simulations that lead to plastic deformation, i.e. compression, tension, or nanoindentation 
~\cite{zhu2004predictive}. In this respect, ML potentials can offer the required accuracy for modelling of these phenomena. However, in large scale simulations, computational costs may become prohibitively high, limiting the use of the most accurate machine learning potentials. Thus, the development of more efficient, but still sufficiently accurate ML potentials for large scale simulations (length and time)  is highly desirable. 

In this work, we train ML potentials for elemental copper, aluminium and nickel, with the specific focus on large-scale simulations and far-from-equilibrium conditions such as those found in radiation damage events. The potentials are validated and ultimately used to study both the threshold displacement energies as well as single crystal uniaxial compression. For the compression tests, in contrast to most of the previous studies, we consider the simulation cells with the saturated level of defects. 
The latter were introduced by means of Frenkel pair insertion with subsequent annealing. This technique allows to obtain uniform distributions of defects of different kind, including point defects, defect clusters and dislocations. Uniaxial compressive loading is then performed along various crystallographic orientations with both the pristine cells and the cells with saturated level of radiation defects. 
 
\newpage

\section{Methods}
\label{sec:methods}

\subsection{Gaussian approximation potentials}

All potentials presented in this work are Gaussian approximation potentials (GAP) trained with different combinations of descriptors~\cite{bartok_gaussian_2010}. GAP is a ML framework based on Gaussian process regression and some combination of descriptors for the encoding of local atomic environments. The total energy of models used in this work are defined as:
\begin{equation}
E_{\mathrm{tot.}} = E_{\mathrm{rep.}} + E_{\mathrm{ML}},
\end{equation}
where the energy of the system is separated into an external repulsive potential, $E_{\mathrm{rep.}}$, and a machine-learned part, $E_{\mathrm{ML}}$, trained as a GAP. The repulsive part is included in order to more accurately describe short-range interactions, as shown in more detail in Ref.~\citenum{byggmastar_machine-learning_2019}. The repulsive part in our case is a Ziegler-Biersack-Littmark-type (ZBL~\cite{ziegler_stopping_1985}) repulsive potential:
\begin{equation}
E_{\mathrm{rep.}} = \sum_{i<j}^{N} \frac{1}{4 \pi \epsilon_0} \frac{Z_i Z_j e^2}{r_{ij}} \phi (r_{ij}/a) f_{\mathrm{cut}}(r_{ij}),
\end{equation}
where
\begin{equation}
    a = \frac{0.46848}{Z_i^{0.23} + Z_j^{0.23}}.
\end{equation}
The screening function $\phi (r_{ij}/a)$ is not the universal ZBL one, but refitted to repulsive dimer data from all-electron DFT calculations~\cite{Nor96c}. Additionally, the screened Coulomb potential is multiplied by a cutoff function to force it to zero well below the nearest-neighbour distance of each material, to avoid interfering with the near-equilibrium interactions described by the machine-learned part. The range of the cutoff was chosen to be $1.0-2.2$ Å to ensure a smooth transition between the all-electron calculations and short-range DFT data that was part of the training data. 

The ML part of the model can be further divided into two-body and many-body terms. For the models referred to in this work as GAP, the total energy is described using the following equation:

\begin{equation}
\begin{split}
E_\mathrm{GAP} = E_{\mathrm{rep.}} &+  \sum_{i<j}^N \delta^2_\mathrm{2b} \sum_s^{M_\mathrm{2b}} \alpha_{s} K_\mathrm{se} (r_{ij}, r_s)  \\  & + \sum_{i}^N \delta^2_\mathrm{SOAP} \sum_s^{M_\mathrm{SOAP}} \alpha_{s} K_\mathrm{SOAP} (\bm{q}_i, \bm{q}_s)
\end{split}
\label{eq:gap-soap}
\end{equation}

The two-body term consists of pre-factors $\delta^2_\mathrm{2b}$ and regression coefficients $\alpha_{s}$. Furthermore, it uses a squared exponential kernel $K_\mathrm{se}$ and a two-body descriptor that corresponds simply to the distance between two atoms. The many body-term uses to the widely known SOAP (smooth overlap of atomic positions) kernel and descriptor~\cite{bartok_representing_2013}. The inner sums in the equation run over sparsified subsets of the reference systems used for the training of the potential.

In addition to the GAP (with the SOAP descriptor), we created low-dimensional tabulated versions (tabGAP) of the ML potentials~\cite{byggmastar_simple_2022,byggmastar2022multiscale}. The total energy for the tabGAP is defined similarly as for the model described in Eq.~\ref{eq:gap-soap}, but with a different selection of descriptors:
\begin{equation}
\begin{split}
E_\mathrm{3b+eam} = E_{\mathrm{rep.}}  &+  \sum_{i<j}^N \delta^2_\mathrm{2b} \sum_s^{M_\mathrm{2b}} \alpha_{s} K_\mathrm{se} (r_{ij}, r_s)  \\ 
& + \sum_{i, j < k}^N \delta^2_\mathrm{3b} \sum_s^{M_\mathrm{3b}} \alpha_{s} K_\mathrm{se} (\bm{q}_{ijk}, \bm{q}_s) \\ 
& + \sum_{i}^N \delta^2_\mathrm{eam} \sum_s^{M_\mathrm{eam}} \alpha_{s} K_\mathrm{se} (\rho_i, \rho_s)
\end{split}
\label{eq:tabgap}    
\end{equation}
Whereas the model in equation (\ref{eq:gap-soap}) uses the SOAP descriptor, the tabGAP potential uses a three-body cluster descriptor (defined as a three-valued permutation-invariant vector~\cite{bartok_gaussian_2015}) and an embedded atom method (EAM) density descriptor. The EAM density is a scalar pairwise summed radial function, as in standard EAM potentials~\cite{byggmastar_simple_2022,daw_embedded-atom_1984,finnis_simple_1984}.

The tabGAP potential is trained similarly as the GAP potential, but after the initial training of the model, the energy contributions of the different terms of the potential are tabulated onto low-dimensional grids and evaluated using cubic-spline interpolation. The repulsive and two-body terms are tabulated into a one-dimensional grid, the three-body term into a three-dimensional grid and the EAM term into two one-dimensional grids. With dense enough grid sampling, the interpolation errors are negligible~\cite{byggmastar_modeling_2021}. Further details on the tabulation process can be found in Refs.~\citenum{byggmastar_simple_2022,byggmastar2022multiscale}. The simpler descriptors of the tabGAP model allows for this tabulation, which while sacrificing a little accuracy compared to SOAP-GAP gives the model a two-orders-of-magnitude increase in computational efficiency from the GAP, making the model more suitable for large-scale simulations. 

\subsection{Density functional theory calculations}\label{sec:dft-calc}

All the DFT calculations in this work were performed using the \textsc{VASP} DFT code~\cite{kresse_ab_1993,kresse_ab_1994,kresse_efficiency_1996,kresse_efficient_1996}. The calculations used the PBE GGA exchange-correlation function~\cite{perdew_generalized_1996} and projector augmented wave (PAW) pseudopotentials (\texttt{Cu\_pv}, \texttt{Al}, \texttt{Ni}). The plane-wave expansion energy cutoff was 500 eV for all elements. K-points were defined using $\Gamma$-centered Monkhorst–Pack grids~\cite{monkhorst_special_1976} with maximum k-spacing of 0.15 Å$^{-1}$. Additionally, first order Methfessel-Paxton smearing of 0.1 eV was applied~\cite{methfessel_high-precision_1989}. For Ni, collinear spin-polarized calculations with initialised ferromagnetic order were performed while the Cu and Al calculations did not include spin-polarization. Parameters such as energy cutoff, k-spacing and smearing were kept the same for all calculations and elements in order to allow for the possibility to combine data between elements for the creation of alloy potentials in the future.

\subsection{Training and testing data}
Like in most ML applications, the quality of the training data is critical for the quality of the final model. The training data should reflect the expected range of circumstances that can reasonably occur during the use of the model. The potentials presented in this work were designed with the goal of making a good general-purpose model with additional considerations for short-range repulsive interactions, which take place in far-from equilibrium simulations, such as radiation damage simulations. The training data consists of the following classes of structures:

\begin{itemize}
\setlength\itemsep{0.1em}
    \item Randomly distorted unit cells (FCC, BCC and HCP).
    \item FCC, BCC and HCP lattices at finite temperatures and different volumes, initially created by running MD simulations with earlier iterations of the final potentials.
    \item Systems containing various small vacancy and interstitial clusters (from one to three defects with different configurations). 
    \item Liquid systems with various densities prepared by melting MD simulations.
    \item \hkl(100), \hkl(110) and \hkl(111) FCC surfaces. 
    \item Structures with a few disordered surface layers.
    \item Stacking-fault-like structures where a small slab is moved on a atom plane randomly.
    \item Dimers and trimers with various distances between atoms.  
    \item Short-range systems, where an atom is randomly inserted into an FCC lattice fairly close to other atoms ($\geq$ 1 \AA). 
    \item Structures with trajectories of 
    a single atom being moved along a rigid path in various high-symmetry directions, while the other atoms remained fixed. 
\end{itemize}

The training datasets for the GAP and tabGAP were kept identical. However, there are slight differences between the different DFT databases for the different elements. The Al potentials do not include dimers as the inclusion of the dimer data degraded the accuracy and computed properties of the potentials. Likewise, for Cu and Ni trimers were omitted from their respective training databases, due to similar concerns. Before training, a random subset (5-10 \%) of the training data containing systems from each category was removed and put aside and used as testing datasets.

\vfill\eject

\subsection{Hyperparameters}

The GAPs described here have a number of hyperparameters that have to be set before training. Table~\ref{tab:hyper} shows the hyperparamaters used for the different descriptors. The cutoff distance of the three-body descriptor for Al was chosen to be longer, as Al has a significantly larger lattice constant. The results showed that, indeed the longer cutoff in case of Al, improved the model accuracy for this material. Additionally, the SOAP descriptor used 8 as the number of radial basis functions ($n_{\mathrm{max}}$) and number of spherical harmonic ($l_{\mathrm{max}}$). Another set of hyperparameters are the regularisation terms $\sigma$ for the energies, forces and virials. The default values for these parameters were 1 meV/atom, 0.04 eV Å$^{-1}$ and 0.1 eV, respectively. For systems such as liquid structures, dense structures, and short-range interactions (including trajectories) these regularisation terms were increased by a factor of ten ($10 \sigma$). The number of grid points used in the tabulation was chosen after convergence testing to be 5000 for both two-body and EAM descriptors and for the three-body grid to be $80 \times 80 \times 80$.

\begin{table}[]
    \centering
        \caption{Hyperparameters used for the different descriptors: cutoff radius $r_\mathrm{cut}$, width of the cutoff region $r_{\Delta \mathrm{cut}}$, energy prefactor $\delta$, and the number of sparse descriptor environments from the training structures $M$.}
    \begin{tabular}{lllllr}
         \toprule
         Formalism & Descriptor & $r_\mathrm{cut}$ (Å) & $r_{\Delta \mathrm{cut}}$ (Å) & $\delta$ & $M$ \\
         \midrule
         \multirow{3}{*}{tabGAP} & 2b & 5.2 & 1.0 & 10 & 20 \\
         & EAM & 5.2 & 1.0 & 1.0 & 20 \\
         & 3b & 4.0 (4.5 Al) & 0.6 & 1.0 & 700 \\
         \midrule
         
         \multirow{2}{*}{GAP} & 2b & 5.2 & 1.0 & 10 & 20 \\
         & SOAP & 4.5 & 1.0 & 2.0 & 2000 \\
         \bottomrule
    \end{tabular}
    \label{tab:hyper}
\end{table}

\subsection{Threshold displacement energy calculations}

All MD simulations in this work were done using the LAMMPS simulation package.~\cite{LAMMPS}.  The threshold displacement energies were determined by sampling 800 random spherically uniformly distributed lattice directions. A series of simulations were performed for each direction, where an atom is given higher and higher kinetic energy (2 eV increments) in the specific direction until a stable defect is formed. During this simulation, border cooling was applied ($NVT$) with a temperature of 10 K, while the rest of the system was kept in the $NVE$ ensemble. The simulation cell was chosen to be ($12 \times 13 \times 14$) units cells which corresponds to 8736 atoms.
A non-cubic simulation cell size was specifically chosen to reduce the possibility that replacement collision sequences along low-index crystal directions interact with themselves across the periodic boundaries.

\subsection{Uniaxial compressive loading}

For the uniaxial compressive loading simulations, cells in two crystal orientations were created: one with the \hkl<100> lattice orientation aligned with the $z$ axis and one with \hkl<111> alignment. Furthermore, for each direction, a simulation cell containing defects was created in addition to the pristine bulk cells. All side lengths of the simulation cells were roughly equal and they were around 22-27 nm in length depending on element and orientation. This corresponds to a system size of around 1.1 million atoms. These simulation cells were large enough for imaging effects from periodic boundary conditions to be negligible~\cite{tschopp2007atomistic}. The pristine systems were thermalized to 300 K in $NPT$ conditions for 60 ps. The simulation cells containing defects were created from the pristine cells by manually inserting around 10000 randomly placed Frenkel pairs (FPs) into the cell. The created structures were then minimized using the conjugate gradient method in order to ensure that atoms were not accidentally moved too close to each other during the insertion of FPs. The cells were then heated up from 100 K to 300 K in the $NPT$ ensemble for 200 ps after which the system was additionally relaxed for 400 ps at the 300 K temperature 
in the same $NPT$ ensemble. The final configurations were periodic in all directions and were used as starting points for the uniaxial loading simulations. The compression simulations were performed by uniaxial compressive loading of a given orientation with a constant strain rate of $1 \times 10^{8}$/s. This strain rate has been used in previous studies and shown to be sufficiently low with a negligible effect on the maximal yield stress~\cite{spearot2009orientation}. The directions orthogonal to the loading direction were decoupled from the latter and the temperature and pressure in these directions were controlled in the $NPT$ ensemble. The temperature of the system was kept at 300 K and a 0.001 ps timestep was used in all simulations. During compression the stresses were calculated from the per-atom stresses of all the atoms in the system and the atomic configurations were stored at regular intervals. This procedure was continued until the maximal strain of 20\% was reached.

\section{Results and discussion}

\subsection{Train and test errors}

Table~\ref{tab:Test_errors} shows the root-mean-square (RMS) errors of the testing data of each potential. Overall the errors show that the simpler descriptors of tabGAP only slightly reduces the accuracy in regards to energy and force errors compared to GAP with the SOAP descriptor. The largest difference between GAP and tabGAP is seen for Ni, but the RMS errors of the Ni tabGAP are still low, a few meV/atom, and hence acceptable. Furthermore, we divide here the testing data into crystalline and liquid structures in order to highlight the difference between them, as liquids and disordered structures are in general more difficult to describe. 

\begin{table}[ht!]
 \caption{Energy $E$ and force $F$ RMS errors of the testing data for crystalline and liquid structures separately.}
 \label{tab:Test_errors}
 \begin{threeparttable}
  \begin{tabular}{ll|cc|cc}
   \toprule
   & &  (meV/atom) & &  (eV/Å) &   \\
   & & $E_{\mathrm{crystalline}}$ & $E_{\mathrm{liquid}}$  & $F_{\mathrm{crystalline}}$ & $F_{\mathrm{liquid}}$ \\
   \bottomrule
    Cu & & & & & \\
    &GAP   &0.8 &3.0&0.015&0.06\\
    &tabGAP&1.1 &2.0&0.015&0.03\\
        Al & & &  & &\\
    &GAP   &0.6 &2.4&0.014&0.06\\
    &tabGAP&1.1 &3.0&0.02&0.06\\
        Ni & &  & & &  \\
     &GAP  &0.5 &1.9&0.012&0.1\\
    &tabGAP&1.6 &3.9&0.07&0.12\\
   \bottomrule
  \end{tabular}
 \end{threeparttable}
\end{table}

\vfill\eject

\subsection{Computational efficiency}

Computational efficiency was measured by running simulations containing 32\,000 atoms with the different interatomic potentials. All simulations were done on a single Xeon Gold 6230 CPU core. Table \ref{tab:comp} shows the benchmarking results of different potentials (smaller number equals to a more computationally efficient potential). We can clearly see that tabGAP is about 80-150 times faster than its GAP counterpart and about 17 times faster than the recent performant implementation of the atomic cluster expansion model~\cite{lysogorskiy_performant_2021}. Furthermore, the tabGAP potential is only an order of magnitude slower than classical EAM potentials, which is a reasonable increase of computational costs given the higher accuracy and the consistency with DFT calculations of the ML models compared to the EAM ones.

\begin{table}[ht!]
 \caption{Benchmark of computational efficiency. Computational efficiency given as 1000 $\times$ seconds/(number of atoms $\times$ time steps). All simulations are run on a single CPU core with a system containing 32\,000 atoms.}
 \label{tab:comp}
 \begin{threeparttable}
  \begin{tabular}{llrrrcr}
   \toprule
   && GAP-SOAP  & tabGAP & & EAM  & PACE\\
   \midrule
    Cu & & 3.7 & 0.029 & & 0.0014\tnote{a} & 0.51\tnote{b} \\
    Al & & 2.2 & 0.028 & &0.002\tnote{c} & - \\  
    Ni & & 4.3 & 0.029 & &0.0036\tnote{d}, 0.002\tnote{e} & -   \\
   \bottomrule
  \end{tabular}
  \begin{tablenotes}
  \centering
    \item [a] (Mishin Cu)\cite{mishin2001structural} 
    \item [b] (PACE) \cite{lysogorskiy_performant_2021}
    \item [c] (Mishin Al)\cite{PhysRevB.59.3393}
    \item [d] (Stoller)\cite{stoller2016impact}
    \item [e] (Bonny)\cite{bonny2013interatomic} 
  \end{tablenotes}
 \end{threeparttable}
\end{table}

\subsection{Basic material properties}\label{sec:basic}

To verify the accuracy of the developed potentials, we calculated different material properties. Table~\ref{tab:bulk} shows the good agreement with the available DFT and experimental data of the basic structural, elastic and defect properties obtained with the tabGAP and GAP. Generally, we observe minor differences between the GAP and tabGAP potentials, mainly in the formation energies of defects and elastic constants. The Cu and Al potentials slightly underestimate the formation energies of defects compared to the DFT calculations, while the Ni potentials slightly overestimate them. The surface energies show excellent agreement with the DFT calculations~\cite{tran2016surface}. Furthermore, the agreement with experimental surface energies is good, while it must be noted that experimental values are not for any specific direction and that there is some spread in experimental values~\cite{kumikov1983measurement}. The melting point was estimated using the two-phase method~\cite{Mor94} with 20 K temperature increments. It should be noted that in general, the ML potentials slightly underestimate the experimental melting temperatures~\cite{haynes_crc_2015}.

\begin{table}
 \caption{Calculated material properties compared with DFT and experimental results. $a$: Lattice constant (Å), $E_\mathrm{coh}$: Cohesion energy (eV), $E_\mathrm{\Delta_\mathrm{lattice}}$: Energy difference between relaxed structure of given type compared with FCC (meV/atom), $E_\mathrm{mig}^{v}$: Vacancy migration energy in FCC (eV), $C_{ij}$: Elastic constants (GPa), $E_S(hjk)$: Surface energies($mJ/\mathrm{m}^2$), $E_\mathrm{f}$: Formation energy of a single vacancy and interstitials (eV), $T_{\mathrm{melt}}$: Melting temperature (K).}
 \label{tab:bulk}
 \begin{threeparttable}
  \begin{tabular}{ll|rrrr}
   \toprule
   && GAP  & tabGAP  & DFT  & Expt.\\
   \midrule
    Cu & & & & &\\
    &$a$  & 3.626 & 3.625 & 3.632 & 3.615\tnote{a} \\
    &$E_\mathrm{coh}$  & 3.724 & 3.724 & 3.724 & 3.49\tnote{b}  \\
    &$E_\mathrm{\Delta_\mathrm{bcc}}$  & 38.2 & 37.9 & 36.3 & - \\
    &$E_\mathrm{\Delta_\mathrm{hcp}}$   & 7.9 & 8.0 & 7.5  & - \\
    &$E_\mathrm{mig}^{v}$  & 0.742 & 0.780  & 0.79\tnote{c} &  0.70\tnote{d} \\
    &$C_{11}$  & 163 & 168 & 166 & 169\tnote{a} \\ 
    &$C_{12}$ & 122 & 138 & 119 & 122\tnote{a} \\ 
    &$C_{44}$  & 77  & 81  & 76 & 75\tnote{a} \\ 
    &$E_{S(100)}$ & 1470 & 1486 & 1470\tnote{e} & 1520\tnote{f,*} \\ 
    &$E_{S(110)}$ & 1556 & 1558 & 1560\tnote{e}  & - - \\  
    &$E_{S(111)}$ & 1321 & 1326 & 1340\tnote{e} & - - \\ 
    &$E^\mathrm{vac}_\mathrm{f}$  & 0.92 & 1.03 & 1.08 & 1.28 $\pm$ 0.05\tnote{a}  \\
    &$E^\mathrm{100d}_\mathrm{f}$ & 2.95 & 3.04 & 3.22 & 2.82--4.12\tnote{a} \\    
    &$E^\mathrm{octa}_\mathrm{f}$ & 3.36 & 3.36 & 3.51 & - - \\
    &$E^\mathrm{tetra}_\mathrm{f}$  & 3.56 & 3.59 & 3.90 &  - -\\
    &$T_{\mathrm{melt}}$ & 1270  & 1250  & - & 1356\tnote{g}\\
        Al & & & & &\\
    &$a$  & 4.040 & 4.041 & 4.043 & 4.050\tnote{a} \\
    &$E_\mathrm{coh}$  & 3.695 & 3.695 & 3.693 & 3.39\tnote{b} \\
    &$E_\mathrm{\Delta_\mathrm{bcc}}$  & 93.1 & 93.3 & 88.7 & - \\
    &$E_\mathrm{\Delta_\mathrm{hcp}}$  & 28.5 & 28.7  & 29.2  & - \\
    &$E_\mathrm{mig}^{v}$ & 0.527 & 0.534  &  0.30–0.63\tnote{h} & 0.61\tnote{d} \\ 
    &$C_{11}$  & 105 & 101 & 101 & 108\tnote{a}  \\ 
    &$C_{12}$  & 63  & 70 & 66 & 62\tnote{a} \\ 
    &$C_{44}$  & 32  & 37 & 29 & 28\tnote{a} \\ 
    &$E_{S(100)}$ & 900 & 897 & 910\tnote{e} &  1140\tnote{f,*} \\ 
    &$E_{S(110)}$ & 985 & 975 & 980\tnote{e}  & - - \\  
    &$E_{S(111)}$ & 772 & 810 & 770\tnote{e} & - - \\ 
    &$E^\mathrm{vac}_\mathrm{f}$  & 0.67 & 0.66 & 0.68  &0.67 $\pm$ 0.03\tnote{a} \\
    &$E^\mathrm{100d}_\mathrm{f}$ & 2.63 & 2.66 & 2.66 &3.0\tnote{a} \\
    &$E^\mathrm{octa}_\mathrm{f}$ & 2.83 & 2.84 & 2.88 & - - \\
    &$E^\mathrm{tetra}_\mathrm{f}$  & 3.16 & 3.15 &3.23 & - -\\
    &$T_{\mathrm{melt}}$ & 910  & 870  & - & 933\tnote{g}
    \\
        Ni & & & & &\\
    &$a$  & 3.518 & 3.518 & 3.519 & 3.524\tnote{a}\\
    &$E_\mathrm{coh}$  & 4.715 & 4.715 &  4.713 & 4.44\tnote{b} \\
    &$E_\mathrm{\Delta_\mathrm{bcc}}$ & 94.6 & 94.7 & 98.3 & - \\
    &$E_\mathrm{\Delta_\mathrm{hcp}}$  & 27.8 & 27.6  & 31.9 & -  \\
    &$E_\mathrm{mig}^{v}$  & 1.0 & 1.06  & 1.12\tnote{c} &  1.04\tnote{d}\\
    &$C_{11}$  & 279 & 281 &273 & 247\tnote{a} \\ 
    &$C_{12}$  & 153 & 156 &155 & 153\tnote{a} \\ 
    &$C_{44}$  & 125 & 125 &131 & 122\tnote{a}\\ 
    &$E_{S(100)}$ & 2213 & 2208 & 2210\tnote{e} & 1940\tnote{f,*} \\ 
    &$E_{S(110)}$ & 2265 & 2272 & 2290\tnote{e} & - - \\  
    &$E_{S(111)}$ & 1910 & 1910 & 1920\tnote{e} & - - \\ 
    &$E^\mathrm{vac}_\mathrm{f}$  & 1.46 & 1.47 & 1.49 & 1.79 $\pm$ 0.05\tnote{a} \\
    &$E^\mathrm{100d}_\mathrm{f}$ & 4.13 & 4.13 & 4.07 & -  \\
    &$E^\mathrm{octa}_\mathrm{f}$ & 4.36 & 4.35 & 4.26  & -\\
    &$E^\mathrm{tetra}_\mathrm{f}$ & 4.73 & 4.73 & 4.67 & - \\
    &$T_{\mathrm{melt}}$ & 1690 & 1690 & -  &1726\tnote{g}\\ 
   \bottomrule
  \end{tabular}
  \begin{tablenotes}[flushleft]
   \item[a] \cite{ma_nonuniversal_2021}
   \item[b] \cite{Kittel} 
   \item[c] \cite{zuo_performance_2020}
   \item[d] \cite{LandoltBornstein1991:sm_lbs_978-3-540-48128-7_61,LandoltBornstein1991:sm_lbs_978-3-540-48128-7_58, LandoltBornstein1991:sm_lbs_978-3-540-48128-7_63}
   \item[e] \cite{tran2016surface} 
   \item[f] \cite{kumikov1983measurement}
   \item[g] \cite{haynes_crc_2015}
   \item[h] \cite{qiu2017energetics}
   \footnotesize
   \item[*] Not specific direction. 
  \end{tablenotes}
 \end{threeparttable}
\end{table}

\begin{figure}[ht!]
    \centering
    \includegraphics[width=0.9\linewidth]{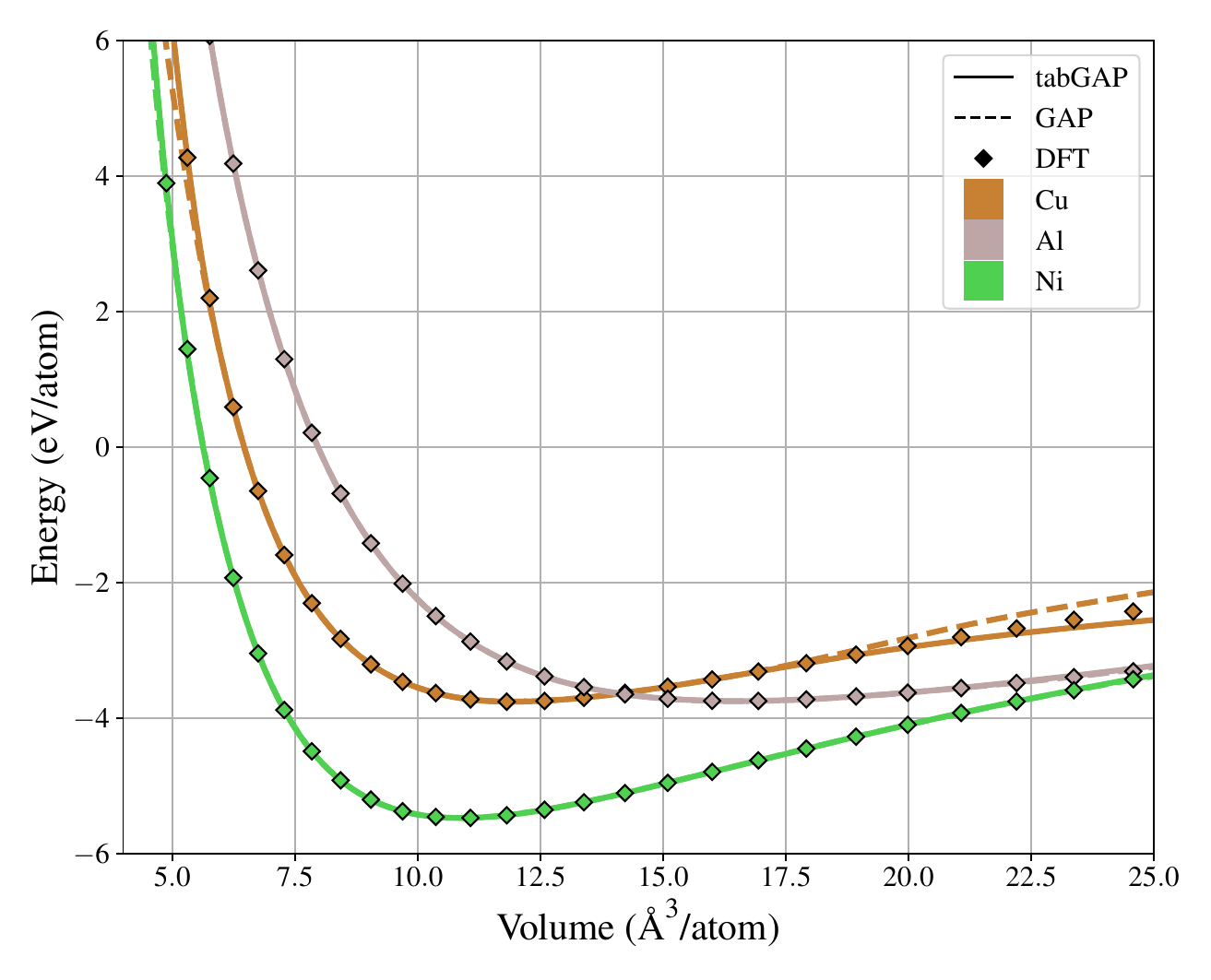}
    \caption{Energy vs. volume compared with DFT calculations for all the potentials in FCC structure.}
    \label{fig:Eng-Vol}
\end{figure}

In Fig.~\ref{fig:Eng-Vol} the energy-volume curves are shown for each of the elements.
Here we see an excellent agreement with DFT calculations in a wide range of atomic volumes. In particular, we are interested to validate that the potentials reproduce the repulsive interactions correctly and that there is a smooth transition to the external repulsive potential. The external repulsive potential is added to describe the interactions at 
very short interatomic distances. However, undesirable issues can still occur in the transition region, where the repulsive potential starts to dominate, especially if the short-range interactions are not properly taken into account. In the case of the Cu GAP potential, we can see some
deviations that appear at larger volumes, which are analogous to those reported for other GAP potentials~\cite{lysogorskiy_performant_2021}. The energy-volume curves for BCC and HCP phases of the materials under the study can be found in Supplementary fig. S1 and the accuracy of those are essentially the same as for the FCC case. 

\begin{figure*}
    \centering
    \includegraphics[width=0.32\linewidth]{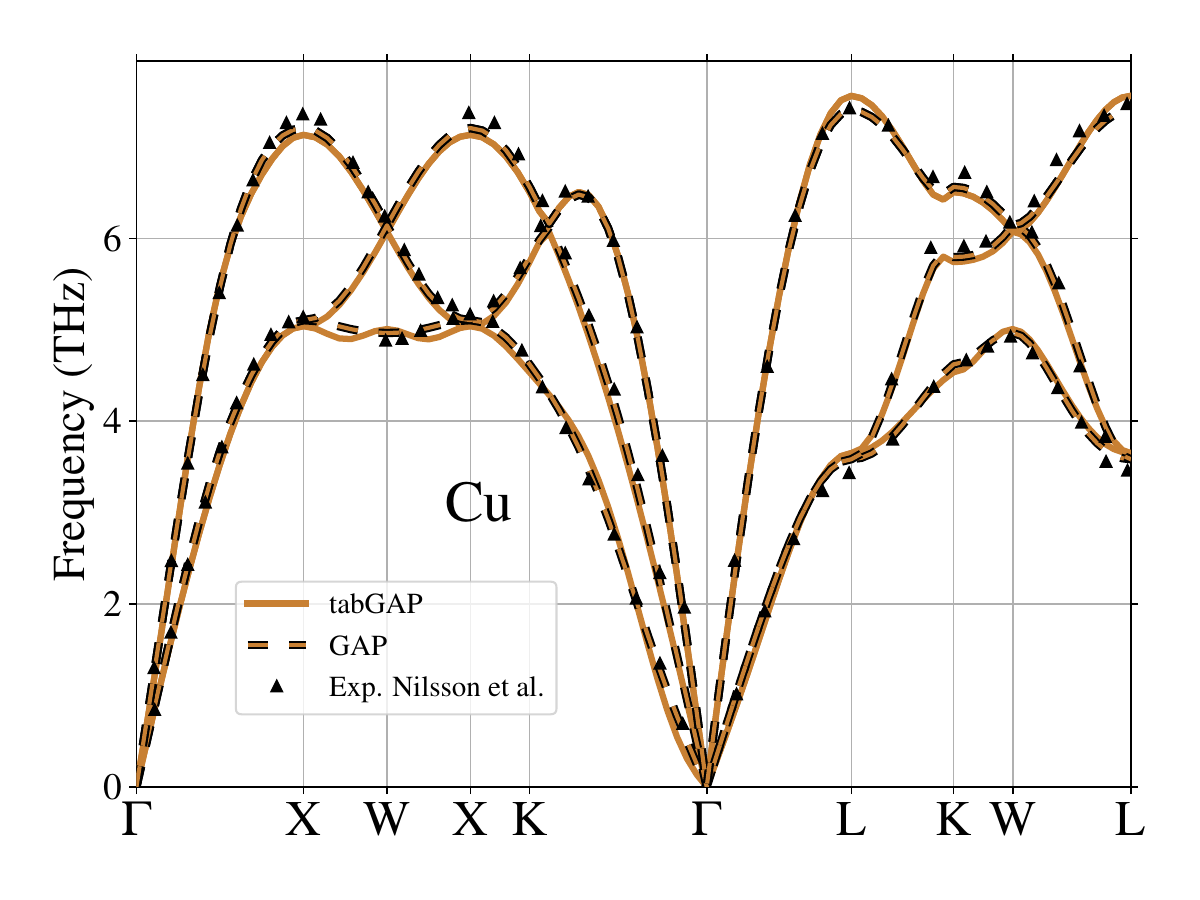}
    \includegraphics[width=0.32\linewidth]{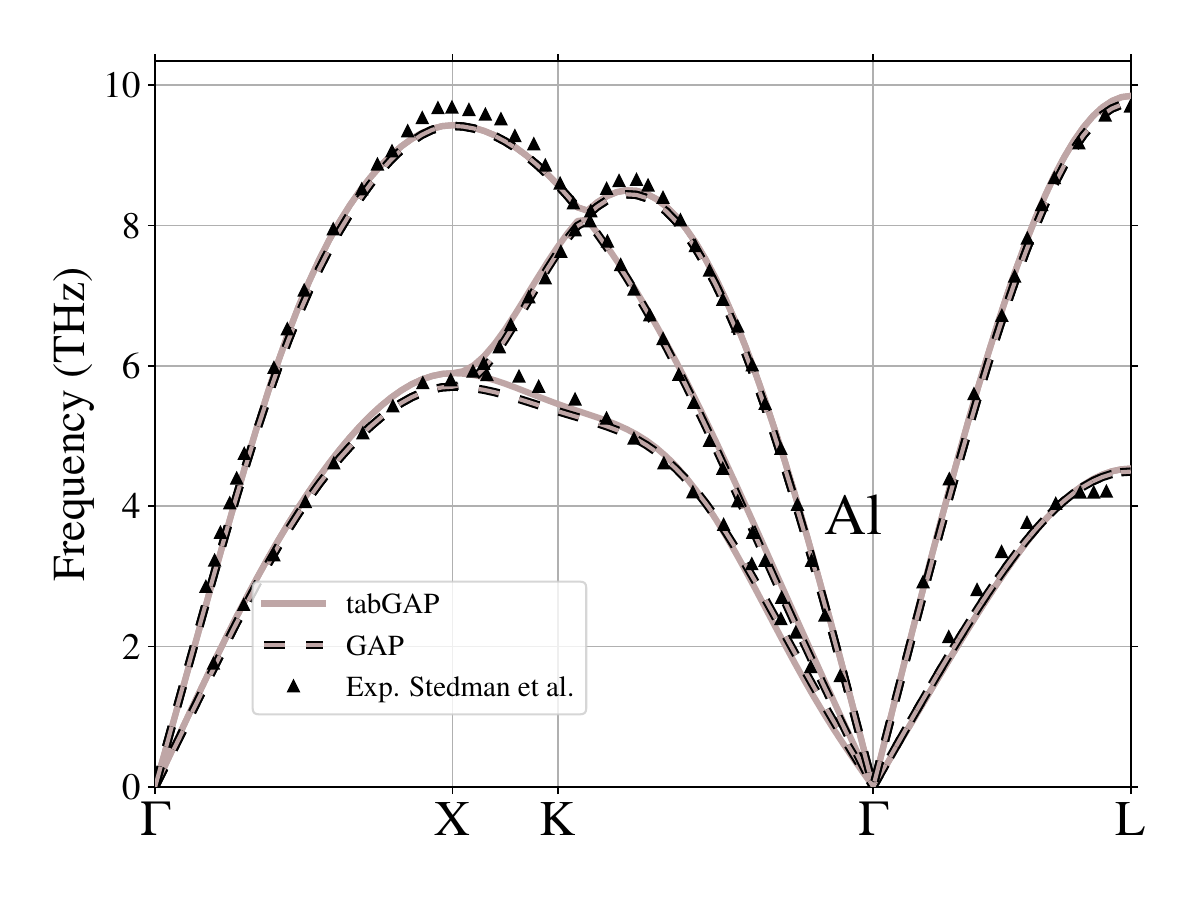}
    \includegraphics[width=0.32\linewidth]{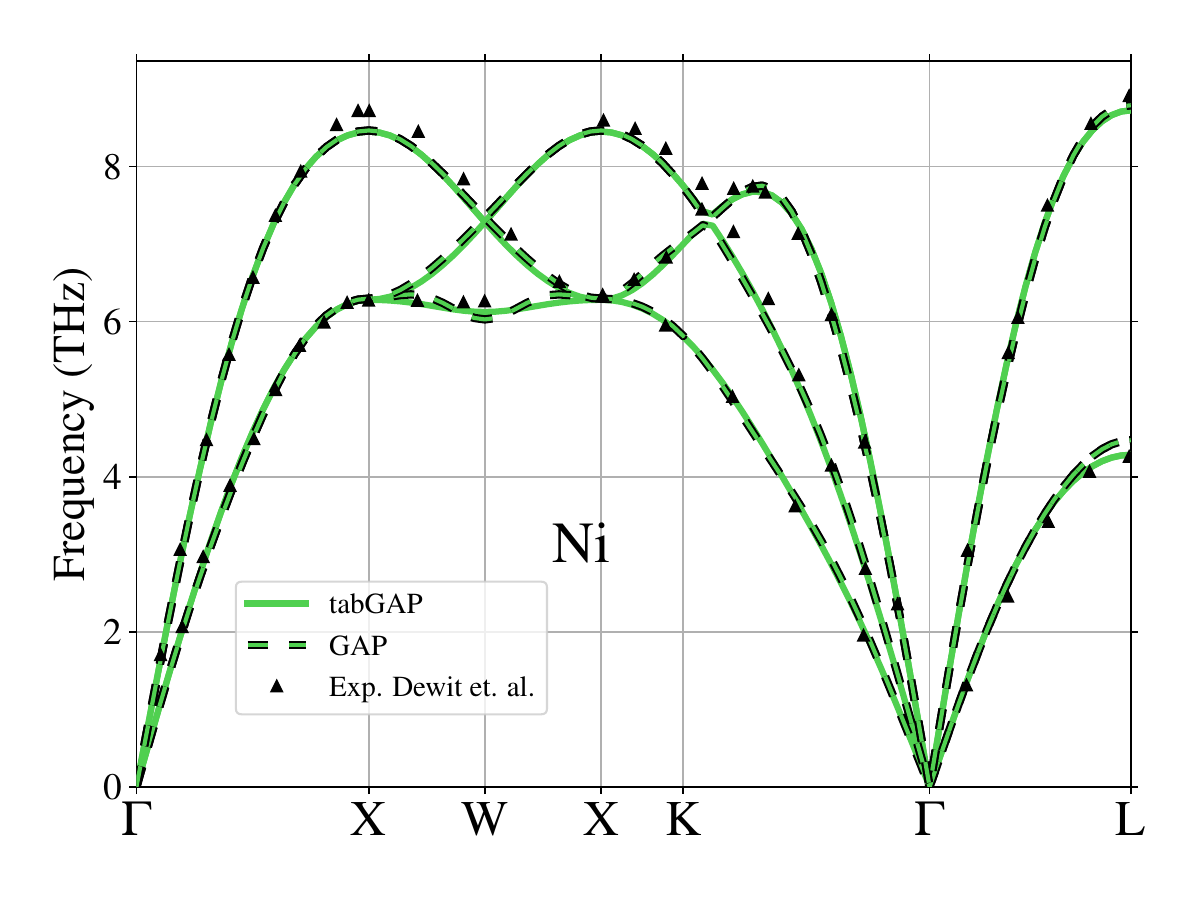}
    \caption{Phonon dispersion curves calculated using the small displacement method compared to experimental results~\cite{nilsson1973lattice,stedman1966dispersion,dewit1968lattice}.}
    \label{fig:phonon}
\end{figure*}

We also calculated the phonon dispersion curves for each element in the FCC phase and compared with the available experimental measurements, as shown in Fig.~\ref{fig:phonon}. The agreement between GAP, tabGAP, and experiment is excellent for all three materials. This further confirms that the developed potentials can accurately capture elastic and thermal properties of the materials.

\begin{figure}[ht!]
    \centering
    \includegraphics[width=0.95\linewidth]{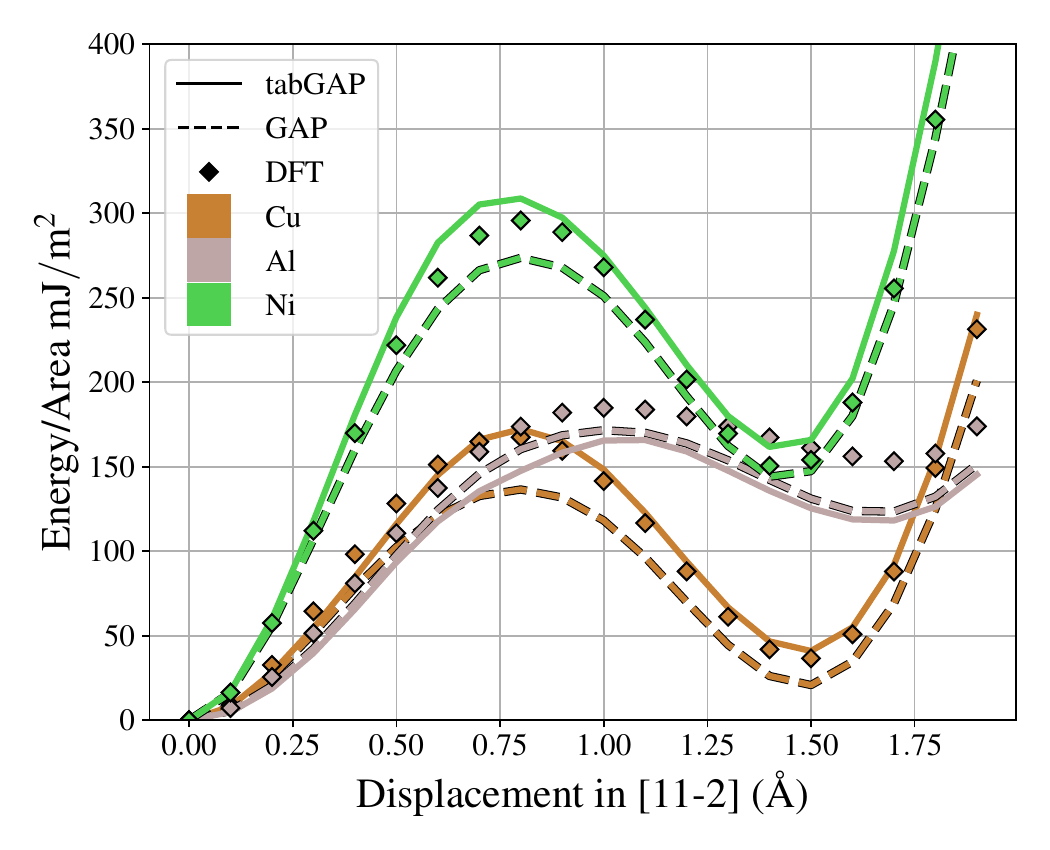}
    \caption{Generalised stacking fault energy curves calculated with the developed ML potentials and DFT.}
    \label{fig:GSFE}
\end{figure}

The generalised stacking fault energy (GSFE) curves shown in Fig.~\ref{fig:GSFE}, were calculated by 
displacing one of the \hkl{111} slip planes 
in a \hkl<112> direction. The DFT results were calculated for the same systems as with the ML potentials, using the parameters described in section~\ref{sec:dft-calc} without relaxation. A good description of stacking faults is important for the simulation of dislocations and grain boundaries. We also observe a good agreement between the potentials and the DFT results. The GAP potentials predict consistently lower stacking fault energies compared to tabGAP and DFT. Somewhat worse agrement is observed for Al, where both GAP and tabGAP underestimate the GFSE value compared to the DFT calculations. Table~\ref{tab:GSFE} shows the stable stacking fault (marked with the subscript "sf") and unstable stacking fault (marked with the subscript "usf") energies compared to the DFT and experiments. The unstable stacking fault energies of Ni and Cu in GAP are somewhat lower in comparison to those in tabGAP and DFT.

Since the accuracy of the reproduced key properties of materials is either very similar or even somewhat better with tabGAP than that with GAP, and also considering the much lower computational cost of the simulations with tabGAP, hereafter we only calculate and include results obtained with the tabGAPs.

\begin{table*}
 \caption{Simulated and experimental values for intrinsic stacking fault energies ($\gamma_{\mathrm{sf}}$) and unstable stacking fault energies ($\gamma_{\mathrm{usf}}$) for the created potentials.}
 \label{tab:GSFE}
 \begin{threeparttable}
  \begin{tabularx}{\linewidth}{@{\extracolsep{\fill}} llrrrr }
   \toprule
   && GAP-SOAP  & tabGAP  & DFT  & Expt.\\
   \midrule
    Cu & & & & &\\
    &$\gamma_{\mathrm{sf}}$ (mJ/m$^2$) & 20 & 40 & 35\tnote{*}, 39\tnote{a}, 39\tnote{b}, 41\tnote{c} & 41\tnote{h}, 45\tnote{d,g} \\
    &$\gamma_{\mathrm{usf}}$ (mJ/m$^2$) & 136 & 170 & 167\tnote{*}, 158\tnote{b}, 164\tnote{a}, 180\tnote{c} &  \\   
    \\
        Al & & & & &\\
    &$\gamma_{\mathrm{sf}}$ (mJ/m$^2$) & 122 & 117 & 153\tnote{*}, 134\tnote{c}140\tnote{a}, 158\tnote{b}, & 120\tnote{e},150\tnote{f}, 166\tnote{g} \\
    &$\gamma_{\mathrm{usf}}$ (mJ/m$^2$) & 168 & 158 & 182\tnote{*}, 169\tnote{c}, 177\tnote{a} 175\tnote{b} &  \\  
    \\
        Ni & & & & &\\
    &$\gamma_{\mathrm{sf}}$ (mJ/m$^2$) & 145 & 161 & 150\tnote{*}, 137\tnote{i}, 145\tnote{a} & 125\tnote{g}  \\
    &$\gamma_{\mathrm{usf}}$ (mJ/m$^2$) & 274 & 308 & 296\tnote{*}, 278\tnote{i} 289\tnote{a} & \\  
   \bottomrule
  \end{tabularx}  
  \begin{tablenotes}
   \item[*] \text{This work} 
   \item[a] \cite{hunter2014core}
   \item[b] \cite{ogata2002ideal} 
   \item[c] \cite{jahnatek2009shear}
   \item[d] \cite{carter1977stacking} 
   \item[e] \cite{rautioaho1982interatomic} 
   \item[f] \cite{mills1989study}
   \item[g] \cite{anderson2017theory}
   \item[h] \cite{stobbs1971weak} 
   \item[i] \cite{brandl2007general}
  \end{tablenotes}
 \end{threeparttable}
\end{table*}

\vfill\eject

\subsection{Equation of state}

\begin{figure}
    \centering
    \includegraphics[width=0.95\linewidth]{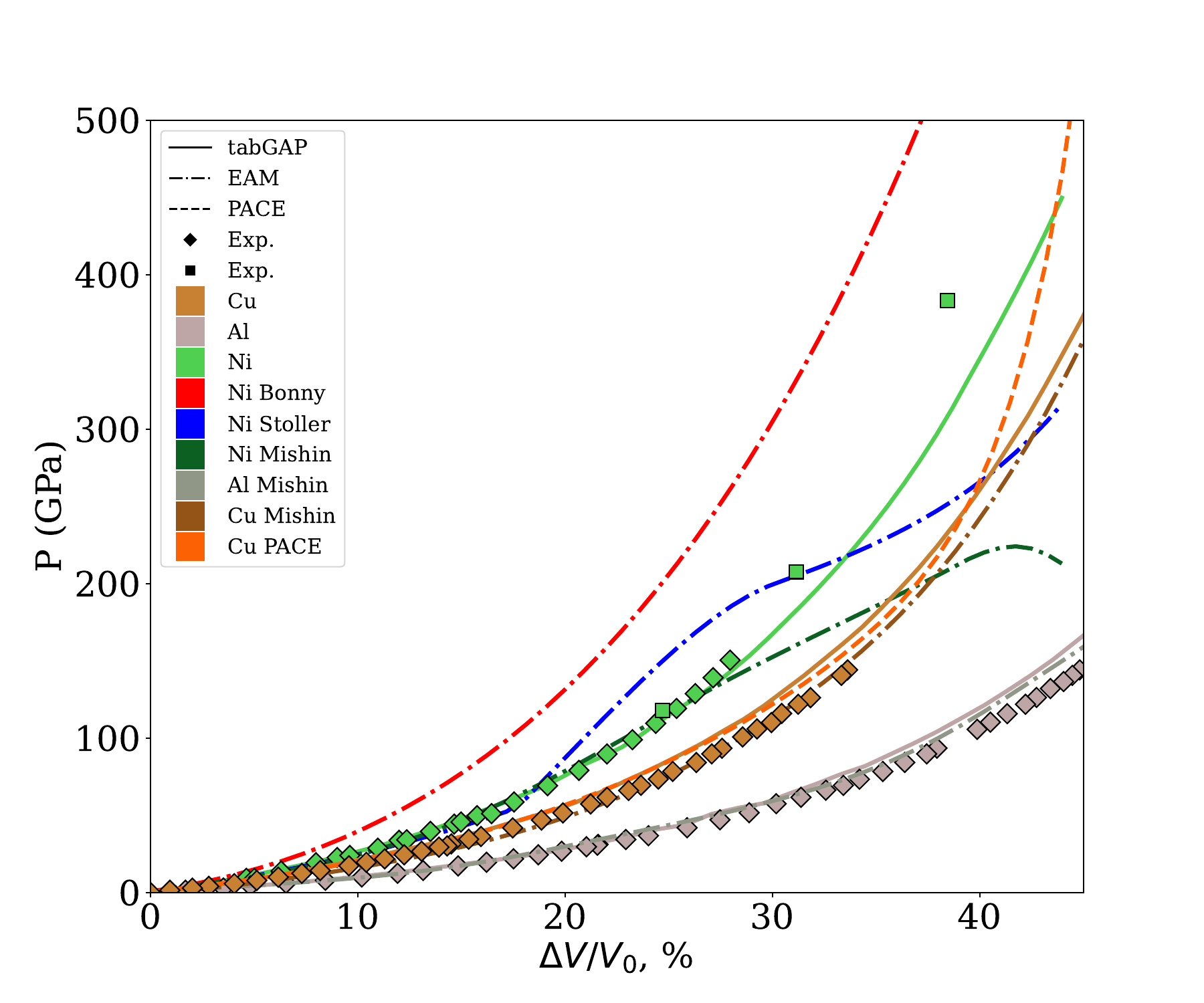}
    \caption{Equation of state calculated with various different interatomic potentials compared to experimental data (Ni~\cite{Ant90}, Cu and Al~\cite{PhysRevB.70.094112}). The different markers for experimental data for Ni are from different experiments~\cite{RICE19581,al1962shock}. Comparison is made to previously developed potentials: EAM~\cite{mishin2001structural,PhysRevB.59.3393,stoller2016impact,bonny2013interatomic}, PACE~\cite{lysogorskiy_performant_2021}. }
    \label{fig:EOS}
\end{figure}

Further validation of the developed potentials versus experiment can be done by comparing the equation of states, as measured in experiment and simulated using the potentials. Fig.~\ref{fig:EOS} shows the pressure-volume relation calculated with various potentials compared with experimental data. From the figure we see that the tabGAPs can reproduce very well the experimental results. Although the agreement with experiment is also good for the potentials previously developed for Al and Cu, the results for Ni in different classical EAM potentials agree to much lesser extent with experiment, especially for higher compression. The good agreement of tabGAP results with experiment for this material is encouraging, since an accurate response to pressure is important for extreme environment simulations, such as the uniaxial compressive loading presented in Sec.~\ref{sec:bulkcomp}.

\subsection{Threshold displacement energies}

\begin{figure*}[t]
    \centering
    \includegraphics[width=0.325\linewidth]{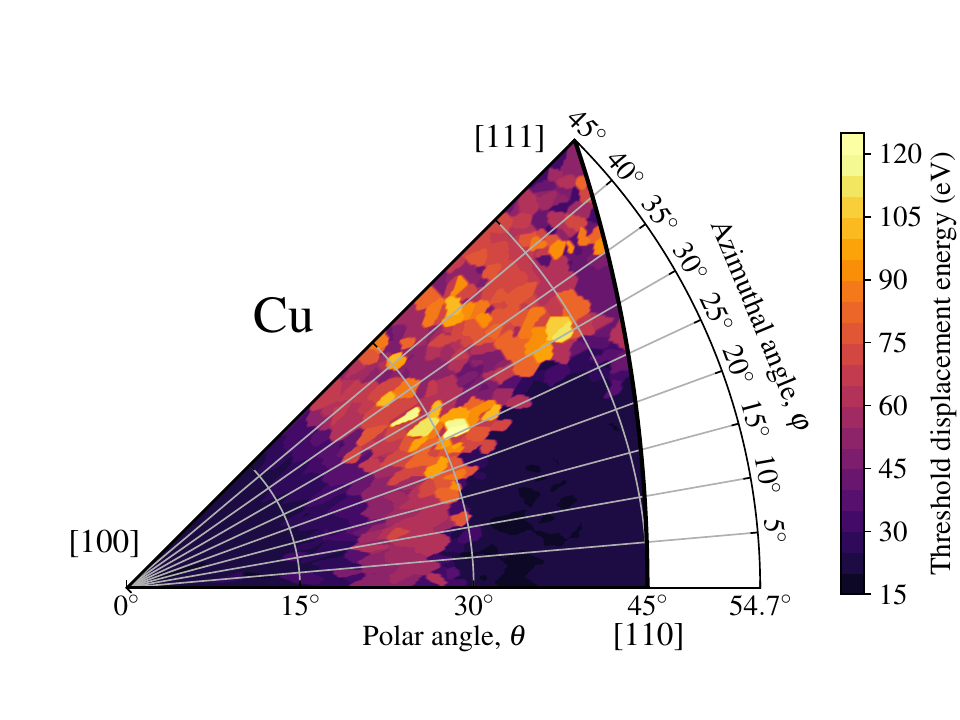}
    \includegraphics[width=0.325\linewidth]{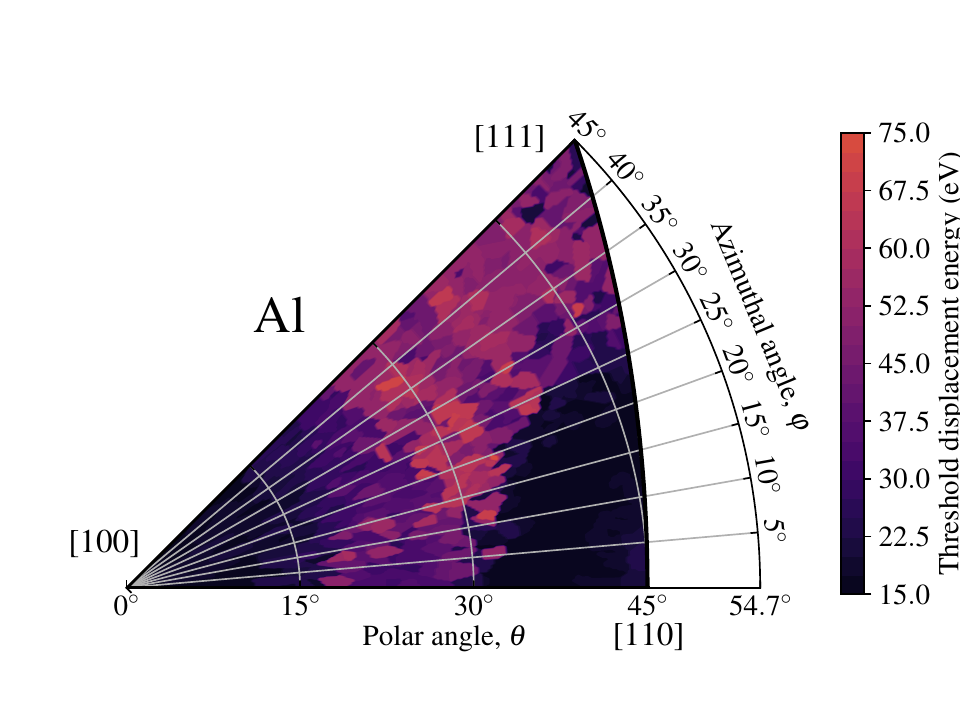}
    \includegraphics[width=0.325\linewidth]{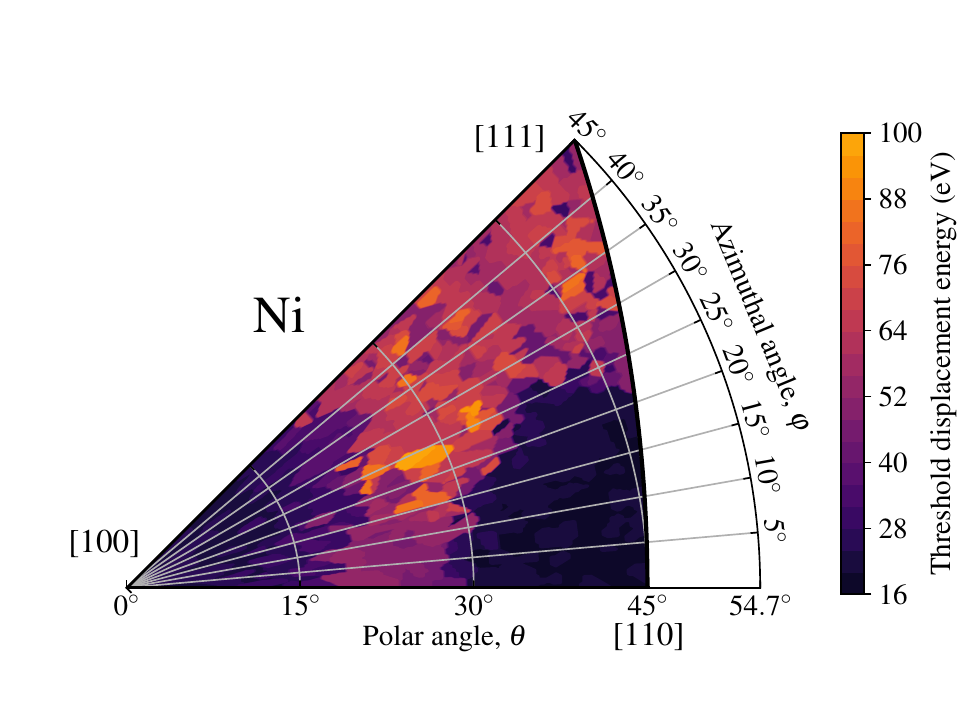}
    \caption{Threshold displacement energy as a function of PKA direction. The same color range (10--120 eV) is used for all materials.}
    \label{fig:tde}
\end{figure*}

The threshold displacement energy (TDE) defines the minimum energy required to displace an atom from its lattice position, creating a stable defect. This is a key parameter for experimental damage dose estimation when using the NRT equation~\cite{nordlund_primary_2018}. Moreover, the accuracy of the TDE values predicted by the interatomic potential is important 
for radiation damage simulations 
since they correlate directly with the amount of defects produced in cascade events~\cite{nordlund2018improving, NORGETT197550}. We have calculated the TDEs as a function of lattice direction for each element with the tabGAP potentials, see Fig.~\ref{fig:tde}. Table~\ref{tab:TDE} shows the TDEs for some high-symmetry directions of the lattice as well as the average TDE over all directions. The TDEs predicted for the high-symmetry directions are in good agreement with experimental results. Comparison with the ab-initio MD results~\cite{yang2021full} shows that the tabGAP predicted TDE values are consistently lower, but closer to experiment than the ab-initio MD values. The Al tabGAP potential has a somewhat higher average TDE when comparing with the previous MD results (17--27 eV) for both EAM and MEAM potentials and a more recent deep-learning potential~\cite{wang2019deep}. The ASTM recommends as the effective TDEs 30, 25 and 40 eV for Cu, Al and Ni, respectively~\cite{ASTM}. The calculated averages for the tabGAP potentials are consistently higher than the ASTM recommended values, while still in fairly good agreement. 

\begin{table}
 \caption{Threshold displacement energy (eV) for high symmetry directions ($\pm 1$ eV) and the average (with standard errors). Values given inside parentheses are the averages of the data points within a five degree cone from the specific direction.}
 \label{tab:TDE}
 \begin{threeparttable}
  \begin{tabular}{llrr|c|rr}
   \toprule
   & & tabGAP &  & AIMD & & Expt.\\
   \midrule
    Cu & & & & & &19\tnote{c}\\
   \hkl<100>& & 19 (20.3)  &  & 25\tnote{a} & & 19 \hkl<100>\tnote{b}  \\
   \hkl<110>& & 21 (22.4) & & &  & 19 \hkl<110>\tnote{b}  \\
   \hkl<111>& & 31 (52.2) & & &  &\\
    (avg.)& &  44.4 $\pm$ 0.8 &&&&29\tnote{d}, 43 $\pm 4$\tnote{c}
    \\
    \midrule
        Al & & & & & &16\tnote{c}\\
    \hkl<100>& & 15 (17) & &  19\tnote{a} & &  \\
    \hkl<110>& & 17 (19)& & & &\\
    \hkl<111>& & 19 (37.5) & & & &\\
    (avg.)& & 34.3 $\pm$ 0.6  &&&&27\tnote{d}, 66 $\pm 12$\tnote{c}
    \\
    \midrule
        Ni & & & & & & 23\tnote{c} \\
    \hkl<100>& & 21 (23.3) & &  27\tnote{a} & &  \\
    \hkl<110>& & 19 (19.5) & & 30\tnote{a} &  & 21 \hkl<110>\tnote{b}\\
    \hkl<111>& & 29 (58.6) & &70\tnote{a} &  &\\
    (avg.)& & 42.9 $\pm$ 0.7  &&&&33\tnote{d}, 69\tnote{c} \\
   \bottomrule
  \end{tabular}
  \begin{tablenotes}
    \item[a] \cite{yang2021full}  
   \item[b] \cite{vajda1977anisotropy}
   \item[c] \cite{jung1981average}
   \item[d] \cite{Luc75}
  \end{tablenotes}
 \end{threeparttable}
\end{table}

\subsection{Frenkel pair insertion}

To verify the stability of the potential as well as investigate the microstructure of defects predicted by the tabGAPs, we now study the dynamics of the system with high concentration of vacancies and interstitials via insertion of large numbers of randomly created Frenkel pairs, which was followed by MD relaxation runs. During the simulations, the number of vacancies was tracked using the Wigner-Seitz (WS) analysis~\cite{stukowski_visualization_2010}. Additionally, DXA dislocation analysis~\cite{0965-0393-18-8-085001} as implemented in OVITO~\cite{stukowski2012automated} was performed to find the types of dislocations present in the simulation cell after relaxation.  

\begin{figure}
    \centering
    \includegraphics[width=0.95\linewidth]{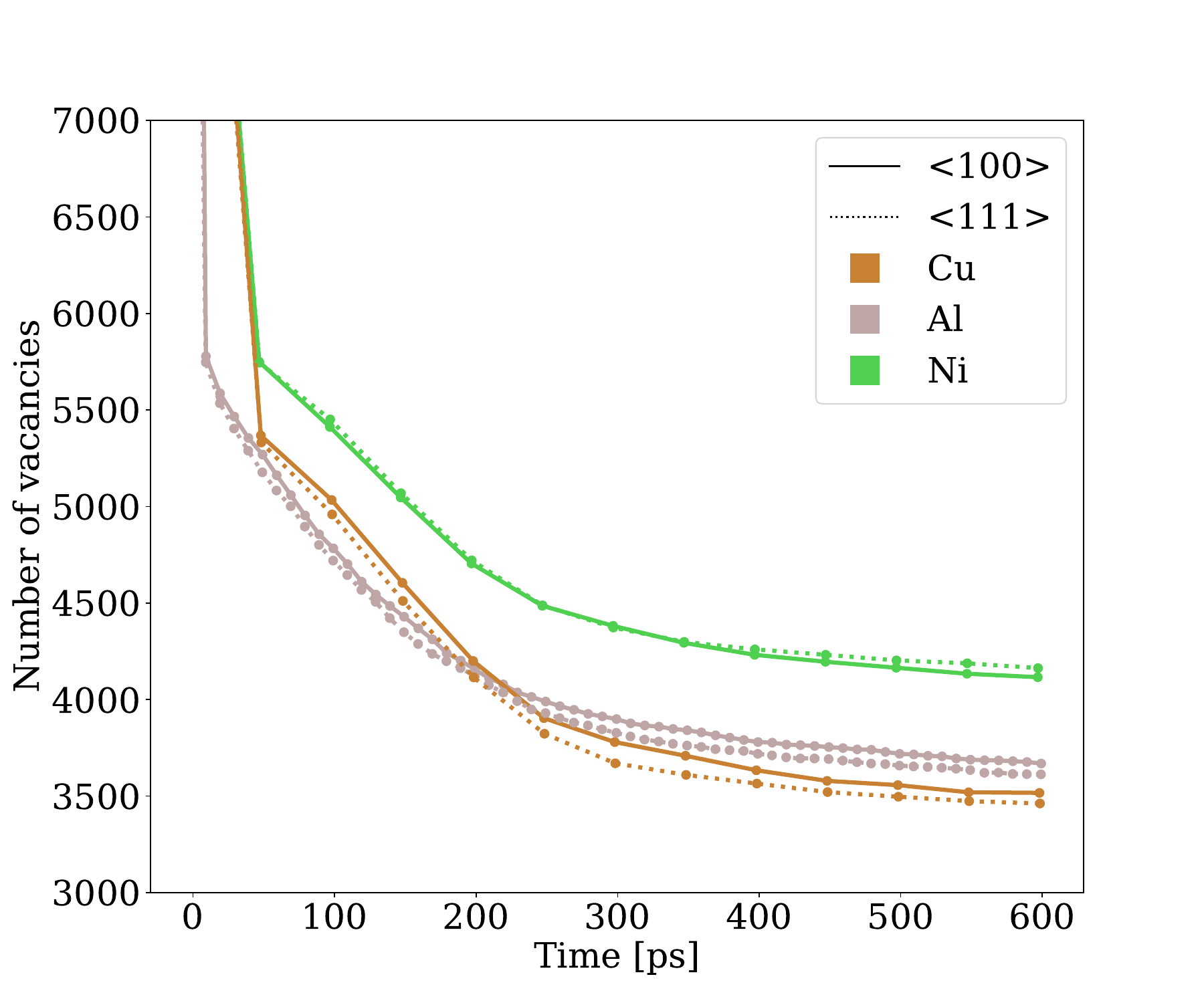}
    \caption{Number of vacancies as a function of time during relaxation of inserted Frenkel pairs. The number of FPs in the initial configurations was of the order of 10000. The two differently oriented simulation cells (\hkl<100> and \hkl<111>) are depicted in the figure.}
    \label{fig:FP-vac}
\end{figure}

\begin{figure*}
    \centering
\begin{subfigure}[t]{0.3\textwidth}
\label{fig:final:Cu}
    \includegraphics[width=\textwidth]{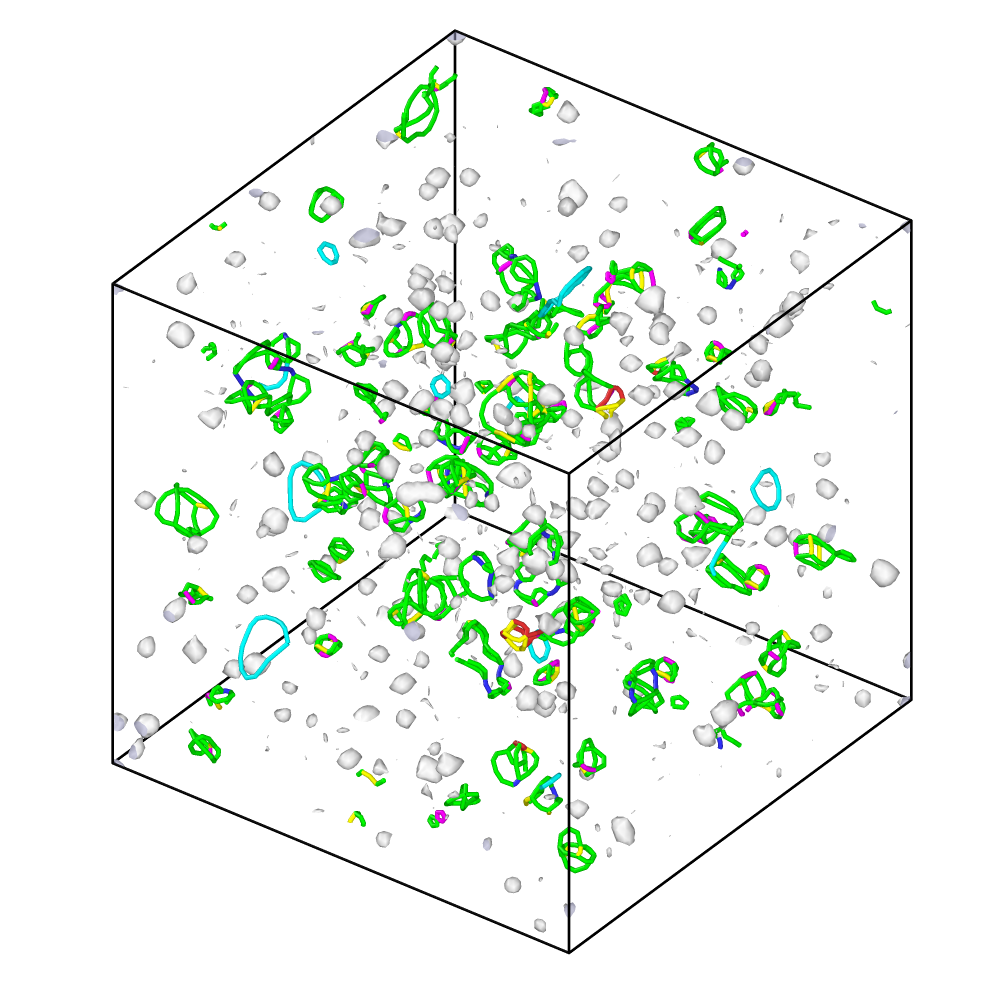}
      \caption{Cu}
\end{subfigure}
\begin{subfigure}[t]{0.3\textwidth}
\label{fig:final:Al}
    \includegraphics[width=\textwidth]{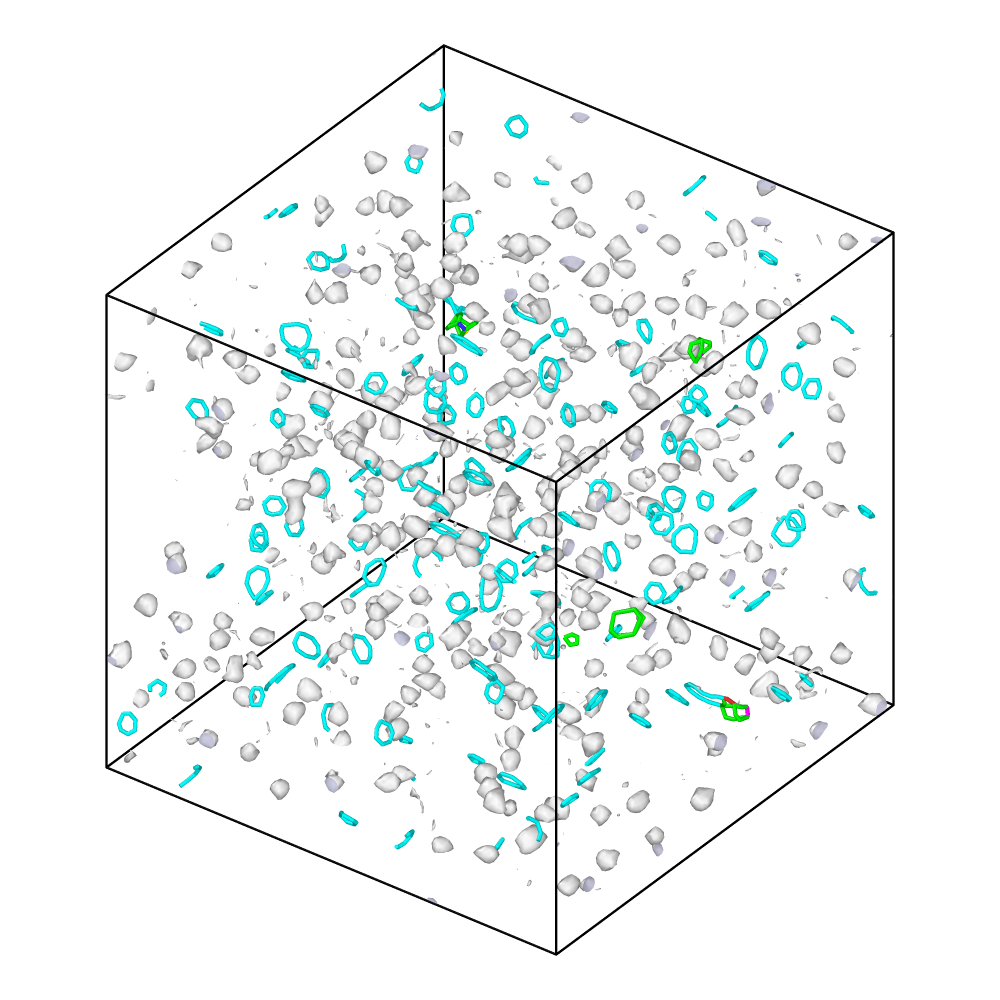}
    \caption{Al}
\end{subfigure}
\begin{subfigure}[t]{0.3\textwidth}
\label{fig:final:Ni}
    \includegraphics[width=\textwidth]{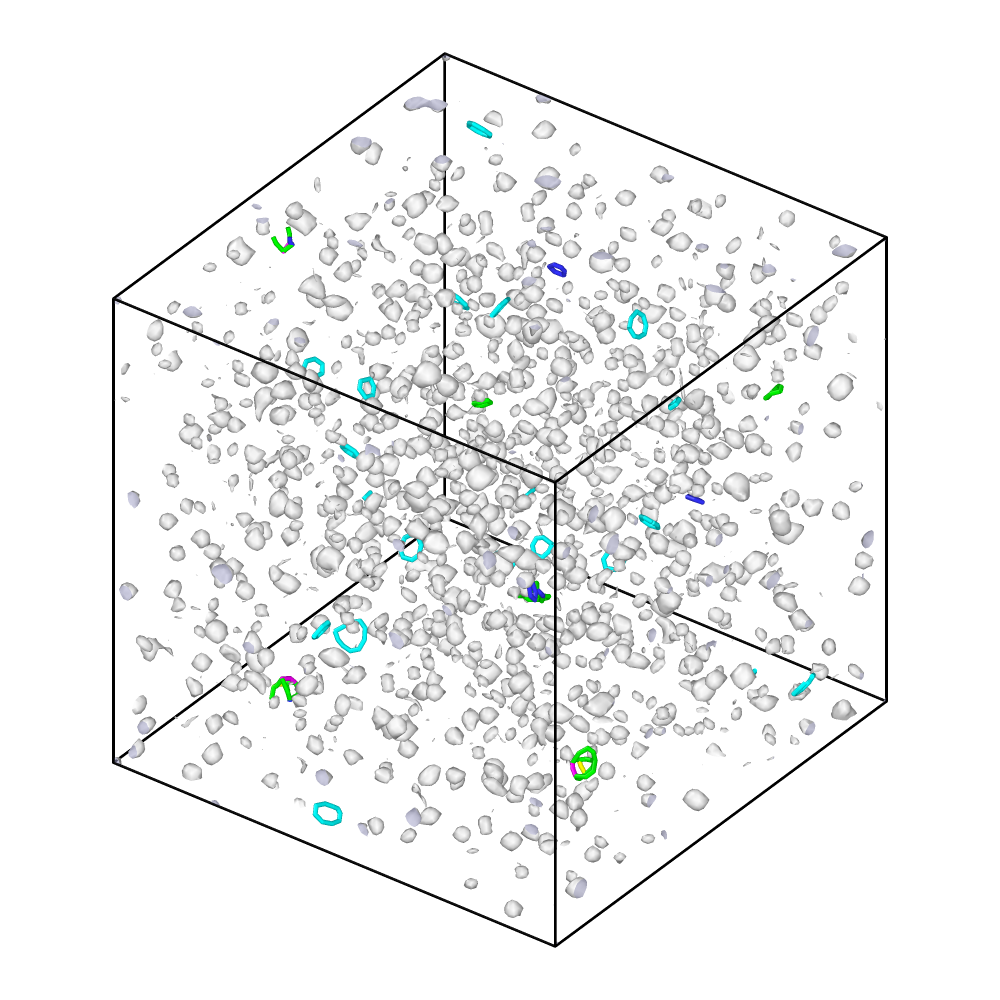}
    \caption{Ni}
\end{subfigure}
    \caption{Final dislocation structures after relaxation of Frenkel pair insertion. Green lines represent Shockley-partial dislocations, cyan lines represent Frank dislocations, pink lines represent stair-rod dislocations, and white blobs are defect clusters not identified as dislocations. The configurations depicted are the \hkl<100> lattice oriented simulation cells.}
    \label{fig:final}
\end{figure*}

Fig.~\ref{fig:FP-vac} shows the evolution of the number of vacancies identified by WS analysis during relaxation. The initial drop of the number of defects is explained by the relaxation of unstable manually inserted defects during the initial relaxation using the conjugate gradient algorithm. During the following relaxation runs in the $NPT$ ensemble, 
the vacancies and interstitials continue recombining but much slower 
until they reach a plateau at about 500 ps. The different saturation points seen are dependent on the defect structures formed during the relaxation. The final dislocation structures visualized using the DXA analysis in OVITO are shown in Fig.~\ref{fig:final}. It must be noted that we do not aim to compare the dislocation structure in different materials, since the final structures do neither represent typical dislocation structures induced by irradiation nor those from simulations of overlapping cascades. Recently, it has been demonstrated that there are significant differences in the structure of defects created via FP insertion method and in radiation damage simulations~\cite{GRANBERG2023111902}. Nevertheless, the FP insertion simulations give insight into the type of dislocations preferred by different potentials. From our results we clearly see that Frank loops are the preferred dislocations in Al, while in Cu Shockley partial dislocations dominate. In Ni we see both types of dislocations but most of the defects are in clusters where no dislocation structure can be identified. From the FP insertion simulations we conclude that a saturated simulation cell has approximately a 0.004 \% vacancy concentration and we will use these structures in the compression simulations.

\vfill\eject

\subsection{Uniaxial compressive loading}\label{sec:bulkcomp}

\begin{figure*}
    \centering
\begin{subfigure}[t]{0.45\textwidth}
\label{fig:NCP:Cu100}
    \includegraphics[width=\textwidth]{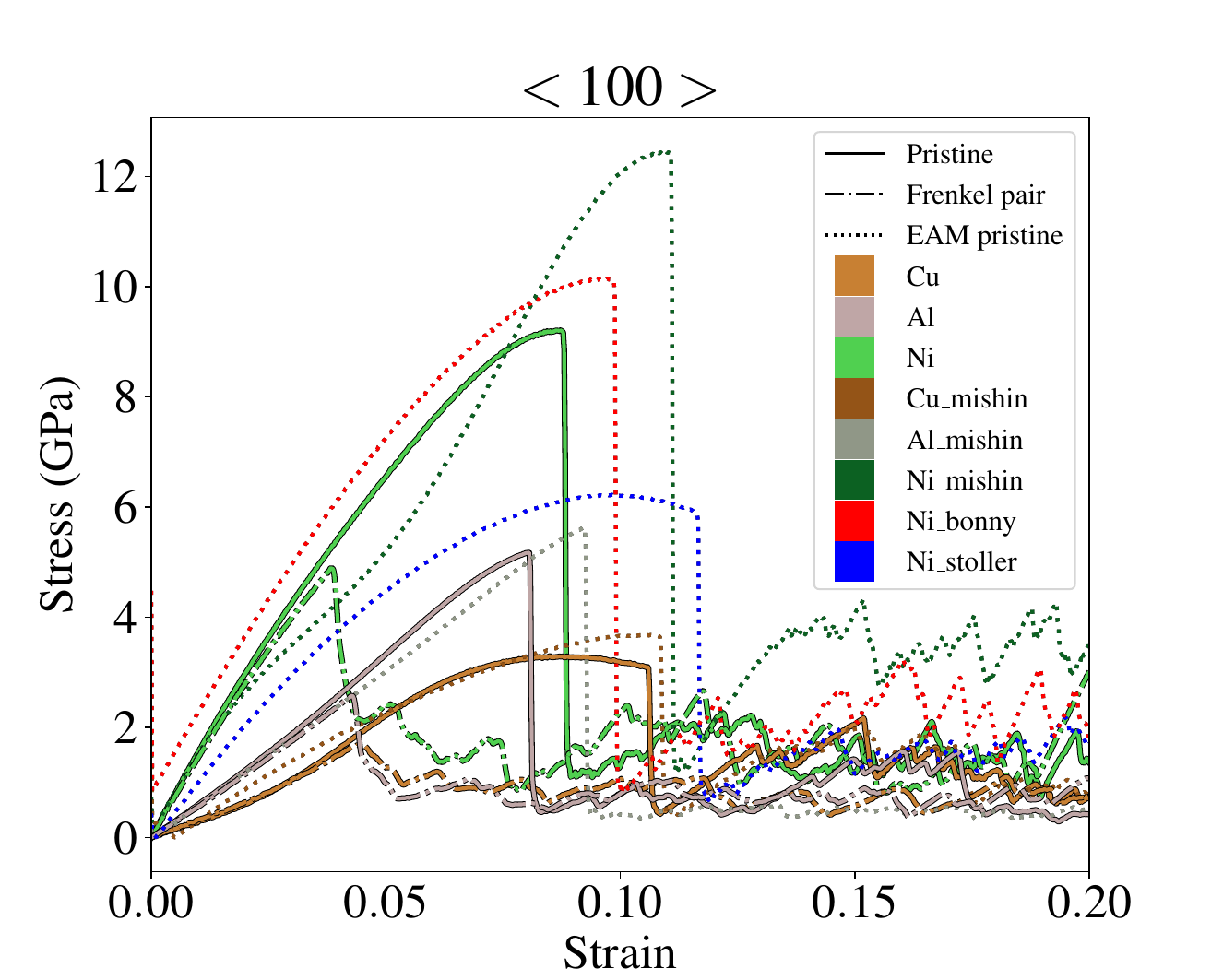}
      \caption{}
\end{subfigure}
\begin{subfigure}[t]{0.45\textwidth}
\label{fig:NCP:Al110}
    \includegraphics[width=\textwidth]{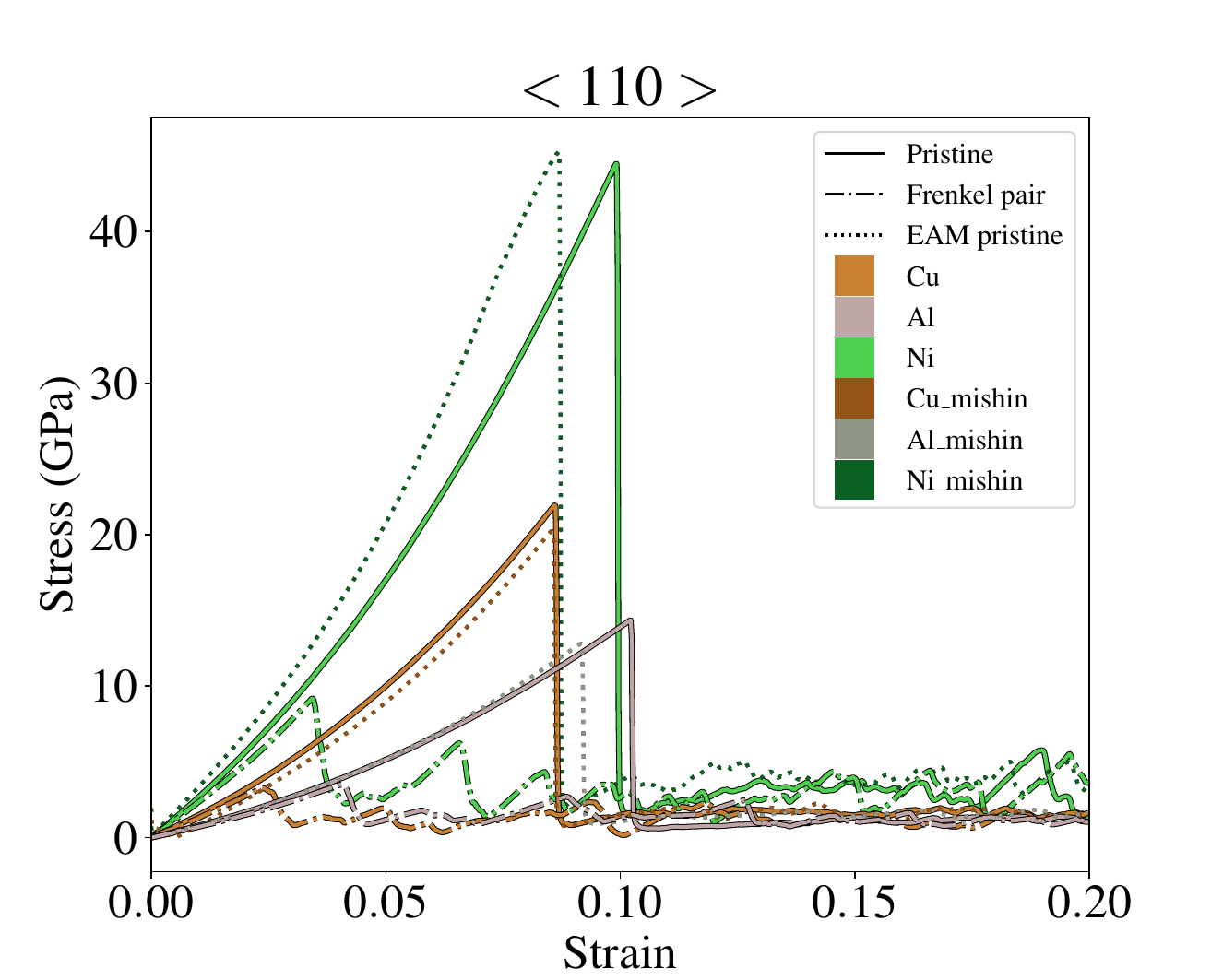}
    \caption{}
\end{subfigure}
\begin{subfigure}[t]{0.45\textwidth}
\label{fig:NCP:Cu111}
    \includegraphics[width=\textwidth]{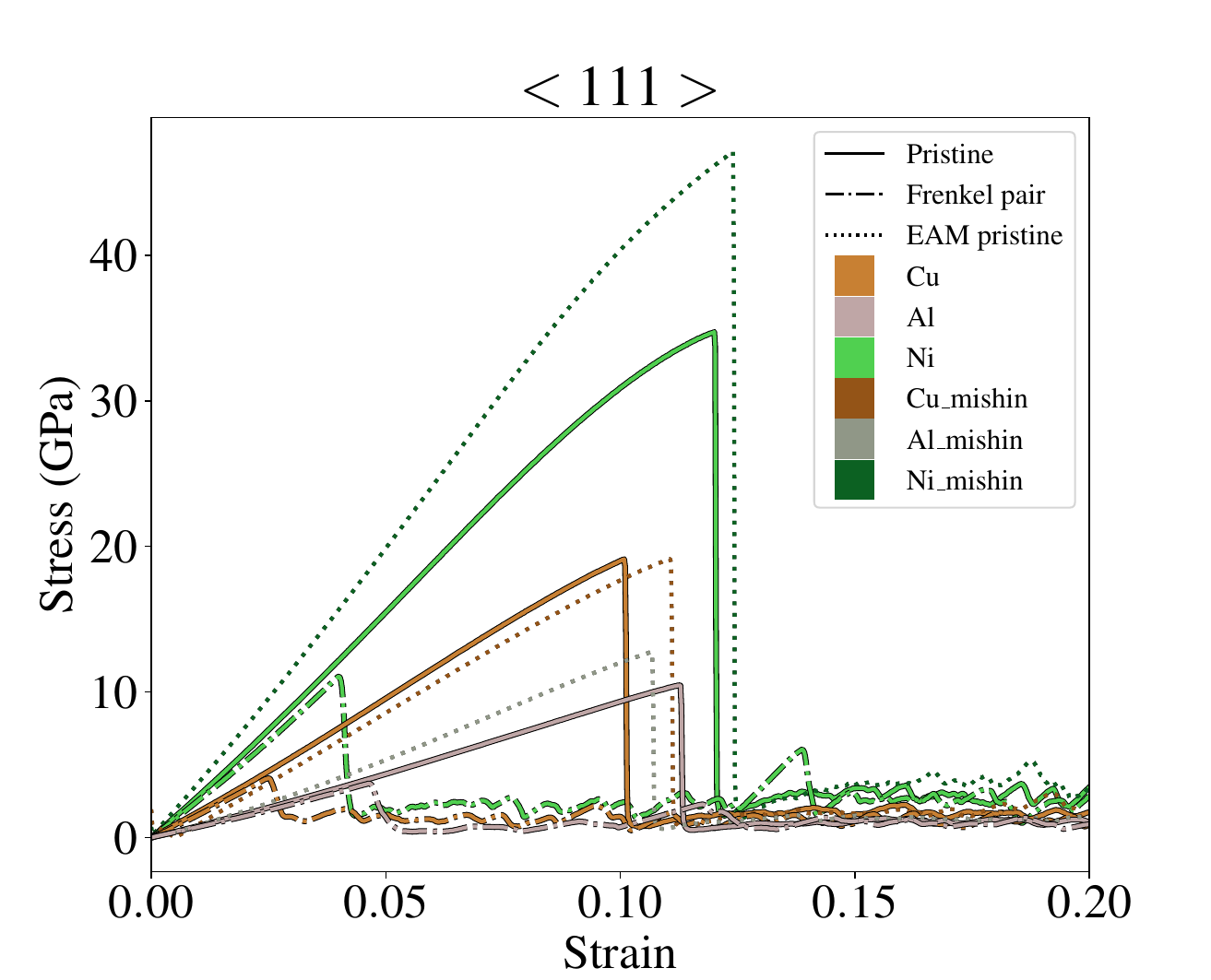}
      \caption{}
\end{subfigure}
\begin{subfigure}[t]{0.45\textwidth}
\label{fig:NCP:Al112}
    \includegraphics[width=\textwidth]{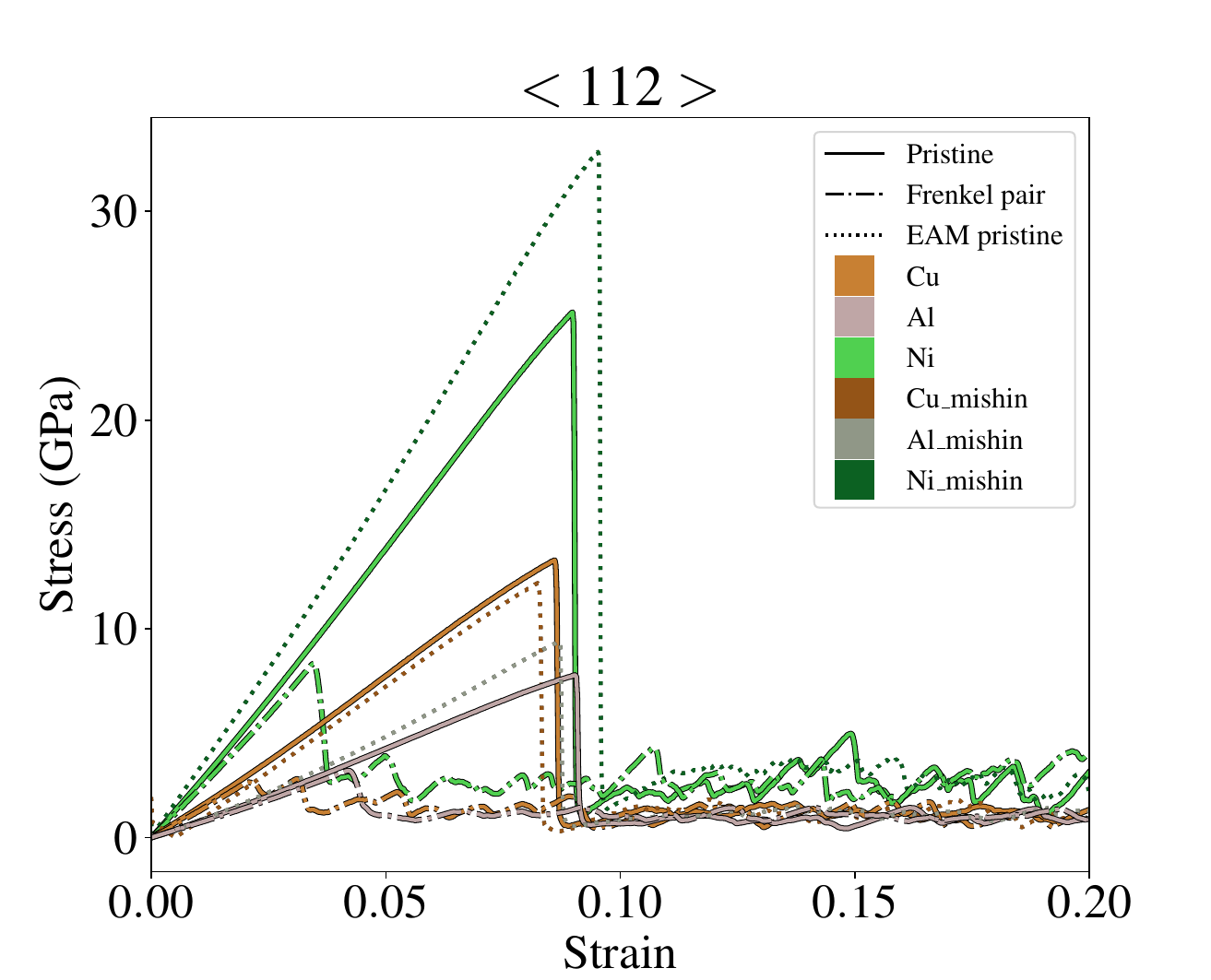}
    \caption{}
\end{subfigure}
    \caption{Stress-strain curves during uniaxial compressive loading in different loading direction for different elements simulated with both tabGAP and EAM potentials. The simulation cells containing defects are here referred as Frenkel pair systems. The results are from simulations done using tabGAP potentials and EAM potentials~\cite{mishin2001structural,PhysRevB.59.3393,stoller2016impact,bonny2013interatomic}.}
    \label{fig:compression}
\end{figure*}

The stress-strain responses of Cu, Al and Ni using the tabGAP potentials during compression are shown in Fig.~\ref{fig:compression}. Additionally, some selected pristine cells were simulated with EAM potentials as a reference. The strain in the figures are defined as the change in length of the simulation cell in the loading direction compared to the initial length. During loading, the stress in the system builds up until the system yields after which the stress drops to a low value of the flow stress. In general, there is a clear order in ranking of materials with respect to the yield stress (in this case defined as the maximal stress before the sudden drop): Ni has the highest and Al the lowest values of the yield stress in pristine cells. As an exception, we observe the lowest yield stress in the \hkl<100> loading direction for Cu. 
The order, which we observe in our simulations is the same as that in the experimental equation of state and is in line with the known properties of the materials such as the Young's modulus of the respective materials. 
Overall, the yield stresses in the pristine cells in Fig.~\ref{fig:compression} are in good agreement with the values reported in previous MD studies of uniaxial compression ~\cite{TSCHOPP20081806,xie2014tension,li2020molecular}. However, in the case of Ni there is a significant difference between tabGAP and the classical EAM potentials, where there is a surprisingly wide spread in the predicted yield stresses (see Fig.~\ref{fig:compression}a, the different potentials in the legend are shown as Ni subscripts). Additionally, the Mishin \etal potential for Ni shows irregular stress response in the \hkl<100> direction compared to the other Ni potentials and predicts significantly higher flow stress in the \hkl<100> case than the others. This irregular shape can also be seen in work on nanopillar compression, where the same EAM potential was used~\cite{ZHANG2022111360}. For the other elements there are no significant differences between tabGAP and the Mishin \etal EAM potentials with regards to the stress-strain relation, other than the strain at which the yield occurs. For all materials, we observe elastic hardening  in the \hkl<110> direction and softening in \hkl<100> (compare in Fig.~\ref{fig:compression}b and Fig.~\ref{fig:compression}a, where the peak sharpens or becomes blunt because of straining right before the yield point). This is a known effect and has been reported previously~\cite{zhang2018nonlinear,TSCHOPP20081806}. In reality, some defects always exist that can act as nucleation sites for dislocations, thus the pristine cases are in practice artificial. In comparison, the systems containing defects (the cells relaxed after the Frenkel pair insertion) yielded with significantly lower strain and with a lower yield stress, at a value between \nicefrac{1}{7} -- \nicefrac{1}{2} of the pristine cells. This is expected, as pre-existing defects serve as nucleation sites for heterogeneous dislocation nucleation, which takes place at lower stresses than the homogeneous dislocation nucleation in the pristine cases. Before yielding the stress-strain curves of the cells with defective structures follow closely their pristine counterparts. Additionally, there are no significant differences in the flow stress between the damaged and pristine cells.

\begin{figure*}
\centering
\begin{subfigure}[t]{0.2\textwidth}
    \includegraphics[width=\textwidth]{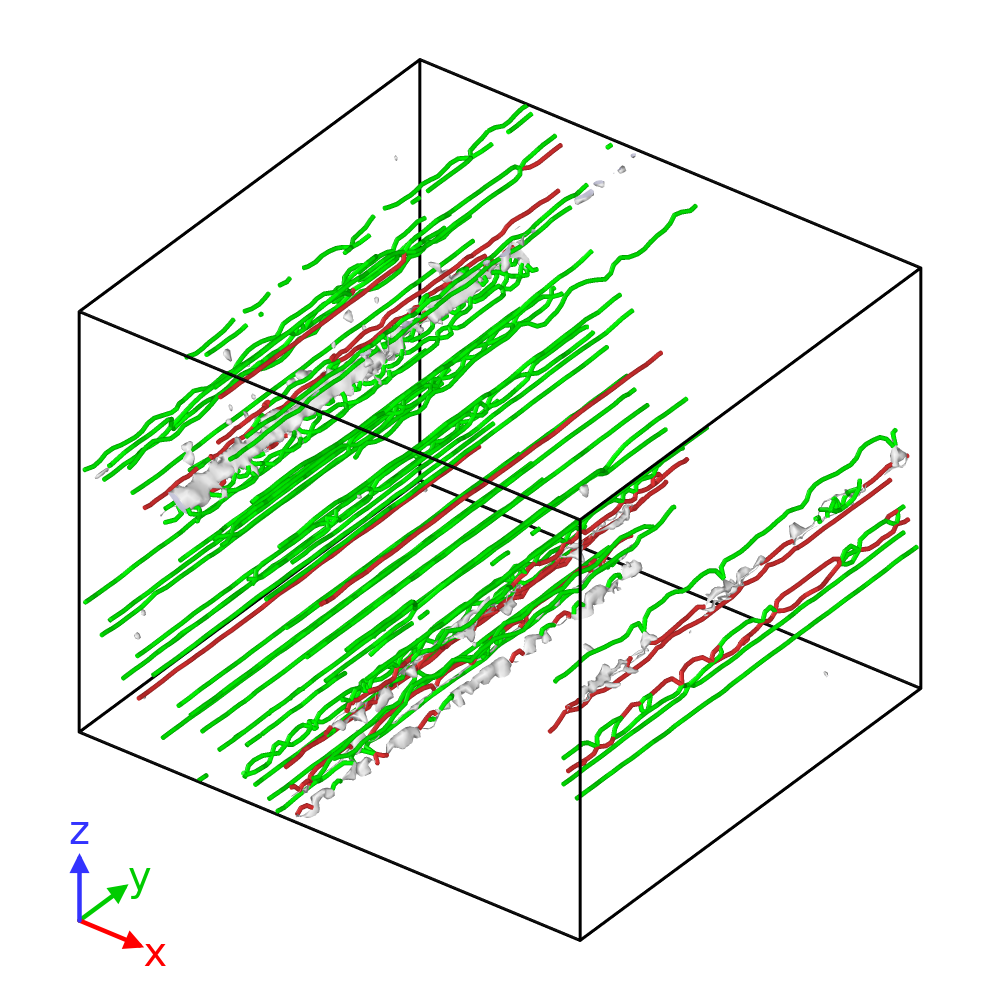}
      \caption{\hkl<100> strain 11\% }
\end{subfigure}
\begin{subfigure}[t]{0.2\textwidth}
    \includegraphics[width=\textwidth]{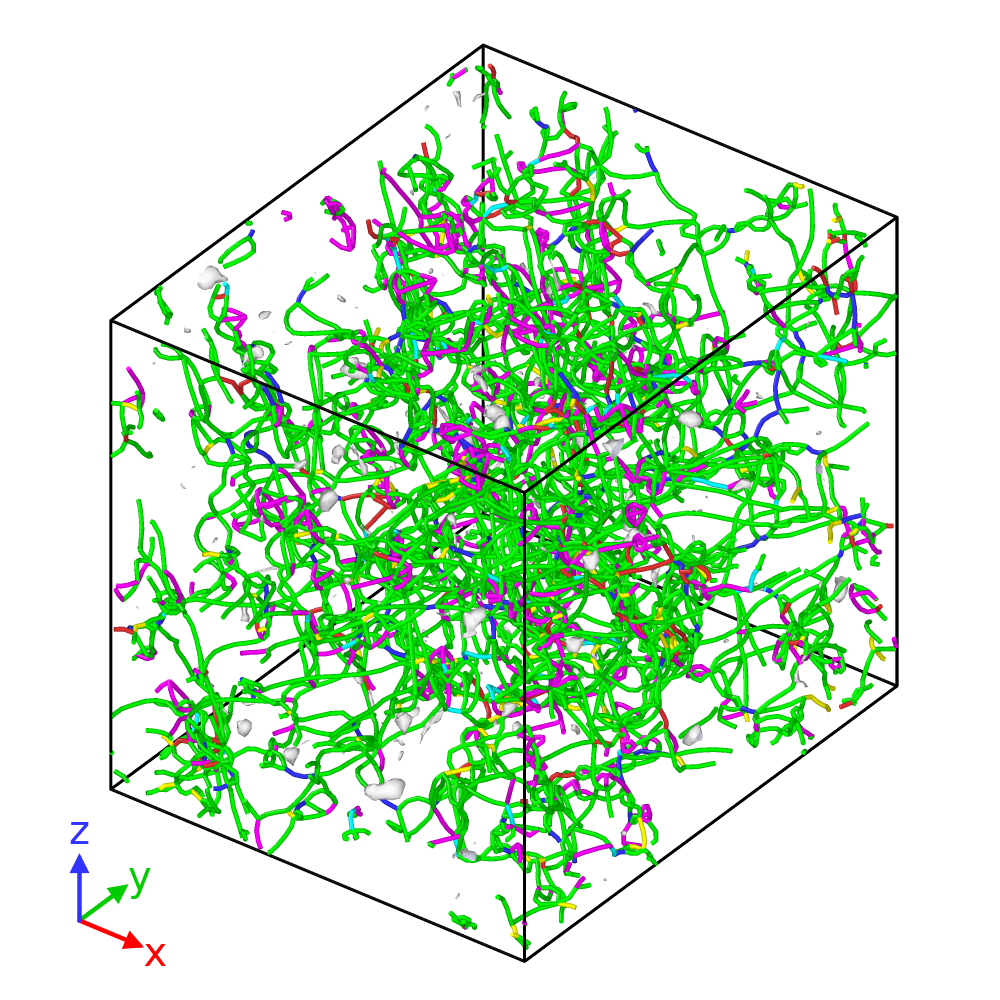}
    \caption{\hkl<110> strain 9\%}
\end{subfigure}
\begin{subfigure}[t]{0.2\textwidth}
    \includegraphics[width=\textwidth]{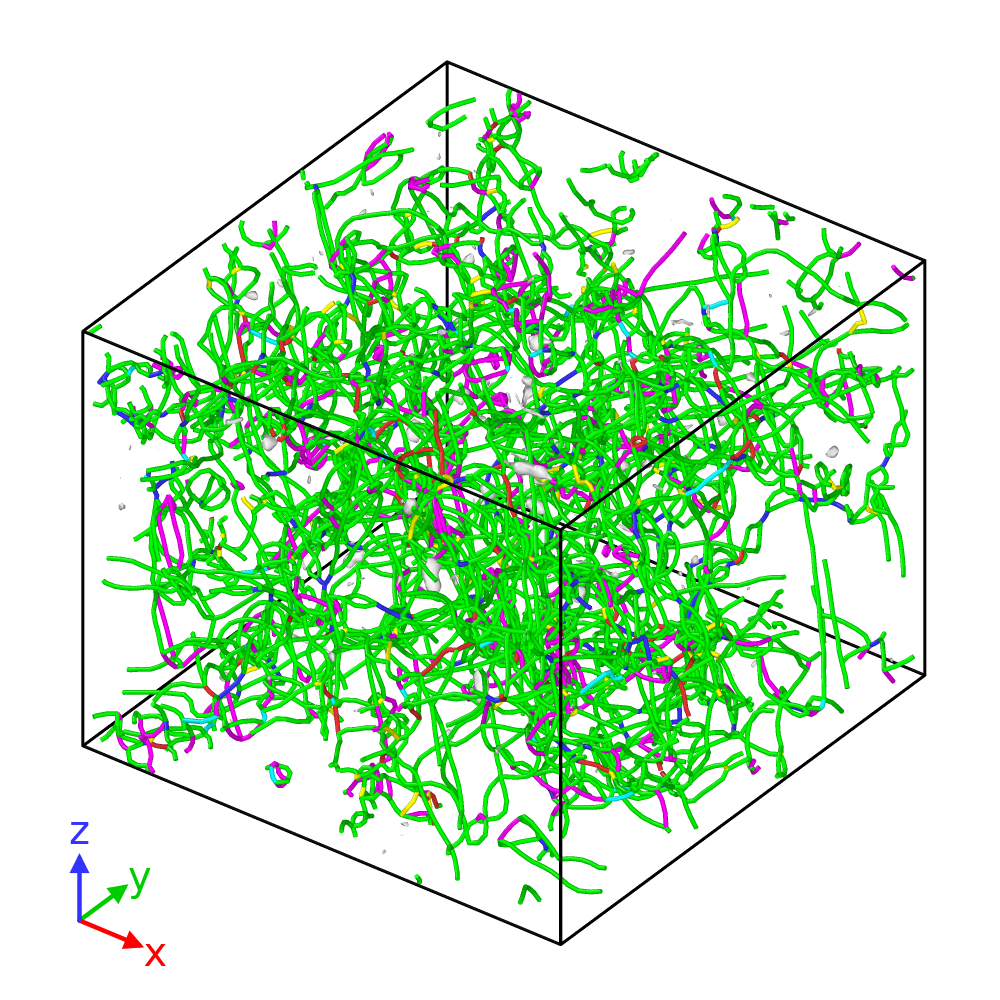}
      \caption{\hkl<111> strain 11\% }
\end{subfigure}
\begin{subfigure}[t]{0.2\textwidth}
    \includegraphics[width=\textwidth]{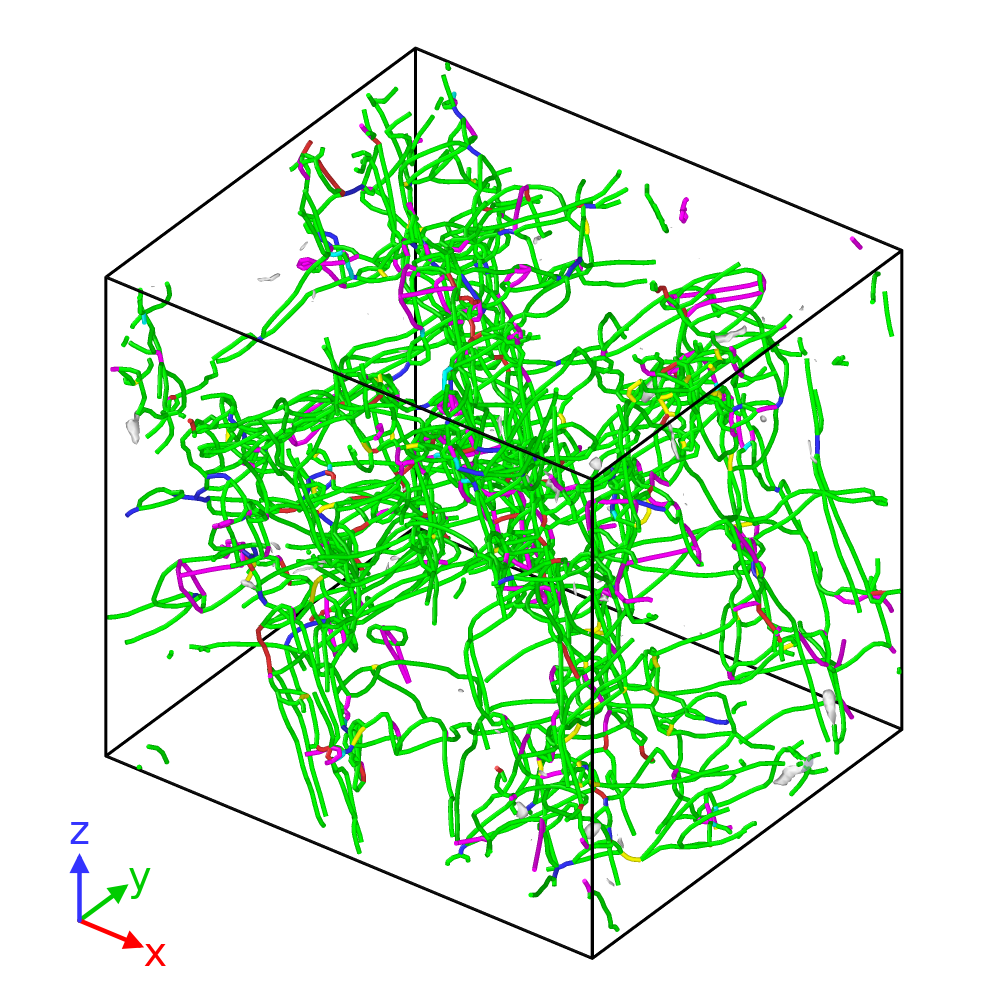}
    \caption{\hkl<112> strain 9\%}
\end{subfigure}
\begin{subfigure}[t]{0.2\textwidth}
    \includegraphics[width=\textwidth]{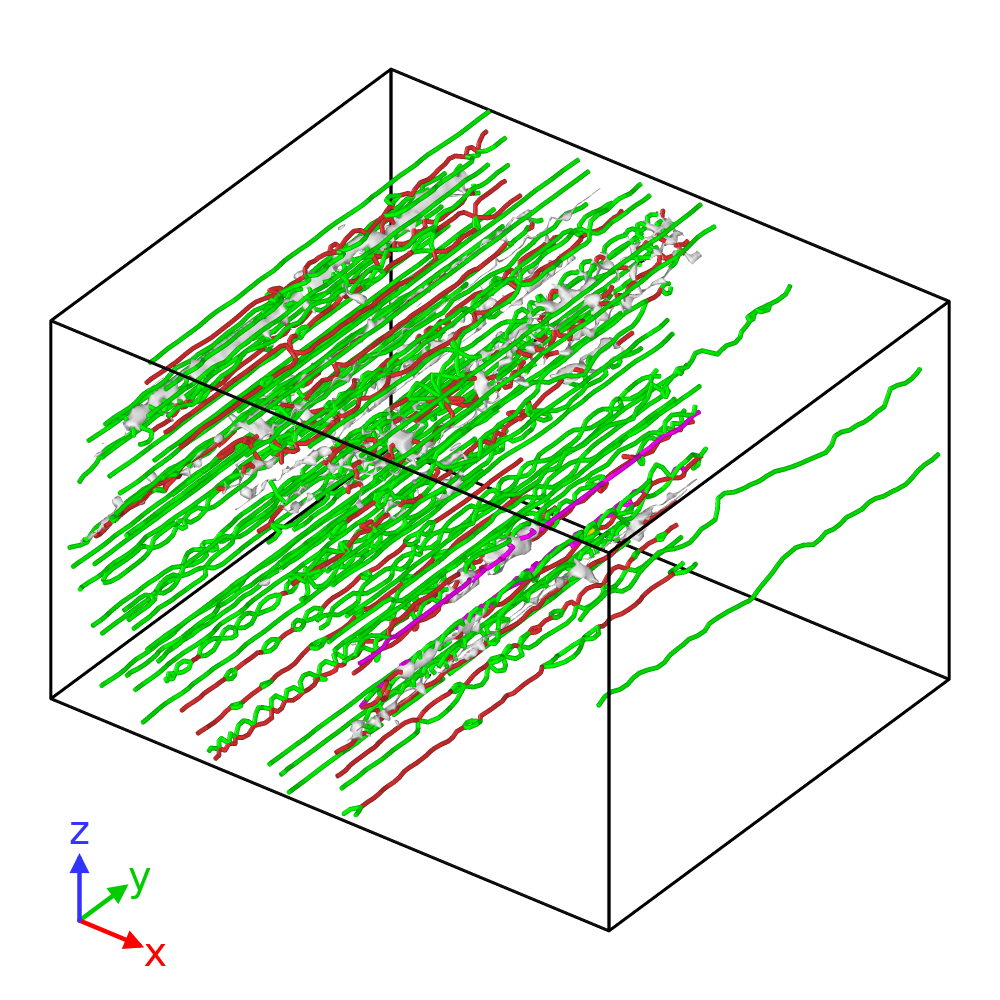}
      \caption{\hkl<100> strain 20\% }
\end{subfigure}
\begin{subfigure}[t]{0.2\textwidth}
    \includegraphics[width=\textwidth]{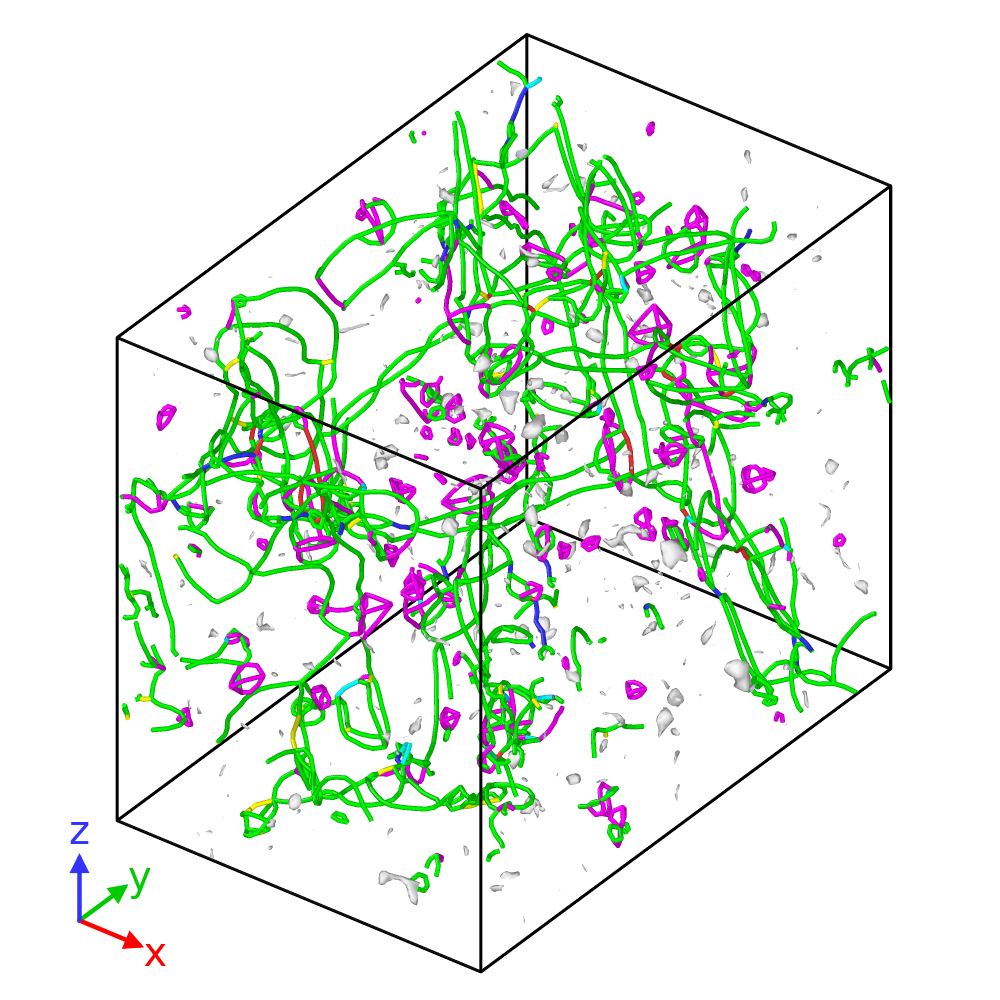}
    \caption{ \hkl<110> strain 20\%}
\end{subfigure}
\begin{subfigure}[t]{0.2\textwidth}
    \includegraphics[width=\textwidth]{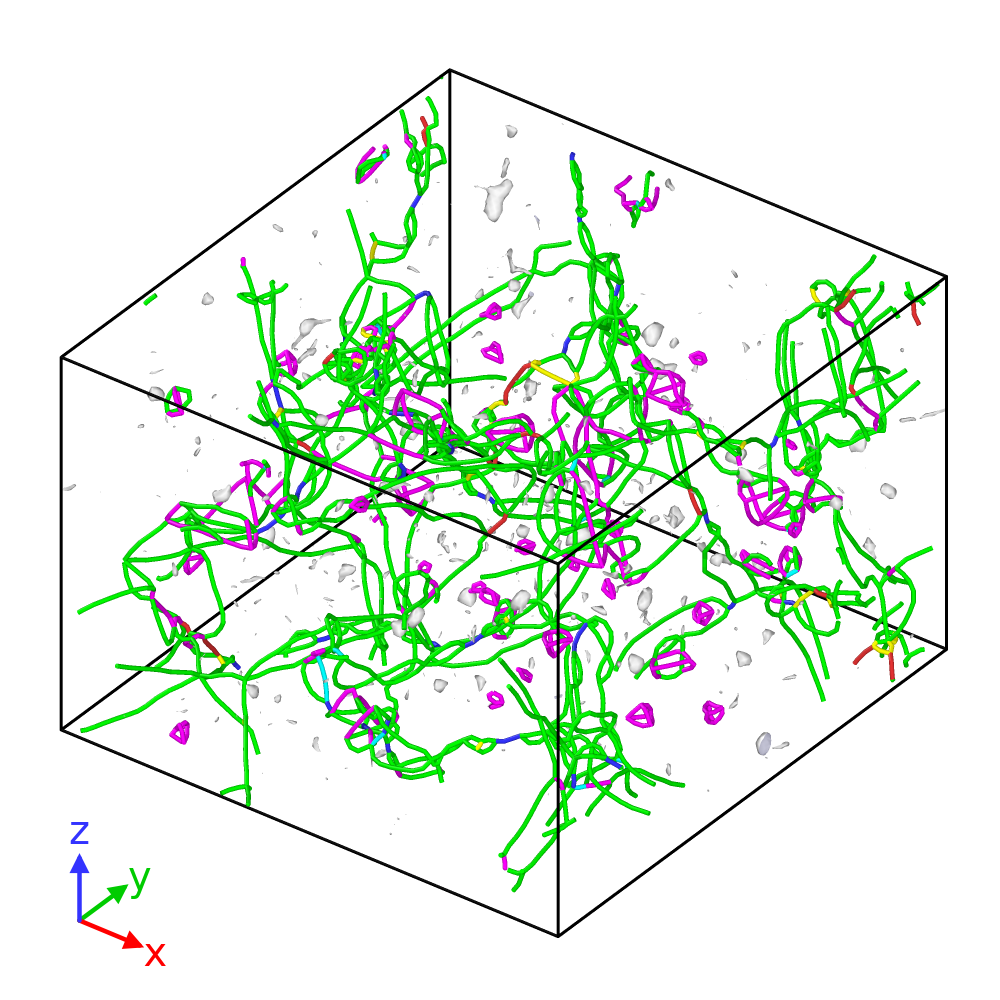}
      \caption{\hkl<111> strain 20\% }
\end{subfigure}
\begin{subfigure}[t]{0.2\textwidth}
    \includegraphics[width=\textwidth]{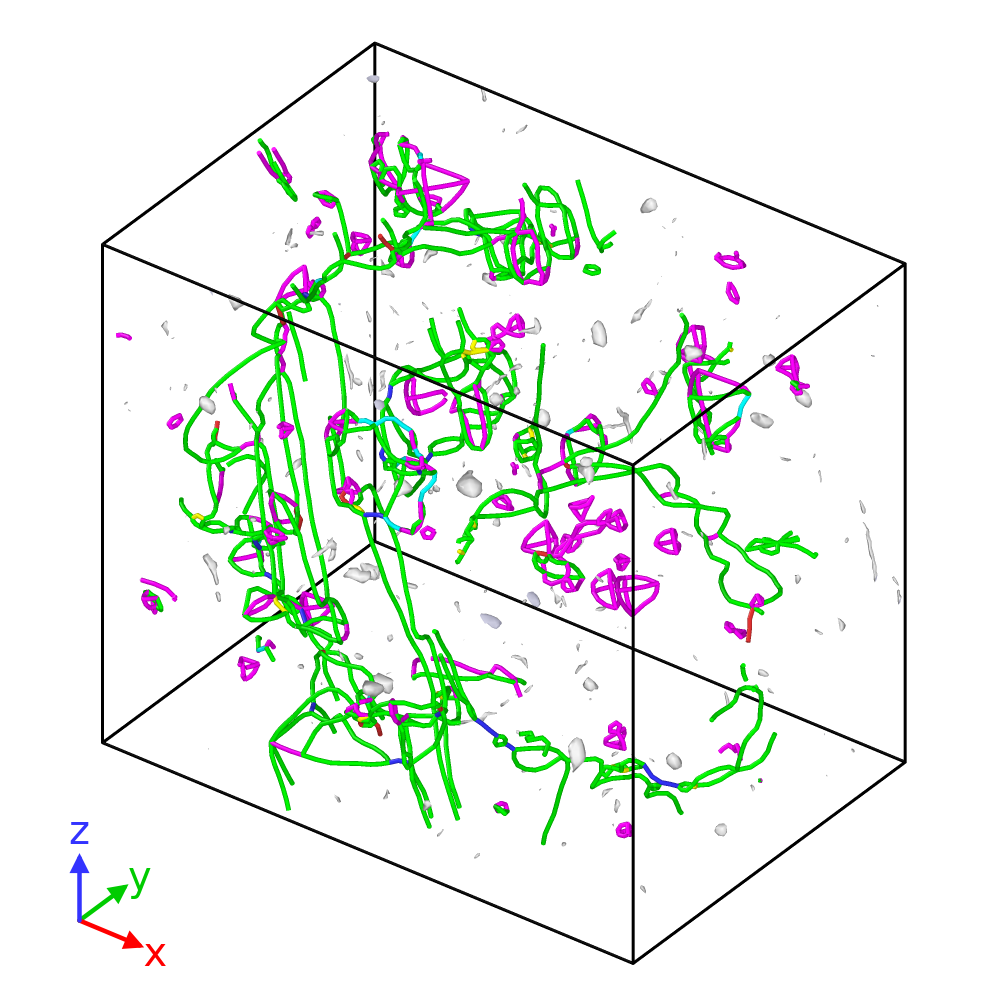}
    \caption{\hkl<112> strain 20\%}
\end{subfigure}
    \caption{Dislocation structures in Cu during compression in different loading directions with pristine cells. First row is immediately after yielding and second row is the final configuration. Green lines represent Shockley partial dislocations, cyan lines represent Frank dislocations, pink lines represent stair-rod dislocations, and white blobs are defect clusters not identified as dislocations.}
    \label{fig:pristine-Cu}
\end{figure*}

\begin{figure*}
    \centering
\begin{subfigure}[t]{0.2\textwidth}
    \includegraphics[width=\textwidth]{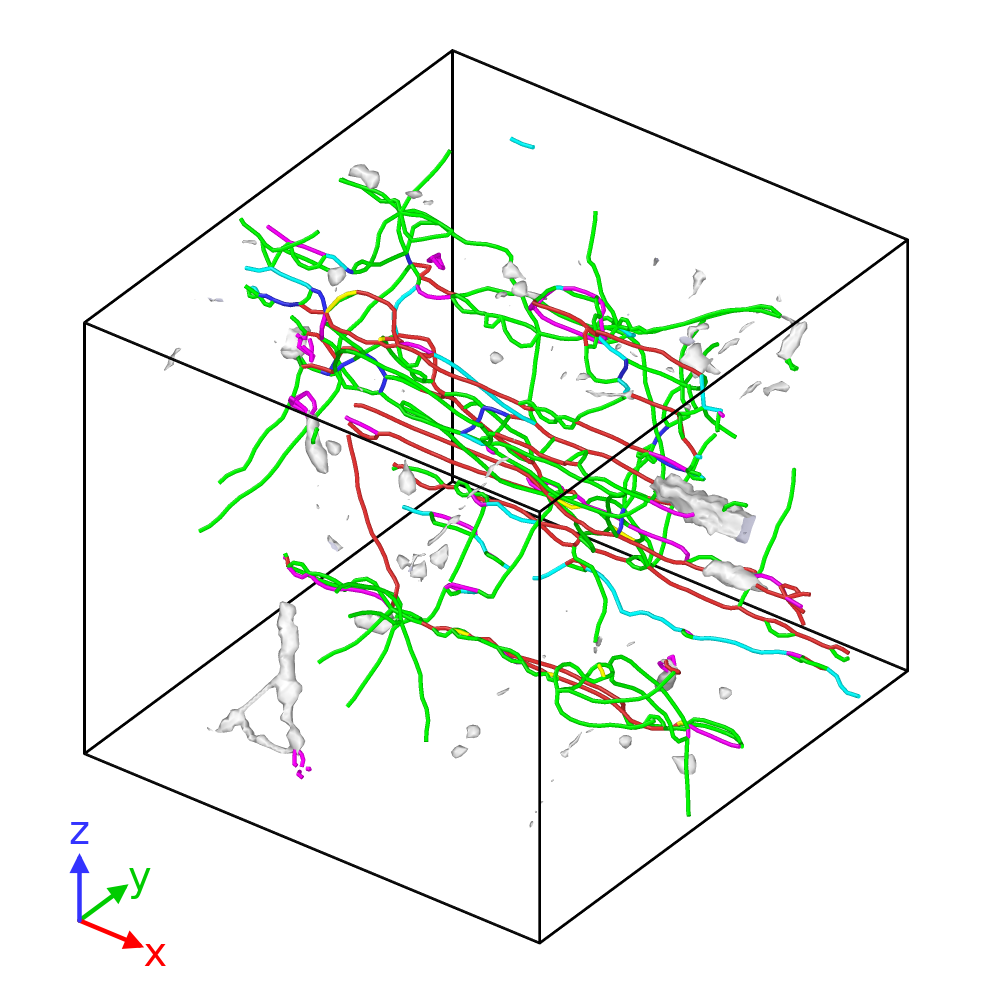}
      \caption{\hkl<100> strain 9\% }
\end{subfigure}
\begin{subfigure}[t]{0.2\textwidth}
    \includegraphics[width=\textwidth]{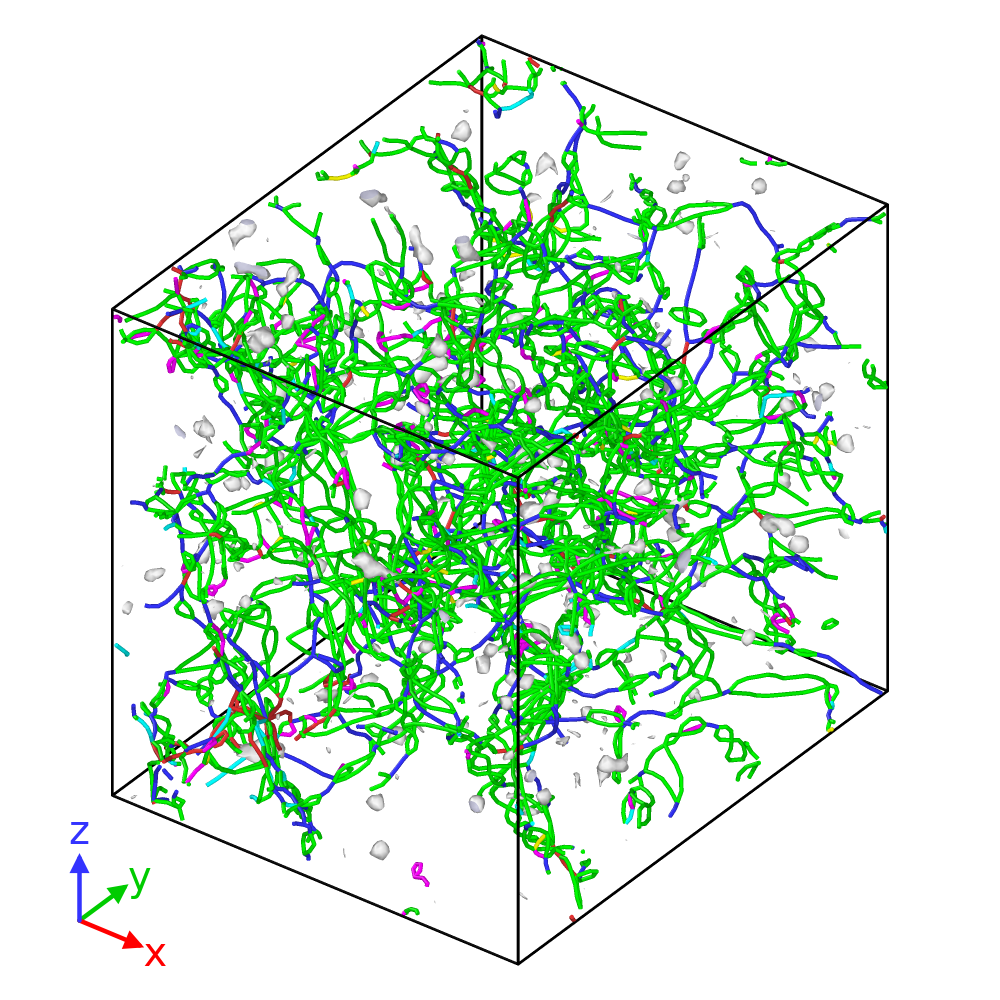}
    \caption{\hkl<110> strain 11\%}
\end{subfigure}
\begin{subfigure}[t]{0.2\textwidth}
    \includegraphics[width=\textwidth]{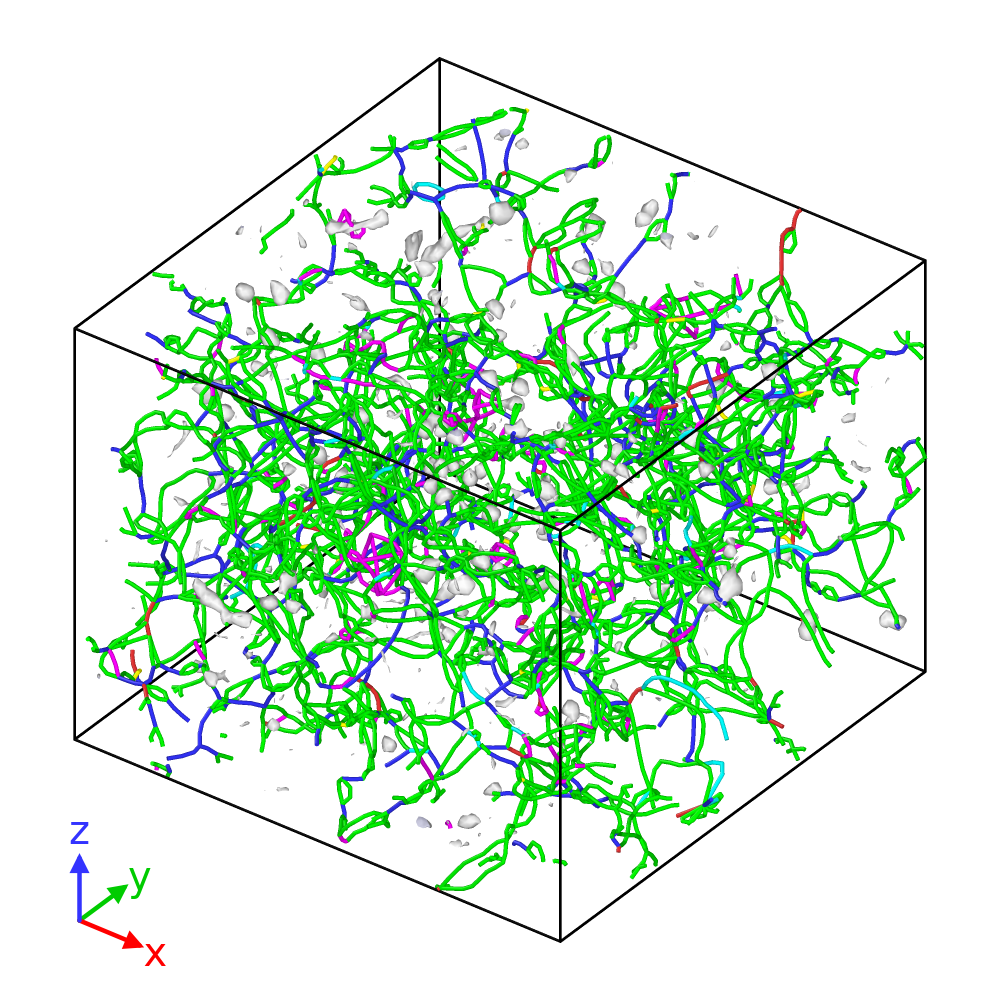}
      \caption{\hkl<111> strain 12\% }
\end{subfigure}
\begin{subfigure}[t]{0.2\textwidth}
    \includegraphics[width=\textwidth]{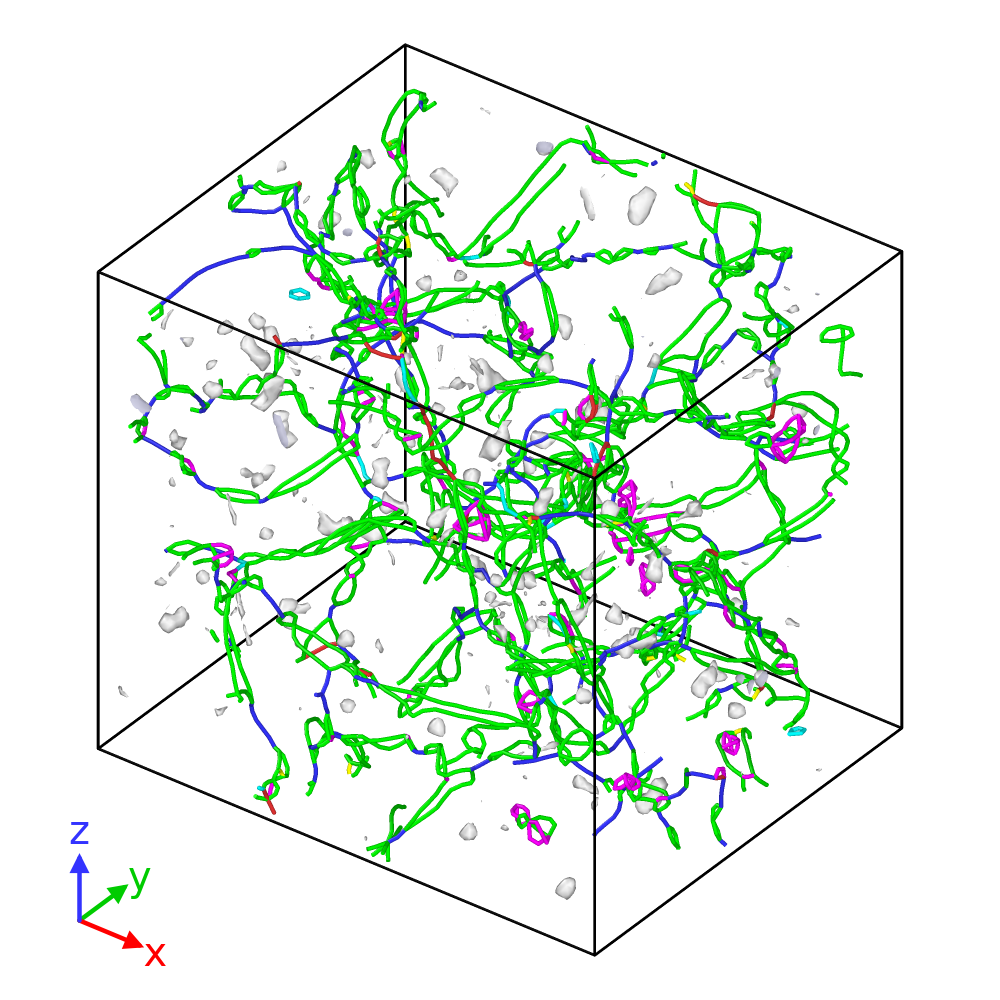}
    \caption{\hkl<112> strain 10\%}
\end{subfigure}
\begin{subfigure}[t]{0.2\textwidth}
    \includegraphics[width=\textwidth]{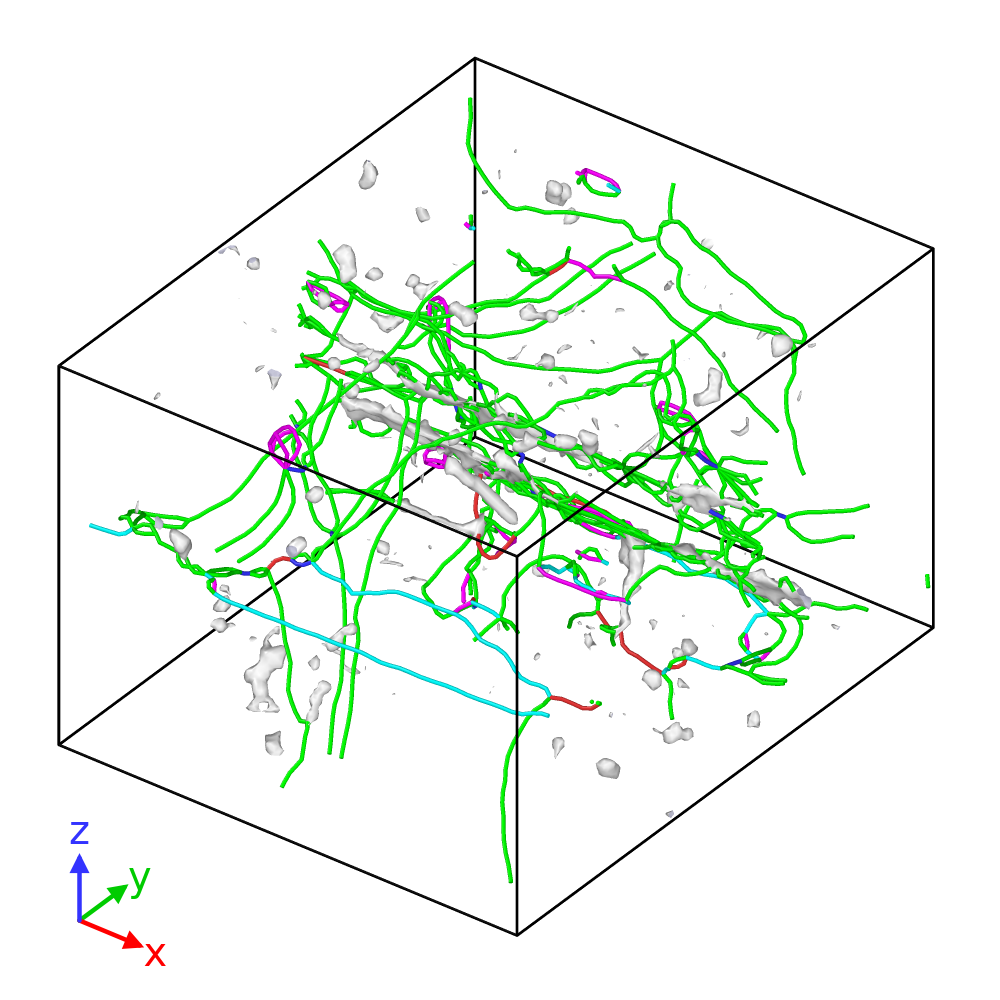}
      \caption{\hkl<100> strain 20\% }
\end{subfigure}
\begin{subfigure}[t]{0.2\textwidth}
    \includegraphics[width=\textwidth]{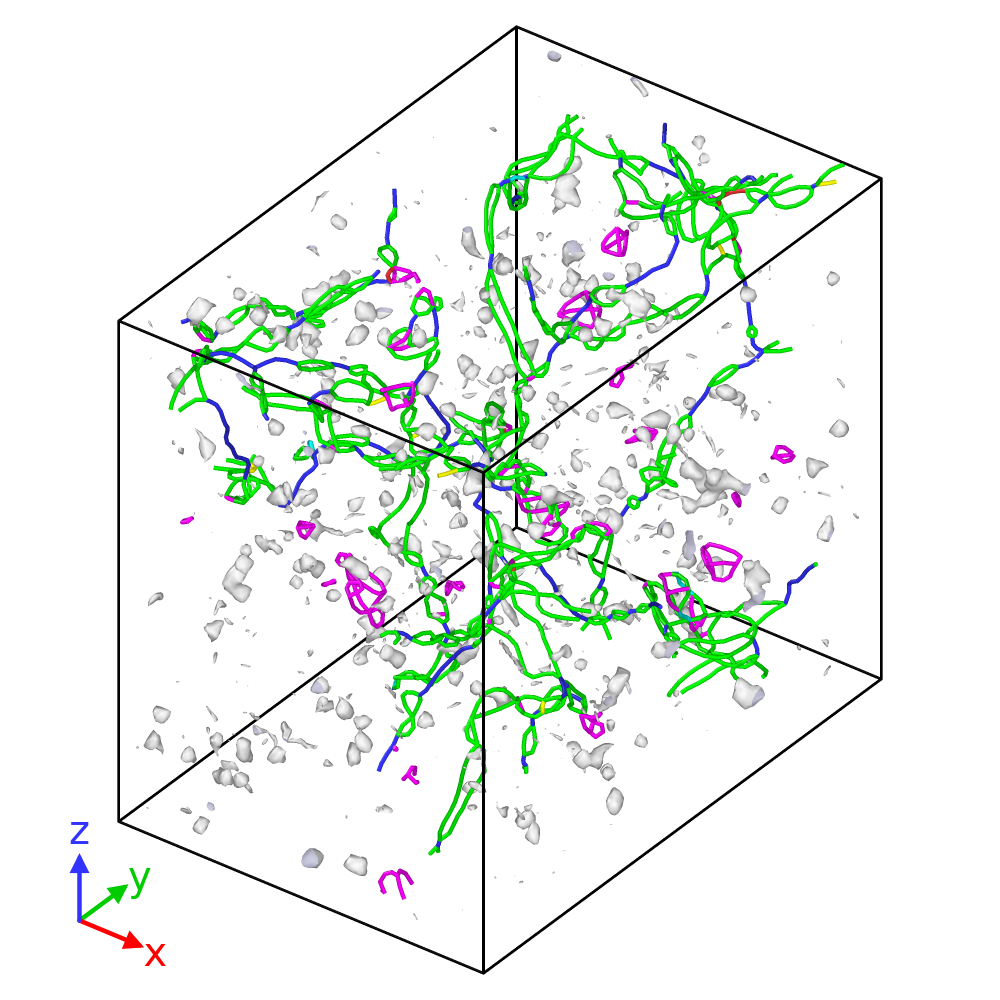}
    \caption{ \hkl<110> strain 20\%}
\end{subfigure}
\begin{subfigure}[t]{0.2\textwidth}
    \includegraphics[width=\textwidth]{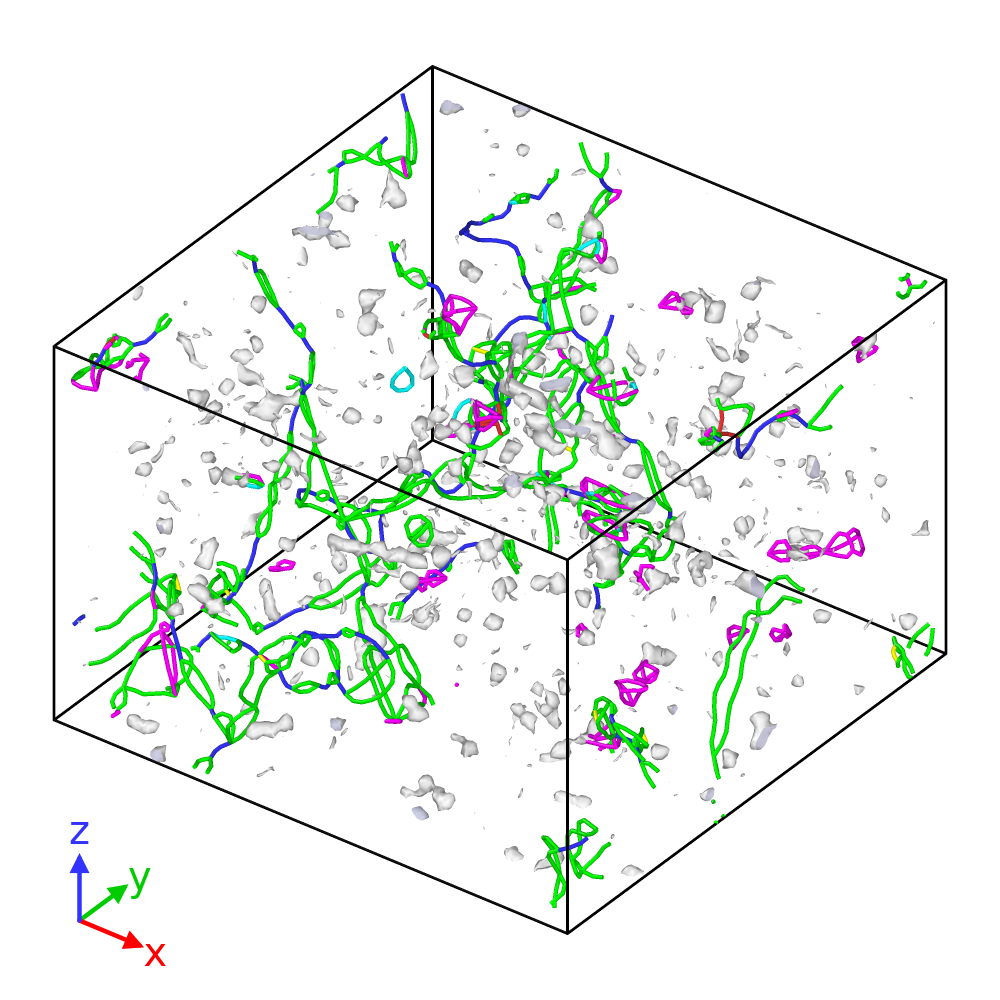}
      \caption{\hkl<111> strain 20\% }
\end{subfigure}
\begin{subfigure}[t]{0.2\textwidth}
    \includegraphics[width=\textwidth]{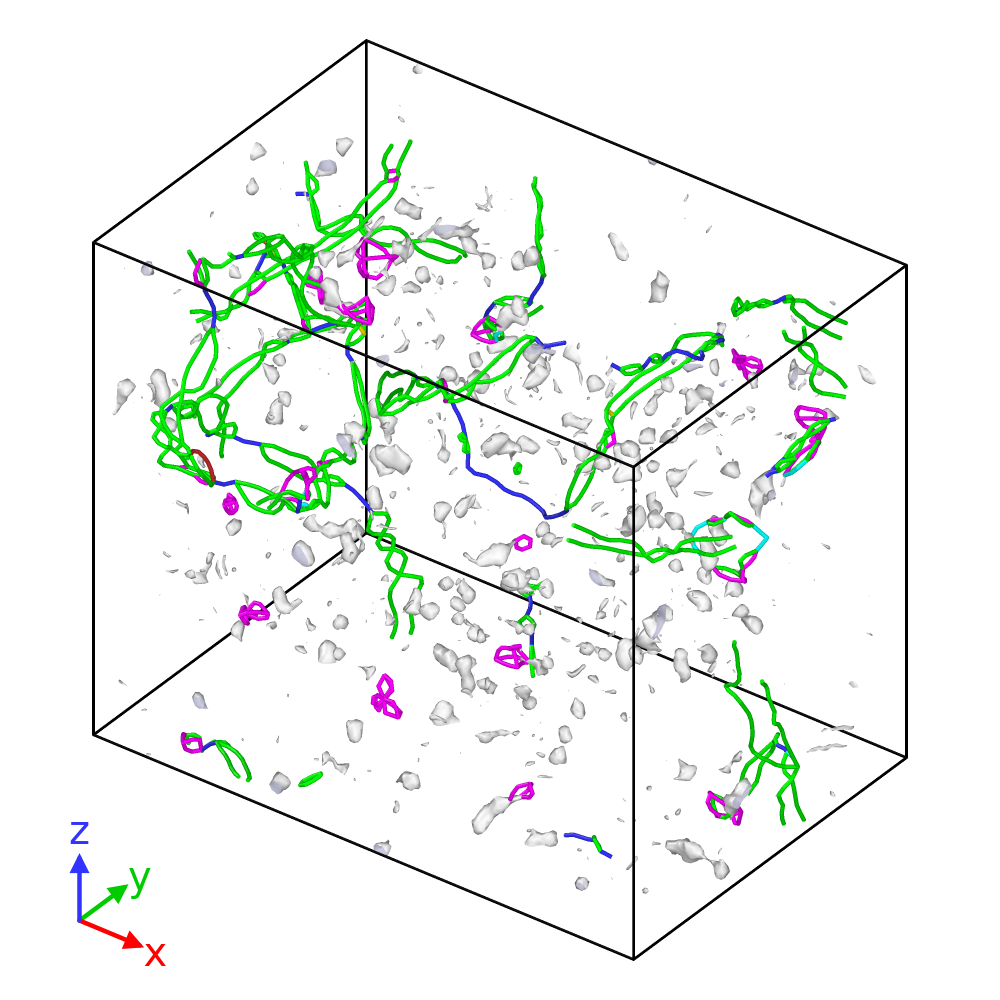}
    \caption{\hkl<112> strain 20\%}
\end{subfigure}

    \caption{Dislocation structures in Al during compression in different loading directions with pristine cells. First row is directly after yielding and second row is the final configuration. Green lines represent Shockley partial dislocations, cyan lines represent Frank dislocations, pink lines represent stair-rod dislocations, and white blobs are defect clusters not identified as dislocations.}
    \label{fig:pristine-Al}
\end{figure*}

\begin{figure*}
    \centering
\begin{subfigure}[t]{0.2\textwidth}
    \includegraphics[width=\textwidth]{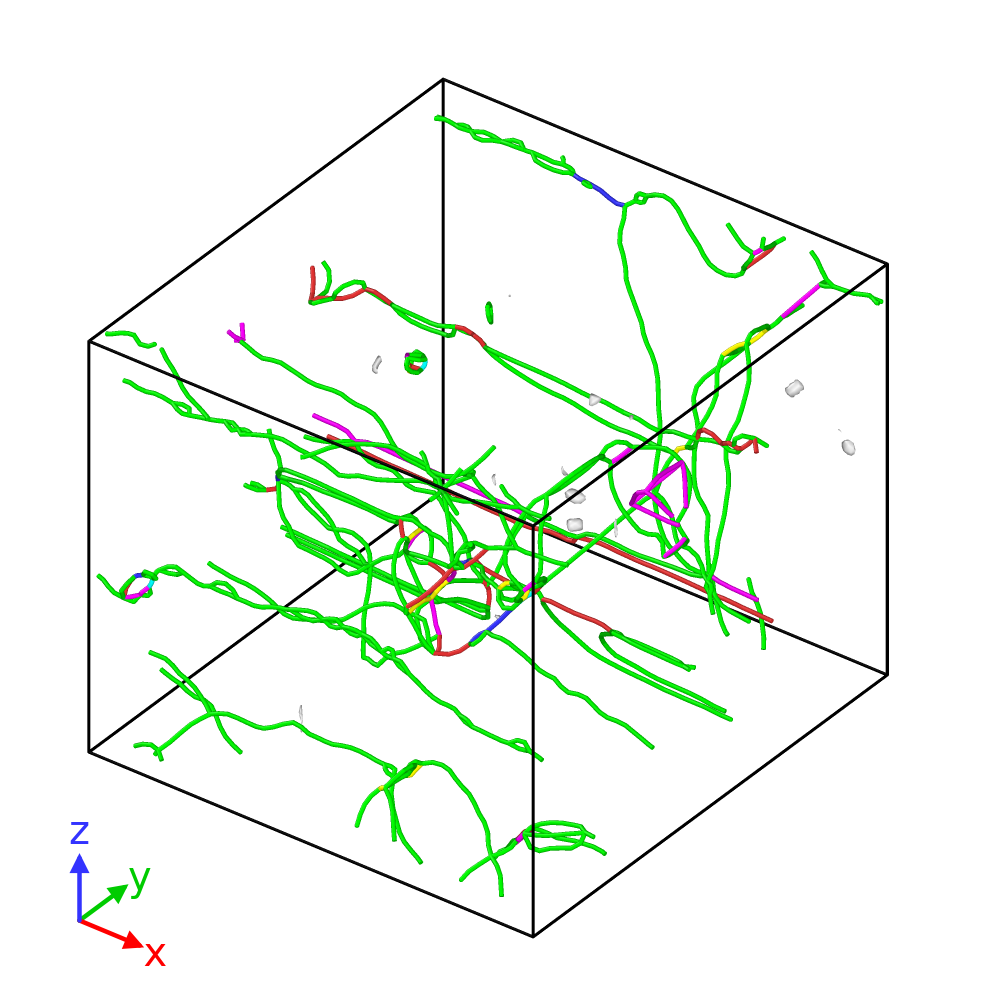}
      \caption{\hkl<100> strain 10\% }
\end{subfigure}
\begin{subfigure}[t]{0.2\textwidth}
    \includegraphics[width=\textwidth]{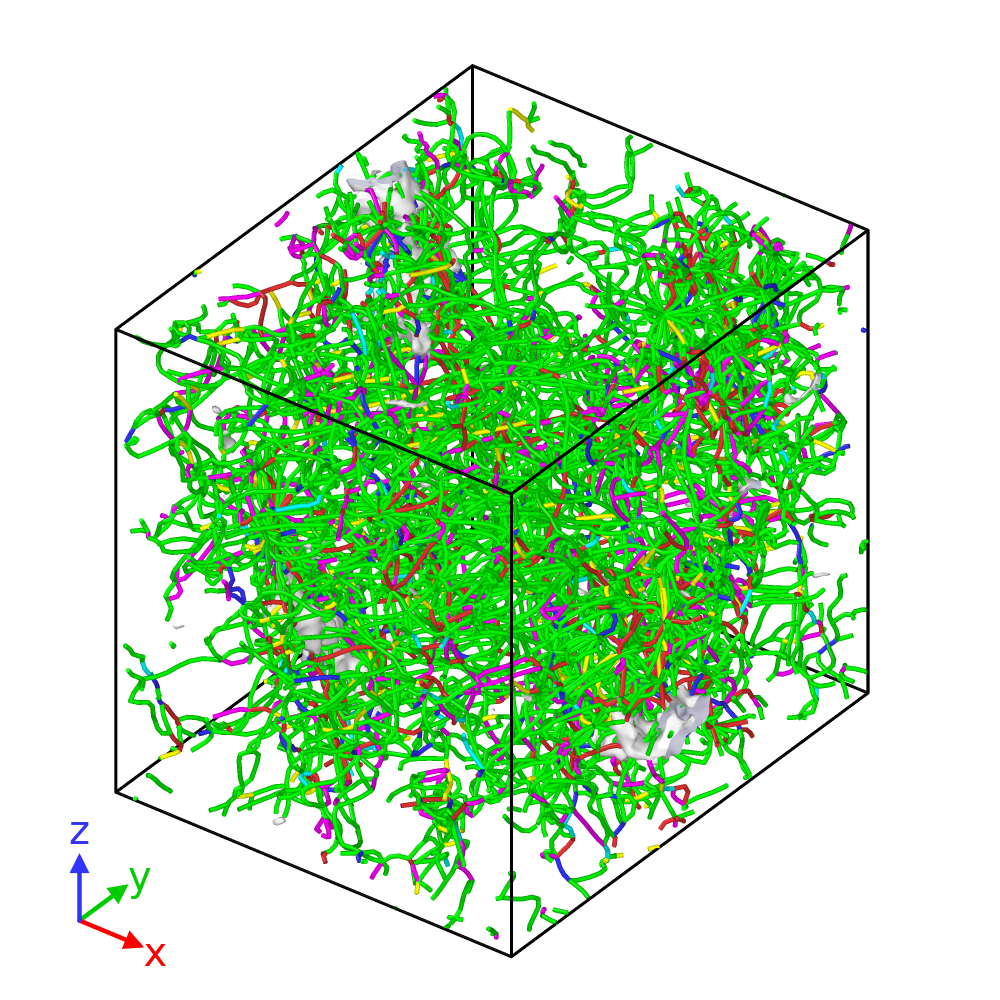}
    \caption{\hkl<110> strain 10\%}
\end{subfigure}
\begin{subfigure}[t]{0.2\textwidth}
    \includegraphics[width=\textwidth]{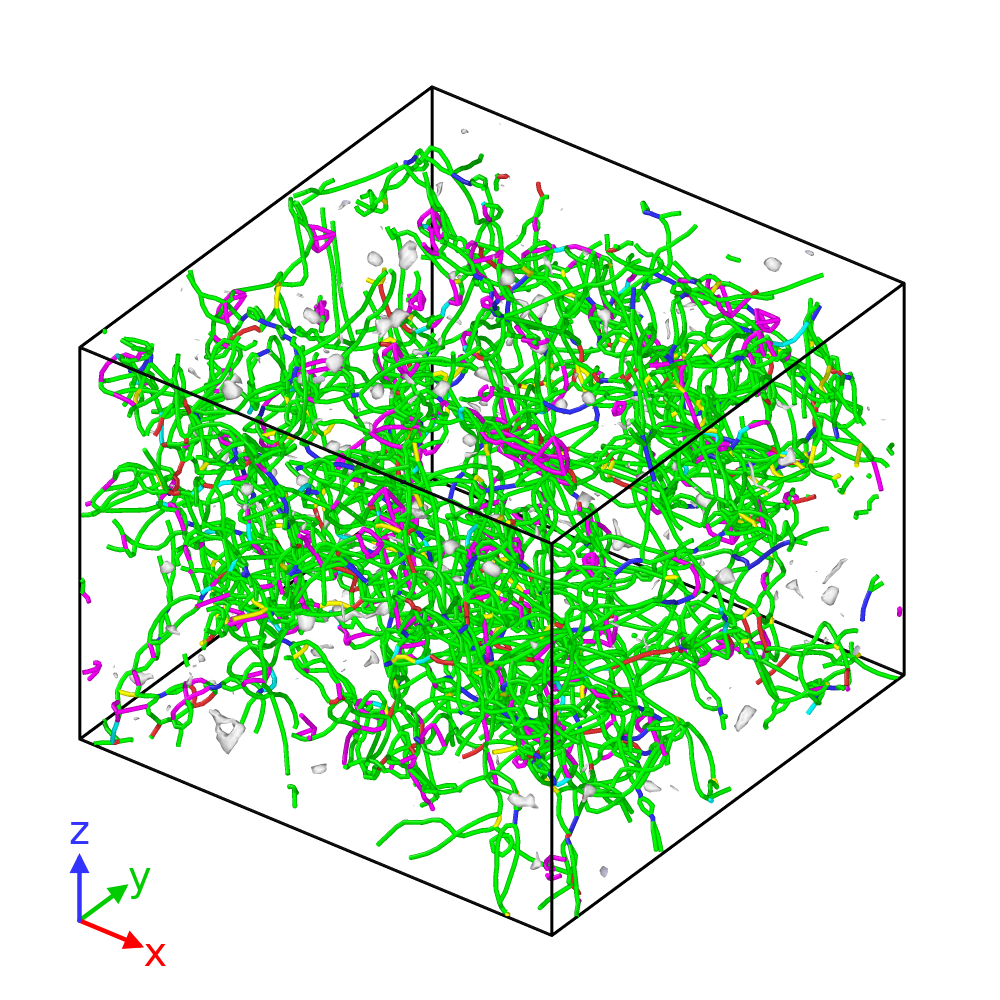}
      \caption{\hkl<111> strain 13\% }
\end{subfigure}
\begin{subfigure}[t]{0.2\textwidth}
    \includegraphics[width=\textwidth]{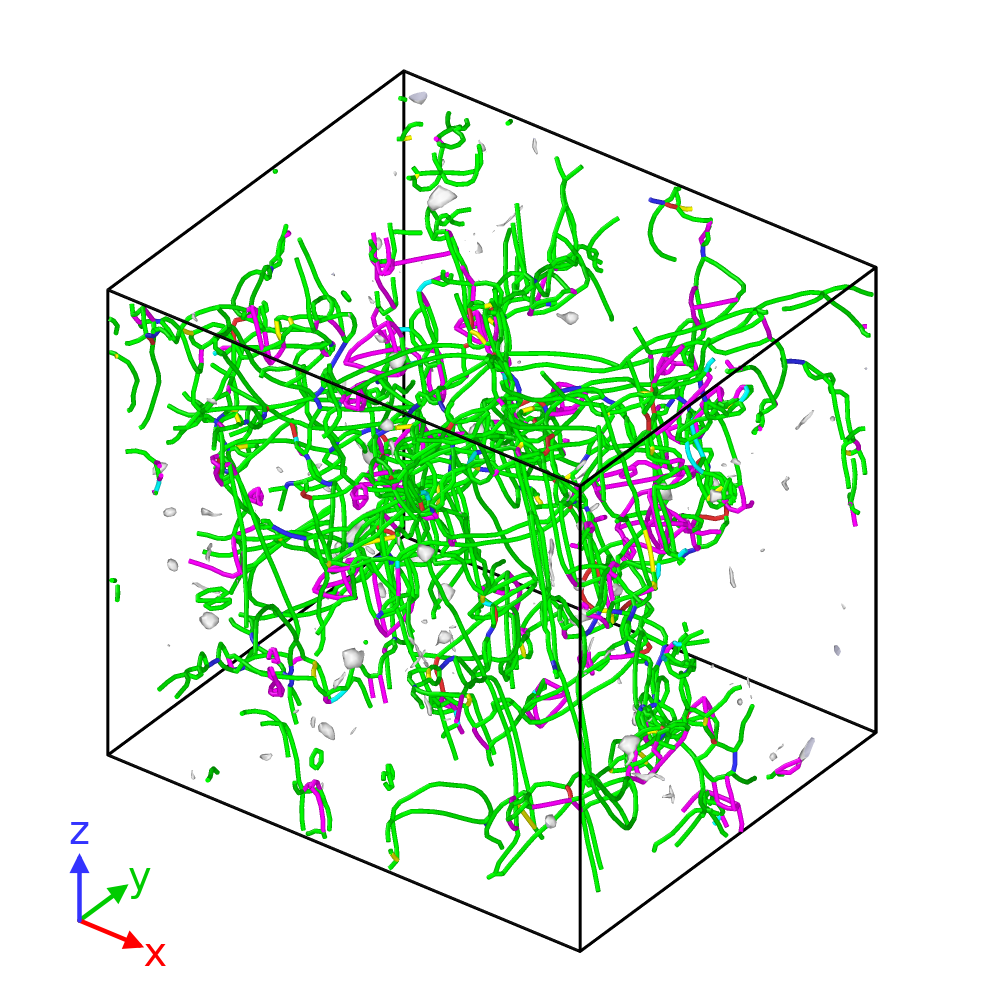}
    \caption{\hkl<112> strain 10\%}
\end{subfigure}
\begin{subfigure}[t]{0.2\textwidth}
    \includegraphics[width=\textwidth]{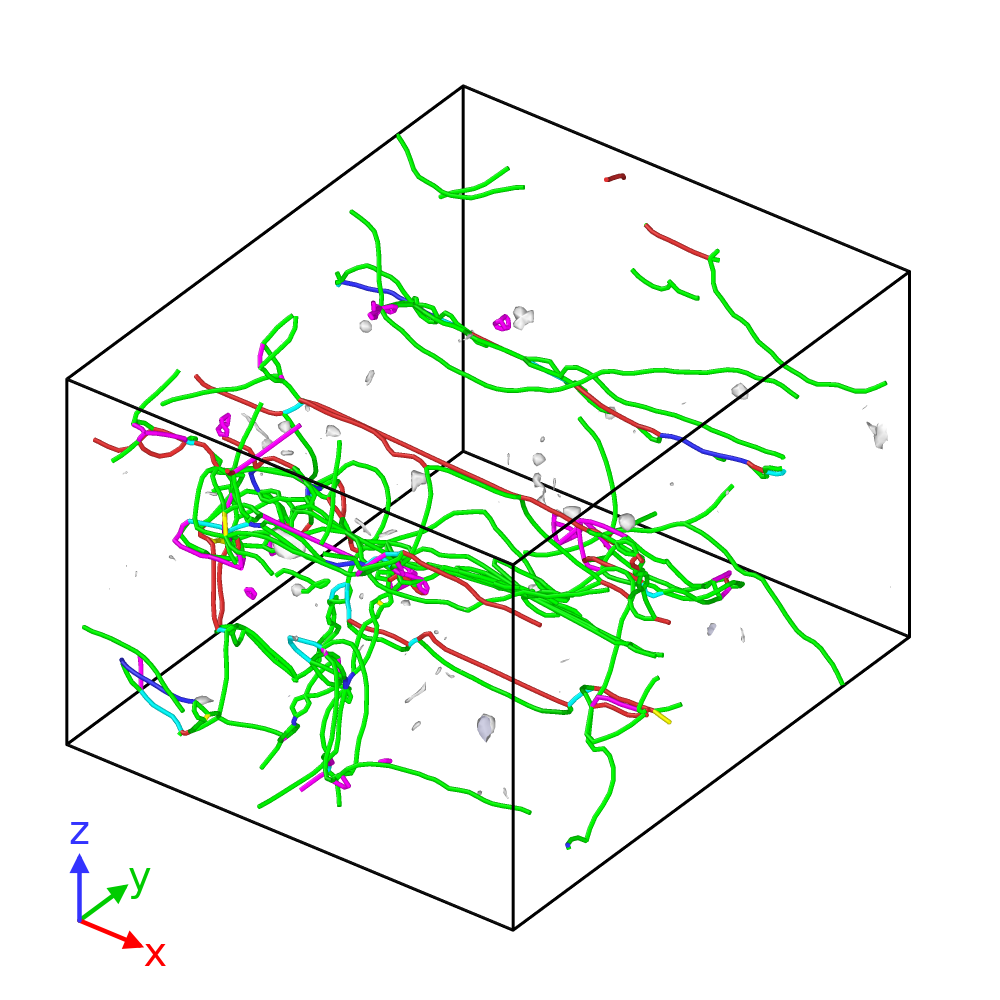}
      \caption{\hkl<100> strain 20\% }
\end{subfigure}
\begin{subfigure}[t]{0.2\textwidth}
    \includegraphics[width=\textwidth]{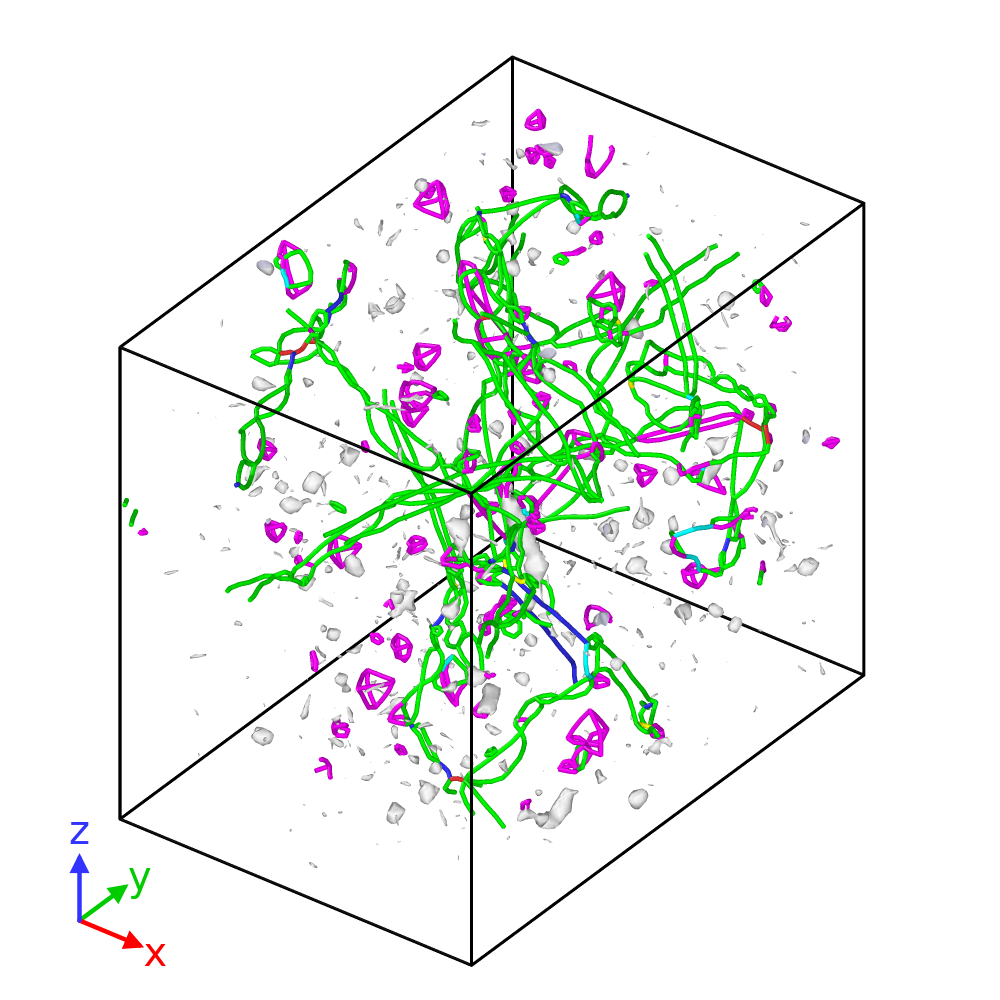}
    \caption{ \hkl<110> strain 20\%}
\end{subfigure}
\begin{subfigure}[t]{0.2\textwidth}
    \includegraphics[width=\textwidth]{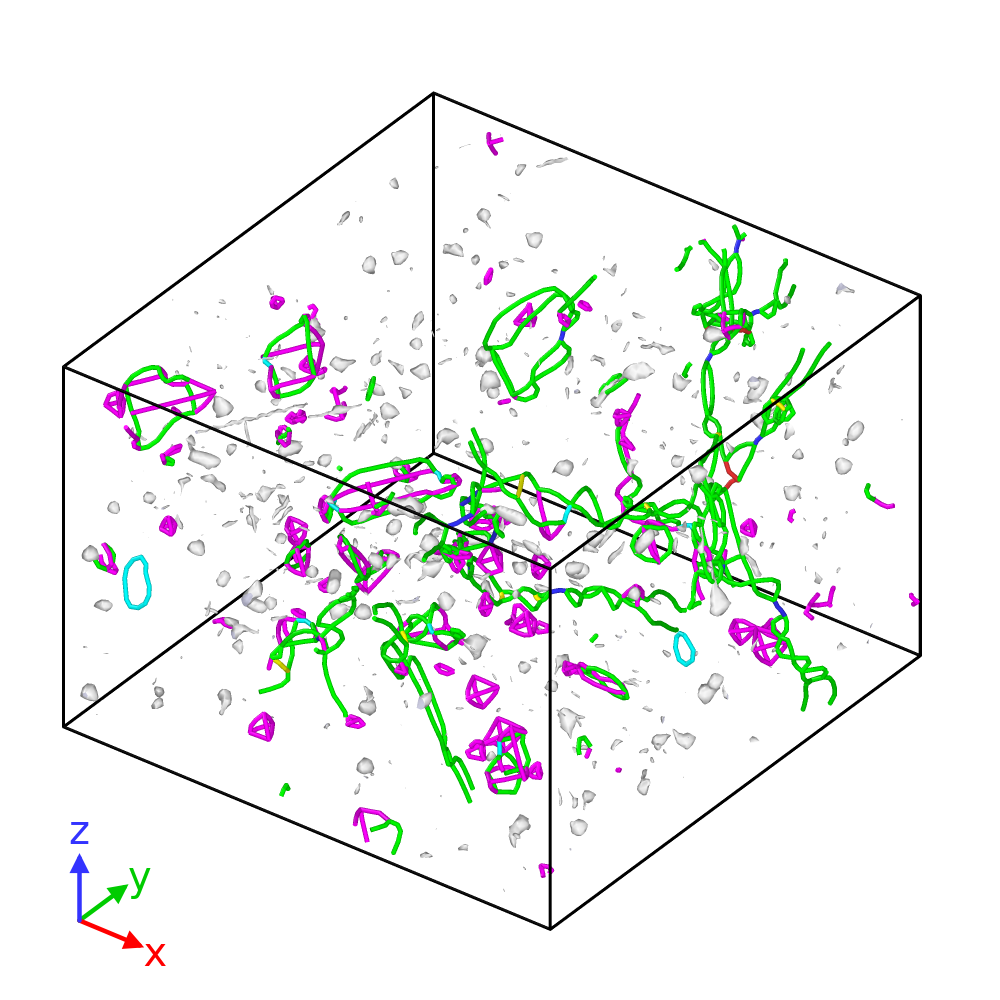}
      \caption{\hkl<111> strain 20\% }
\end{subfigure}
\begin{subfigure}[t]{0.2\textwidth}
    \includegraphics[width=\textwidth]{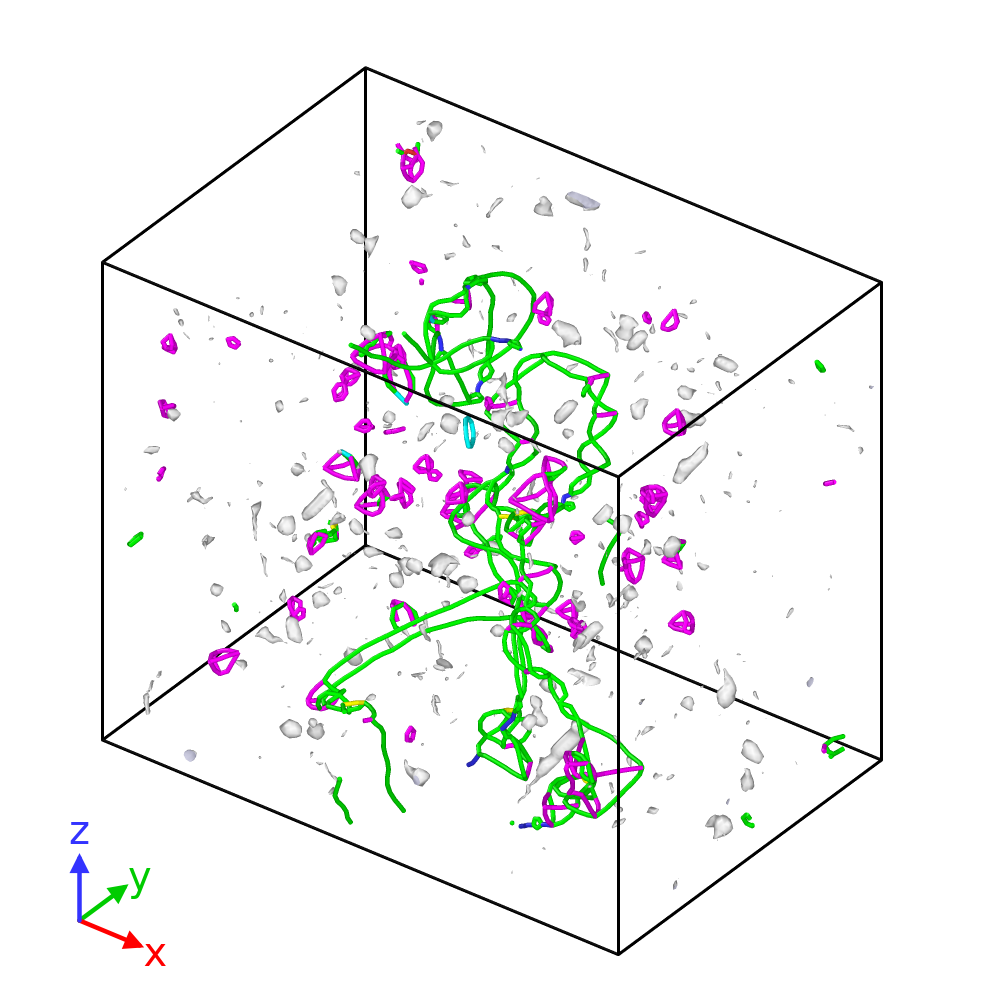}
    \caption{\hkl<112> strain 20\%}
\end{subfigure}

    \caption{Dislocation structures in Ni during compression in different loading directions with pristine cells. First row is directly after yielding and second row is the final configuration. Green lines represent Shockley partial dislocations, cyan lines represent Frank dislocations, pink lines represent stair-rod dislocations, and white blobs are defect clusters not identified as dislocations.}
    \label{fig:pristine-Ni}
\end{figure*}

Figs.~\ref{fig:pristine-Cu}-\ref{fig:pristine-Ni} show the dislocation structures identified by DXA analysis of the cells with all pristine materials and loading directions simulated with the tabGAP potentials. The figures show both the dislocation structures shortly after yielding and the final structures at 20\% strain. We see that shortly after yielding there is a complex dislocation network that mostly consists of Shockley-partial dislocation segments with some stair-rod type dislocations. During continued compression, the dislocation network sparsifies and at the end of the simulations, there are fewer dislocations, but they span the whole simulation cell. In addition, stacking-fault tetrahedra (SFT) can be observed for all materials, although fewer SFTs are found in Al. Furthermore, we also find considerably fewer SFTs in all materials during the compression in the \hkl<100> loading direction.  

\begin{figure*}
    \centering
\begin{subfigure}[t]{0.4\textwidth}
    \includegraphics[width=0.6\textwidth]{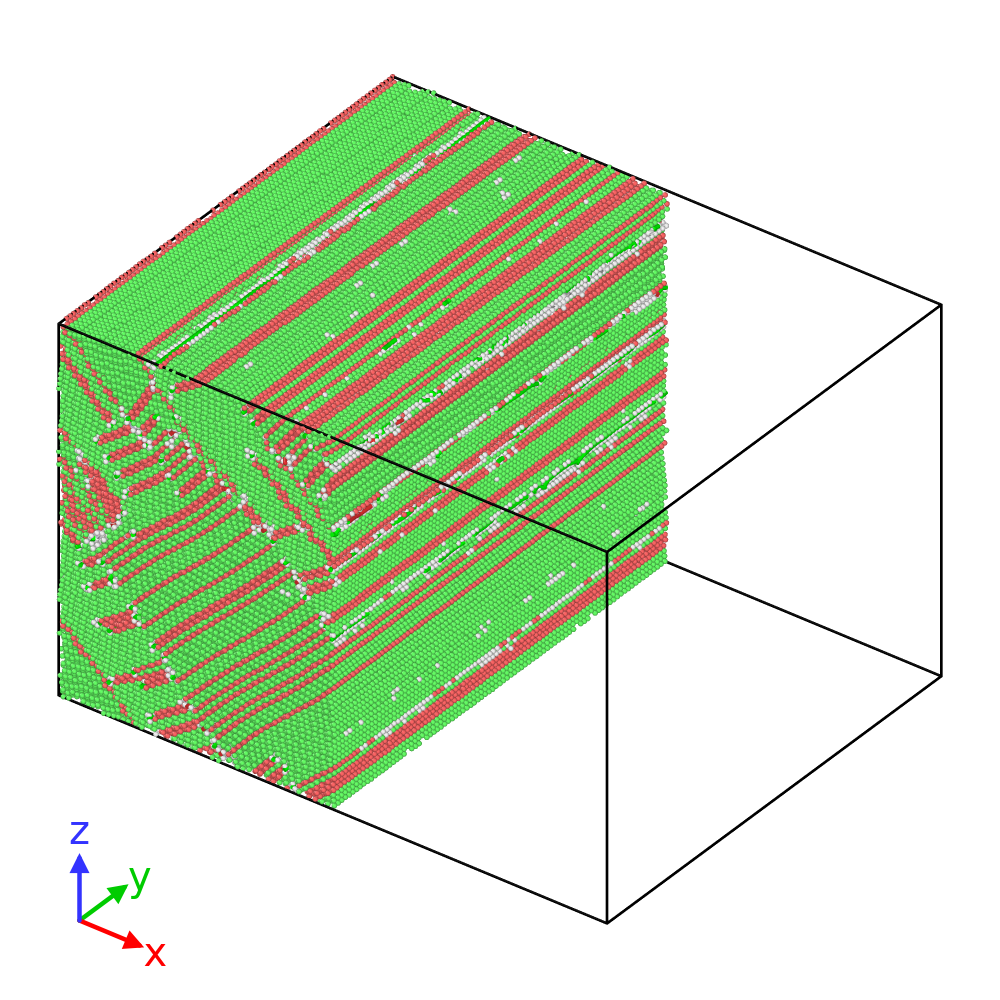}
      \caption{\hkl<100>}\label{fig:CNA:100}
\end{subfigure}
\begin{subfigure}[t]{0.4\textwidth}
    \includegraphics[width=0.6\textwidth]{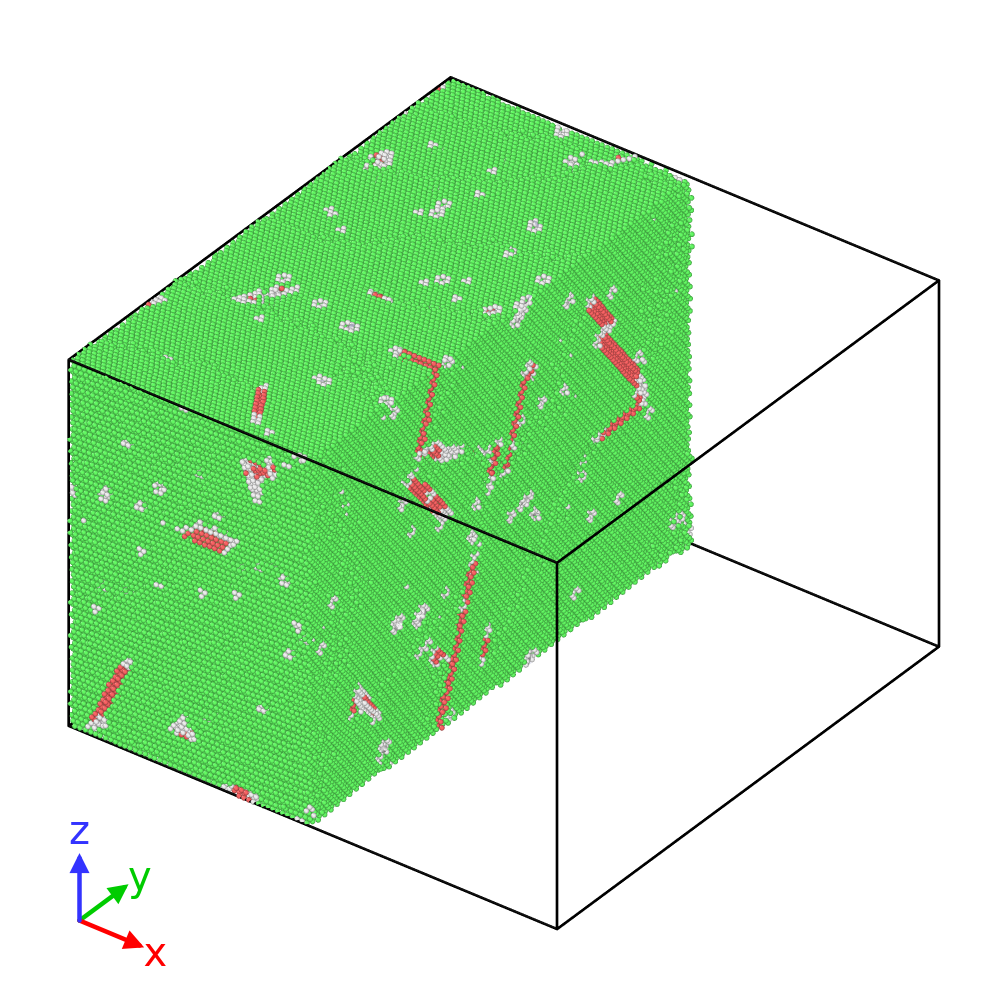}
    \caption{\hkl<111>}\label{fig:CNA:111}
\end{subfigure}
\begin{subfigure}[t]{0.4\textwidth}
    \includegraphics[width=0.6\textwidth]{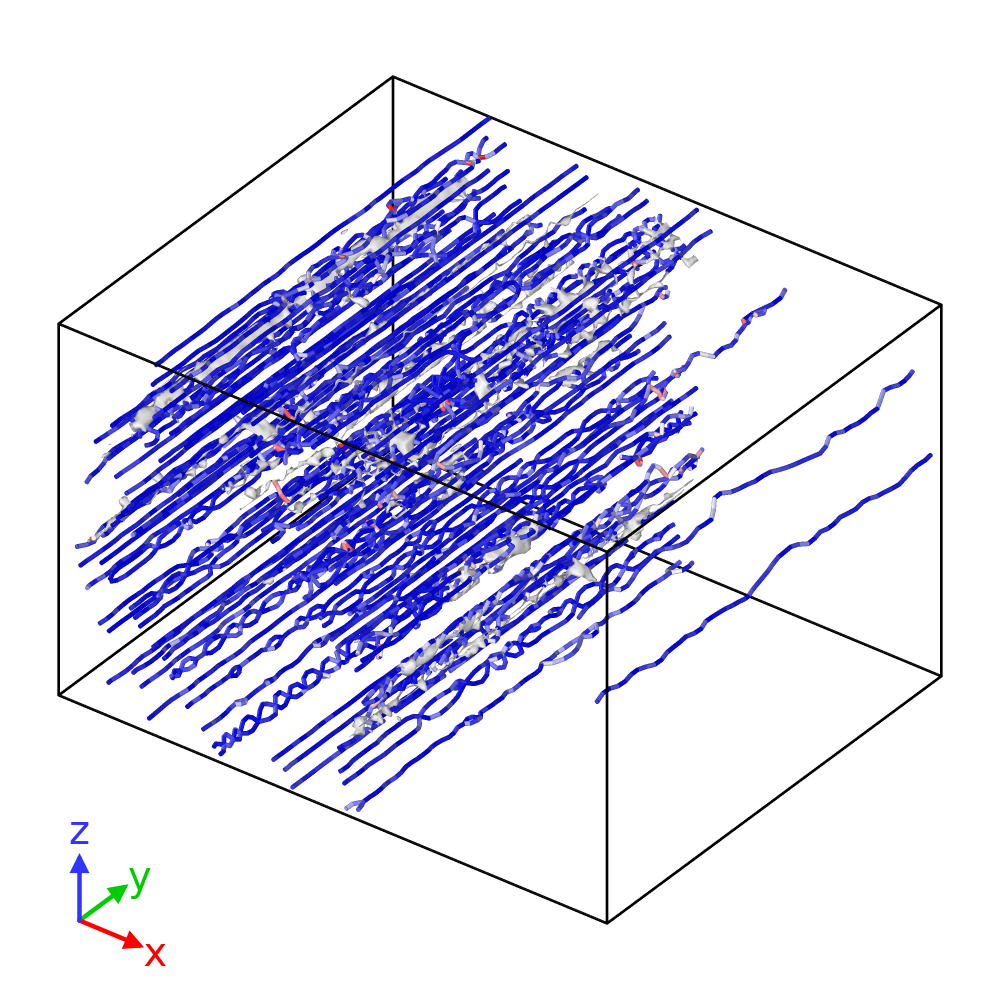}
      \caption{\hkl<100>}\label{fig:SE:100}
\end{subfigure}
\begin{subfigure}[t]{0.4\textwidth}
    \includegraphics[width=0.6\textwidth]{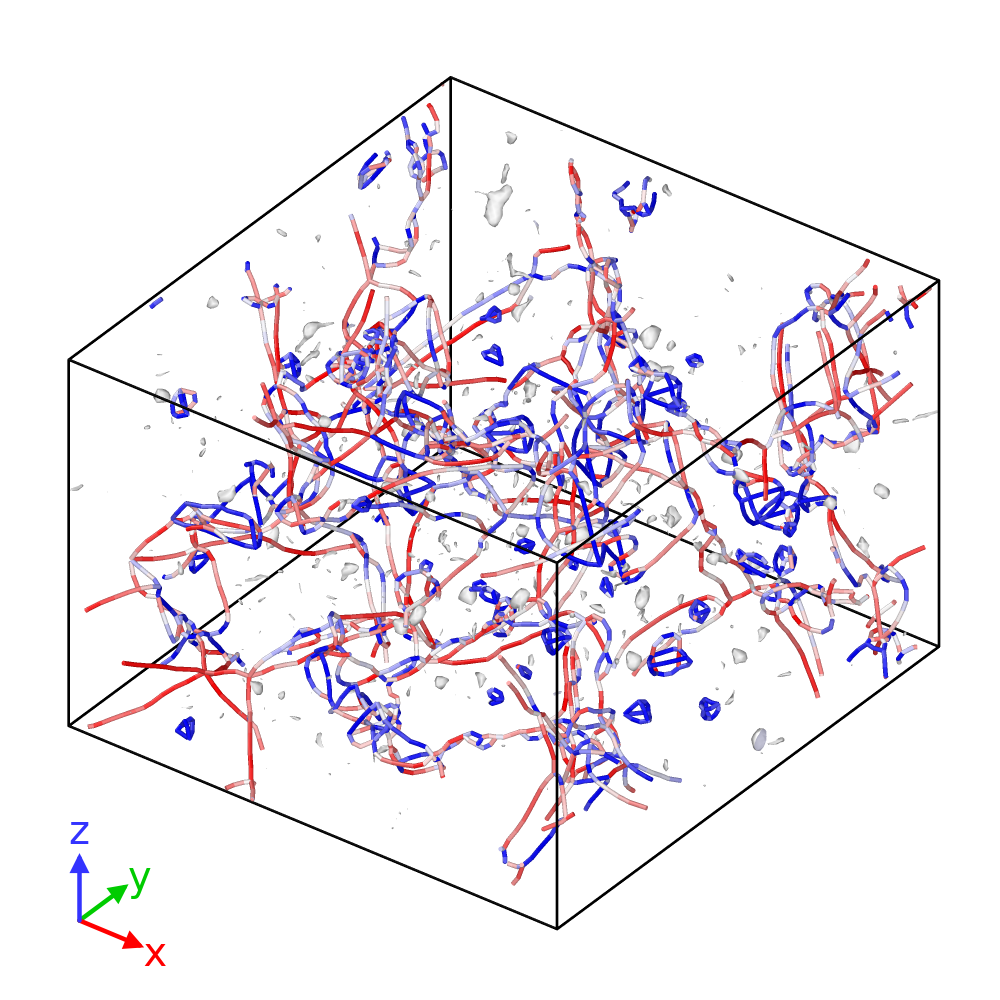}
    \caption{\hkl<111>}\label{fig:SE:111}
\end{subfigure}
    \caption{(a) and (b) shows the atomistic structure of Cu in \hkl<100> and \hkl<111> loading directions at the final strain of 20\%. The coloring of the atoms indicate lattice type of the atom based on common neighbour analysis. Green atoms are FCC, red atoms are HCP and white atoms are without defined lattice structure. Half of the atoms are cut away in order to get a better view of the internal structure. (c) and (d) show the corresponding dislocation structure with screw-type dislocations shown in red and edge-type in blue.}
    \label{fig:CNA}
\end{figure*}

The \hkl<100> loading direction in Cu falls out from the general trend that we observe in our simulations. To clarify the reason for this behavior, we plot in Fig.~\ref{fig:CNA} the resulting atomic structures of the Cu pristine cells after the loading in the \hkl<100> and \hkl<111> directions (Figs. ~\ref{fig:CNA}a and ~\ref{fig:CNA}b. respectively) along with the dislocations as identified by the DXA in Figs.~\ref{fig:CNA}c and ~\ref{fig:CNA}d. We color the dislocations according to their type, i.e. the screw dislocations are shown in red and the edge ones in blue. 

We see in the Cu cell deformed with the loading in the \hkl<100> direction a family of parallel stacking faults has formed, while loading in the \hkl<111> (or any other) direction does not cause similar cascade of the slips parallel to one another. Furthermore analyzing the dislocation structures, we observe that in the cell deformed in the \hkl<100> loading direction only the edge type dislocations are present. In the cell deformed in the \hkl<111> loading direction we observe the mixture of both edge- and screw-type dislocations. For the other elements and loading directions we see a mixture of screw- and edge-type dislocations, similar to that in the Cu deformed  in the \hkl<111> loading direction, but the ratio between the two types varies. Additionally, during compression no SFTs were identified in Cu for the \hkl<100> direction. The dislocation structure of the Cu cell during the \hkl<100> loading was not an artefact of the potential, since similar structures have developed in the cells simulated with the EAM potential. Furthermore, in Ni simulated with the EAM potentials the dislocation structures are formed similar to those in the Cu cells deformed in the \hkl<100> loading direction.

\begin{figure*}
    \centering
\begin{subfigure}[t]{0.2\textwidth}
    \includegraphics[width=\textwidth]{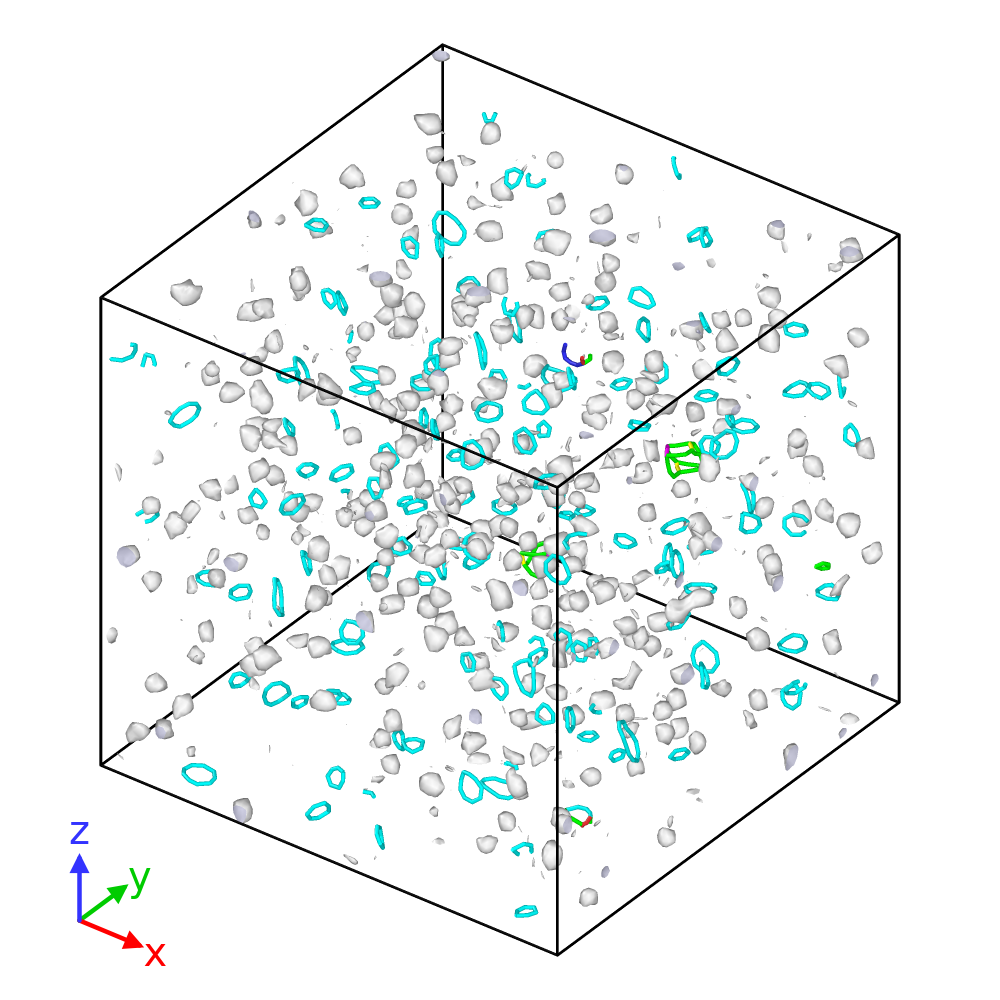}
      \caption{strain 0\% }
\end{subfigure}
\begin{subfigure}[t]{0.2\textwidth}
    \includegraphics[width=\textwidth]{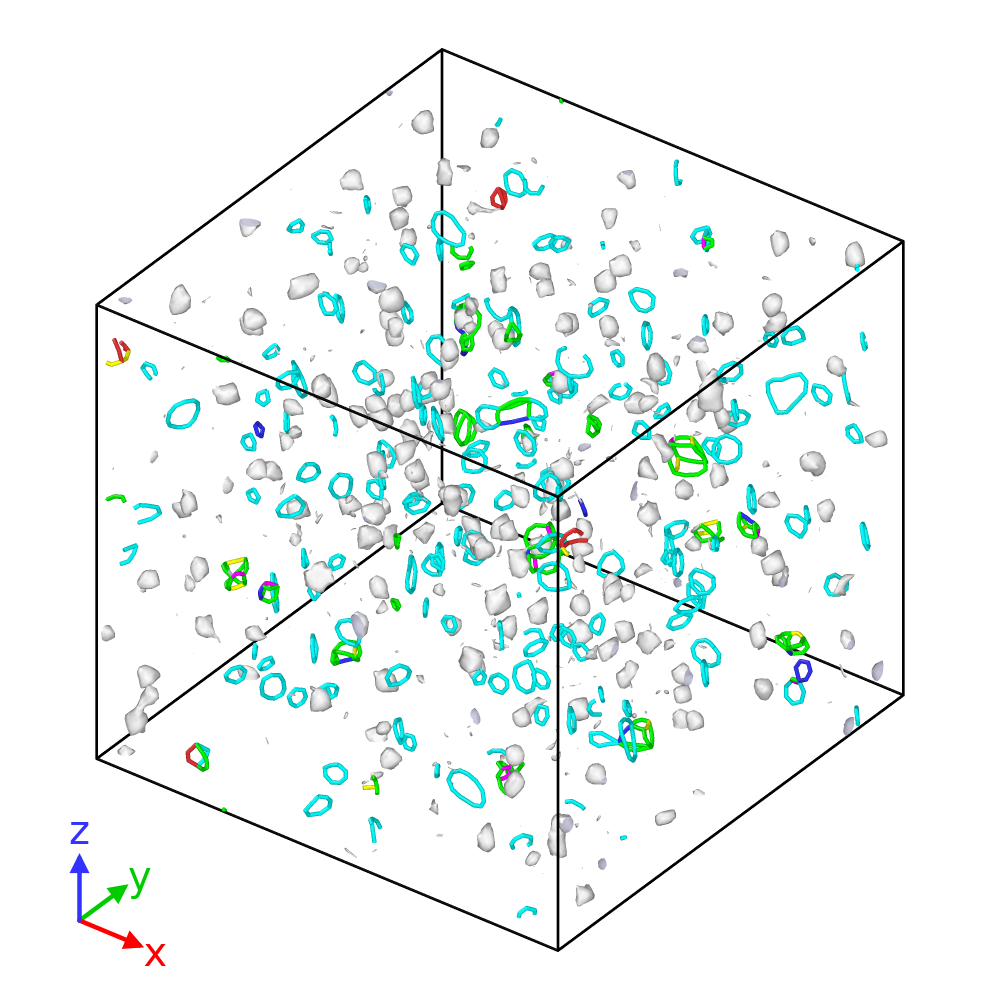}
    \caption{strain 3\%}
\end{subfigure}
\begin{subfigure}[t]{0.2\textwidth}
    \includegraphics[width=\textwidth]{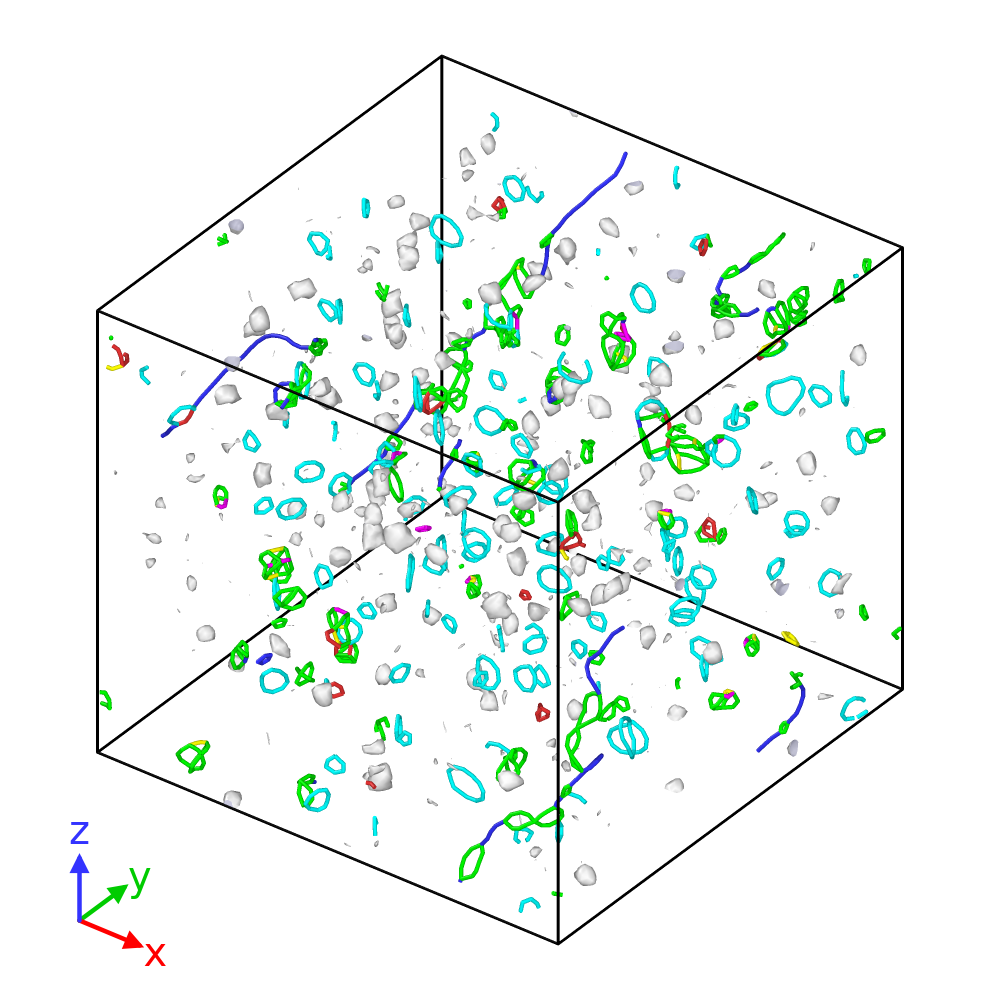}
      \caption{strain 4.7\% }
\end{subfigure}
\begin{subfigure}[t]{0.2\textwidth}
    \includegraphics[width=\textwidth]{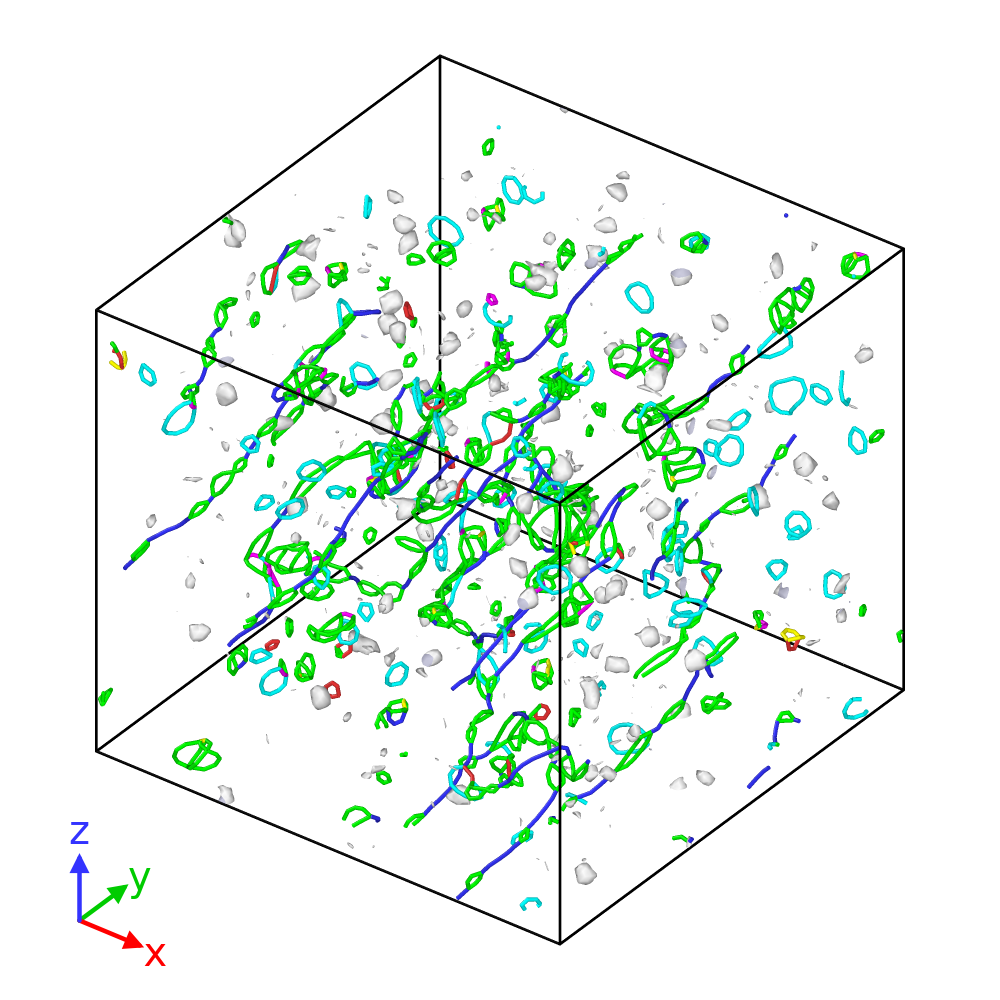}
    \caption{strain 4.8\%}
\end{subfigure}
\begin{subfigure}[t]{0.2\textwidth}
    \includegraphics[width=\textwidth]{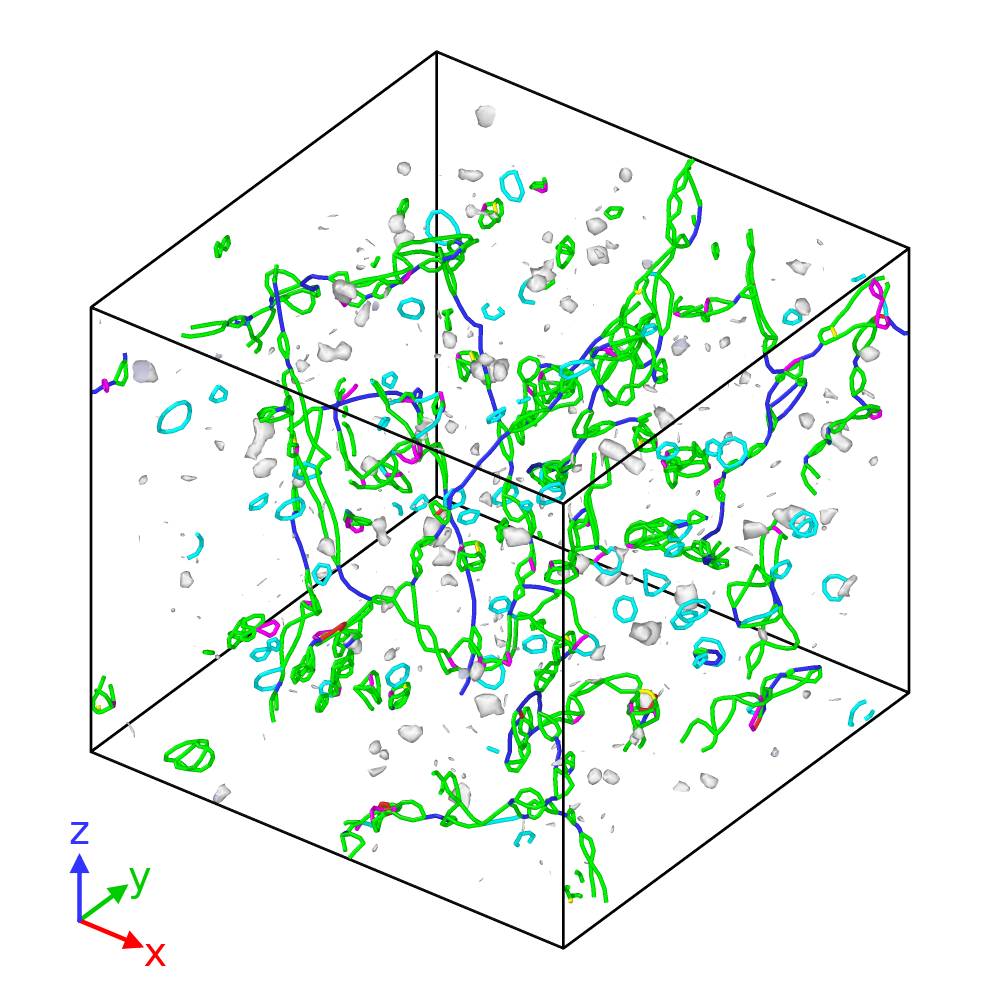}
      \caption{strain 5\% }
\end{subfigure}
\begin{subfigure}[t]{0.2\textwidth}
    \includegraphics[width=\textwidth]{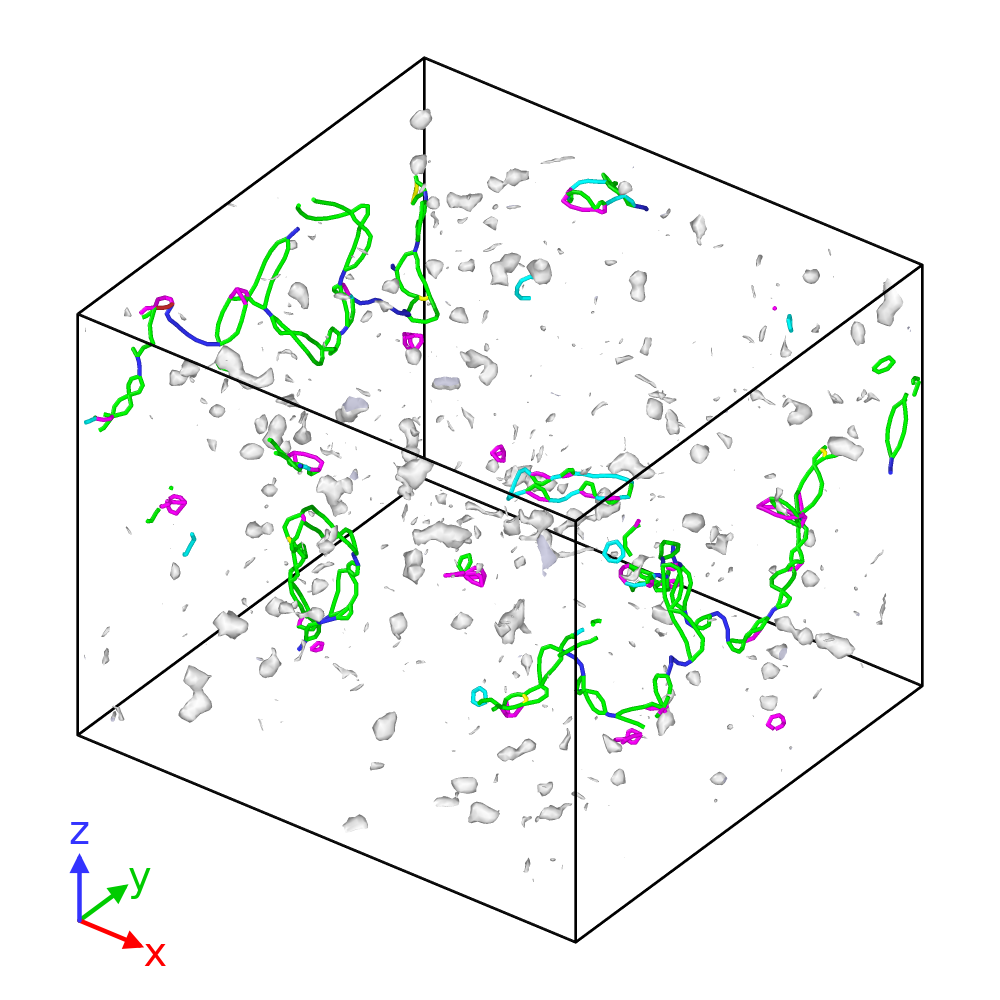}
    \caption{strain 10\%}
\end{subfigure}
\begin{subfigure}[t]{0.2\textwidth}
    \includegraphics[width=\textwidth]{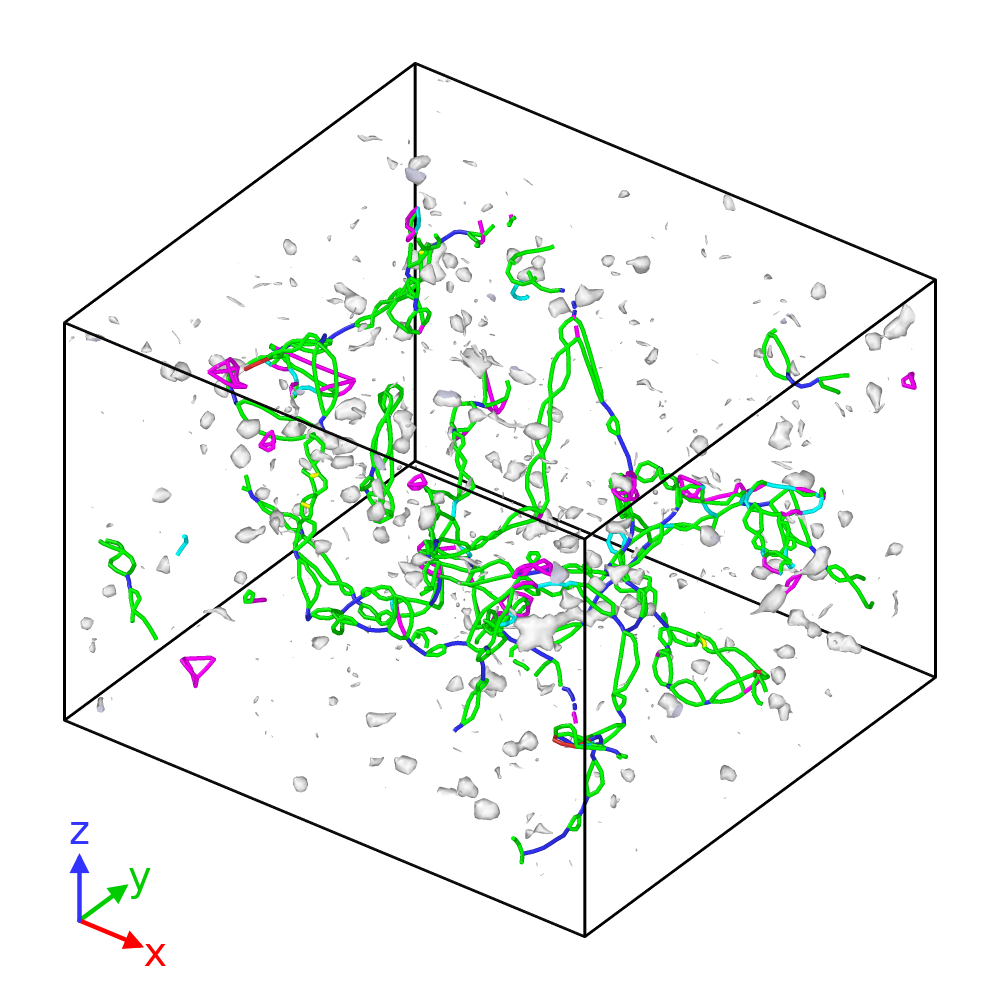}
      \caption{strain 15\% }
\end{subfigure}
\begin{subfigure}[t]{0.2\textwidth}
    \includegraphics[width=\textwidth]{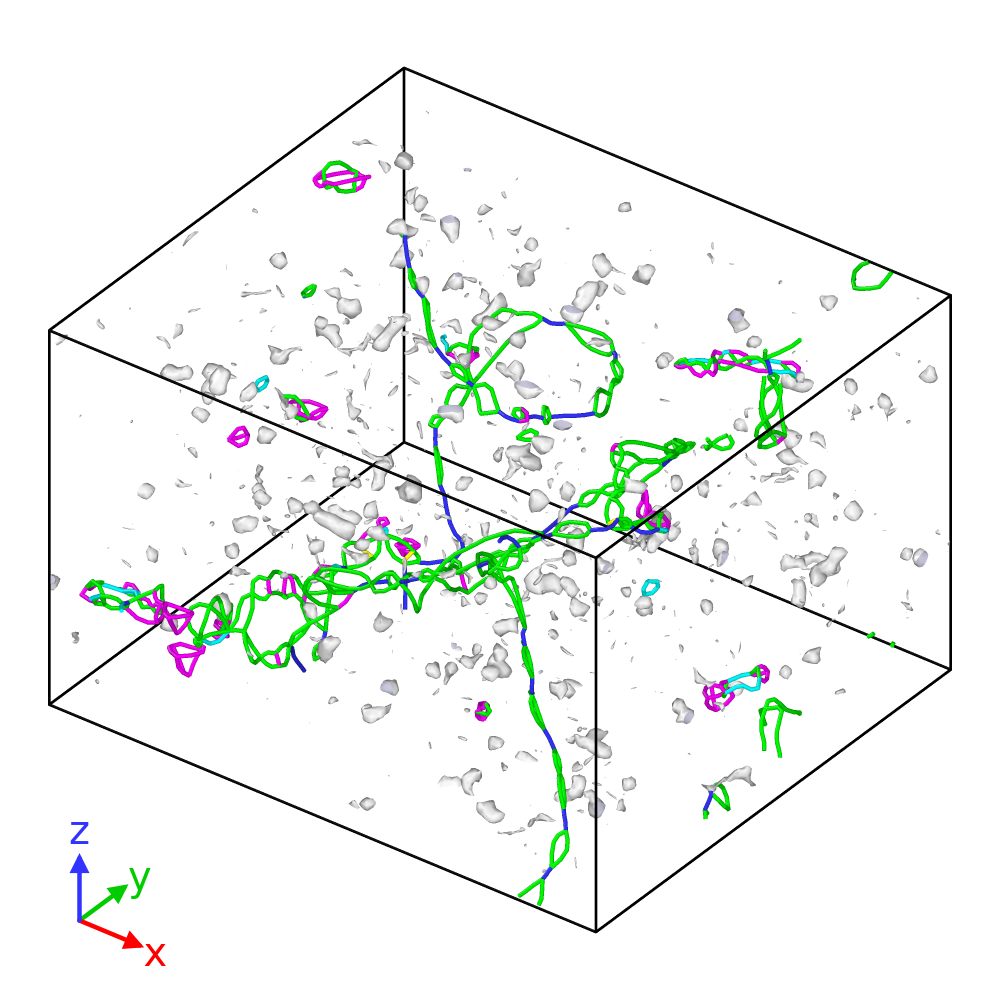}
    \caption{strain 20\%}
\end{subfigure}

    \caption{Al cells containing pre-existing defects from Frenkel pair insertion during compression in \hkl<111> loading direction. Green lines represent Shockley partial dislocations, cyan lines represent Frank dislocations, pink lines represent stair-rod dislocations, and white blobs are defect clusters not identified as dislocations.}
    \label{fig:FP-Al}
\end{figure*}

Figs.~\ref{fig:FP-Al}a-\ref{fig:FP-Al}h demonstrates the evolution of dislocation structures under compression in the \hkl<111> loading direction of the defected Al as relaxed after the random insertion of FPs. We see that at early stages of compression, small dislocations start nucleating at the pre-existing defects in the system. The longer dislocations connecting the small initially disconnected ones appear at the strain of 4.7\%, after which the dislocation network first grows denser and more connected, and then the initial smaller dislocations start fusing forming larger dislocation structures. Eventually, we observe formation of the dislocation network which is rather similar to that formed at deformation of the cell with the pristine structure. The dislocation networks grown in the cells during the loading in other directions as well as in the remaining materials with the pre-damaged structures can be found in the supplementary material. 

\begin{figure*}
    \centering
\begin{subfigure}[t]{0.24\textwidth}
    \includegraphics[width=\textwidth]{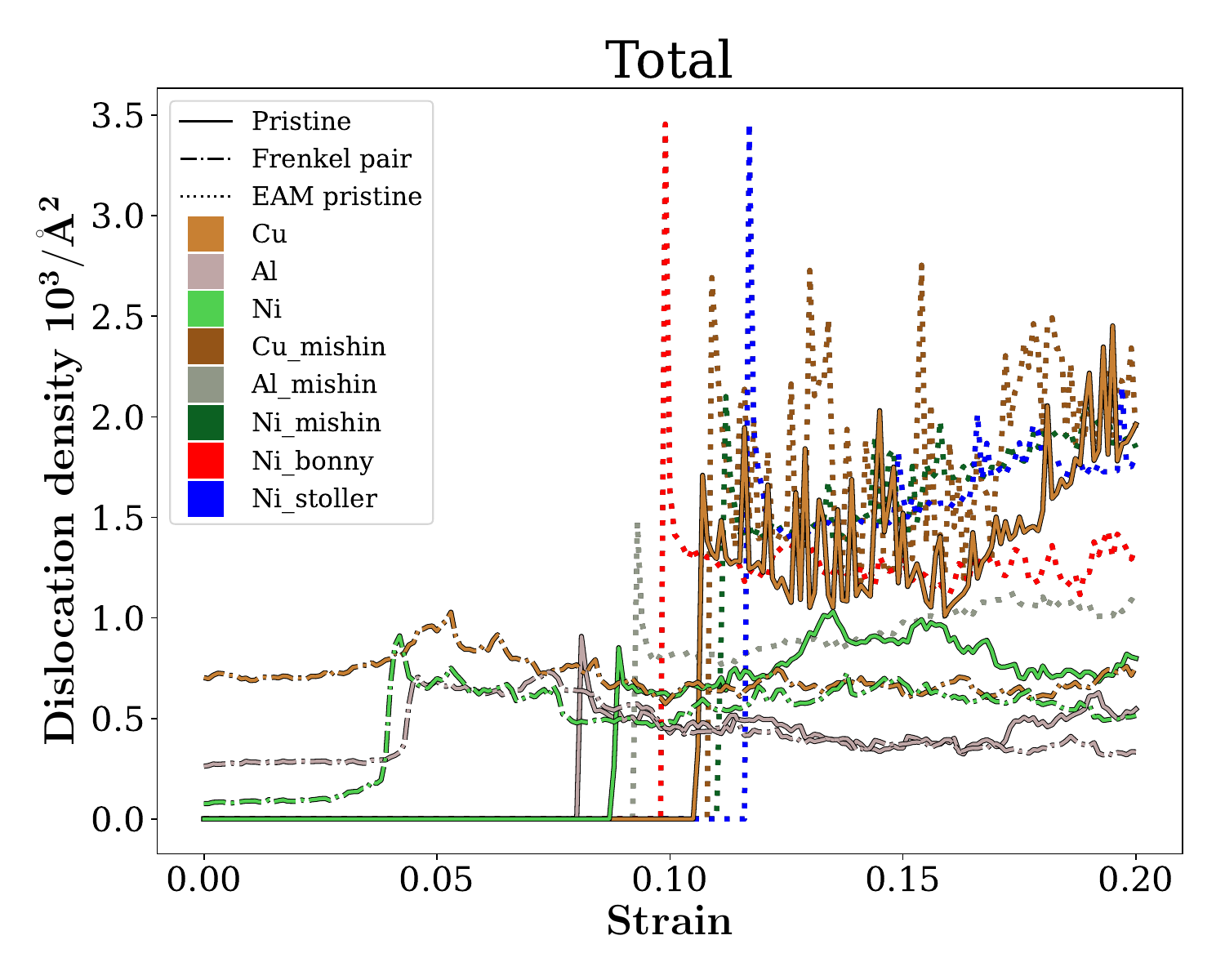}
      \caption{\hkl<100>  }
\end{subfigure}
\begin{subfigure}[t]{0.24\textwidth}
    \includegraphics[width=\textwidth]{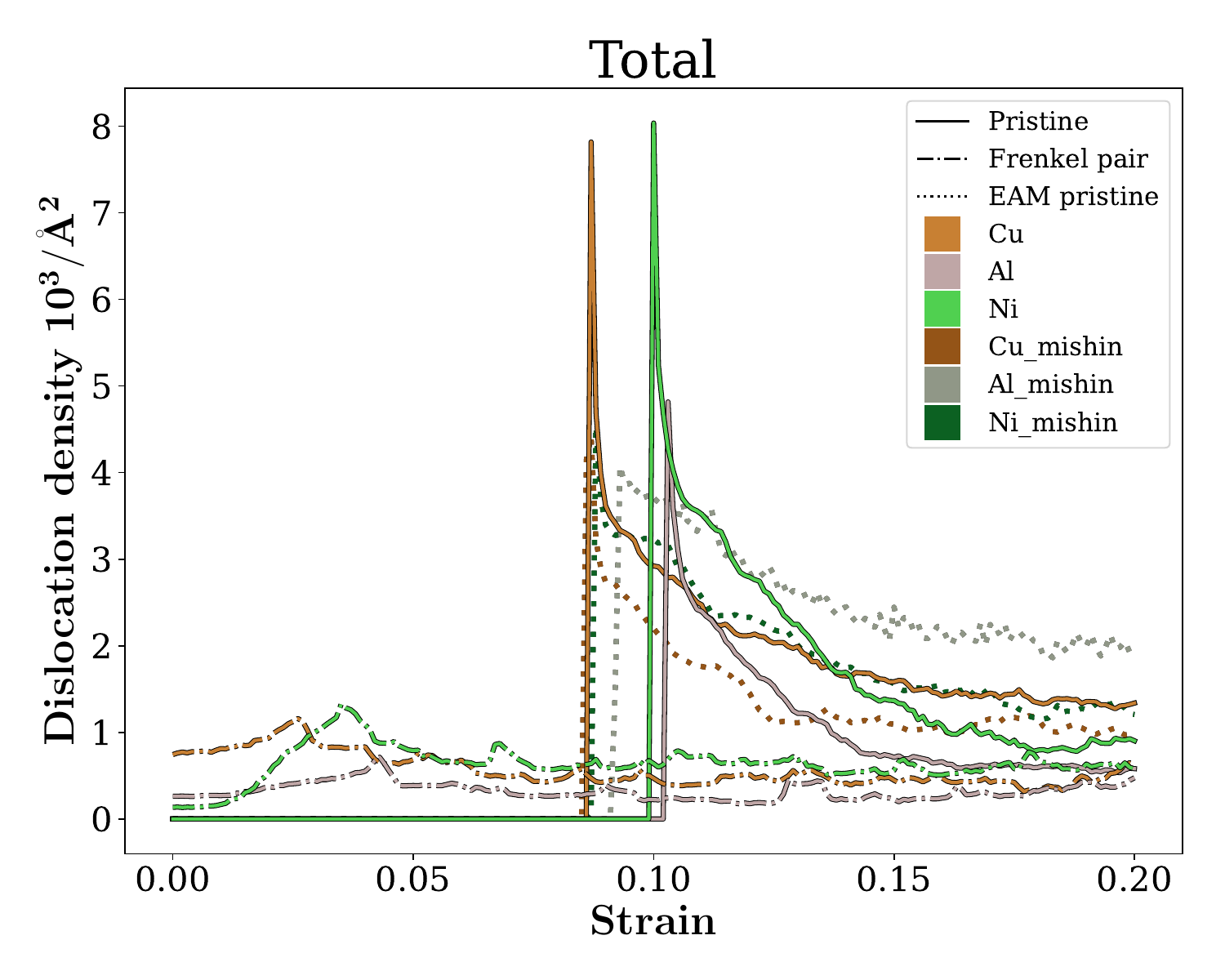}
    \caption{\hkl<110>}
\end{subfigure}
\begin{subfigure}[t]{0.24\textwidth}
    \includegraphics[width=\textwidth]{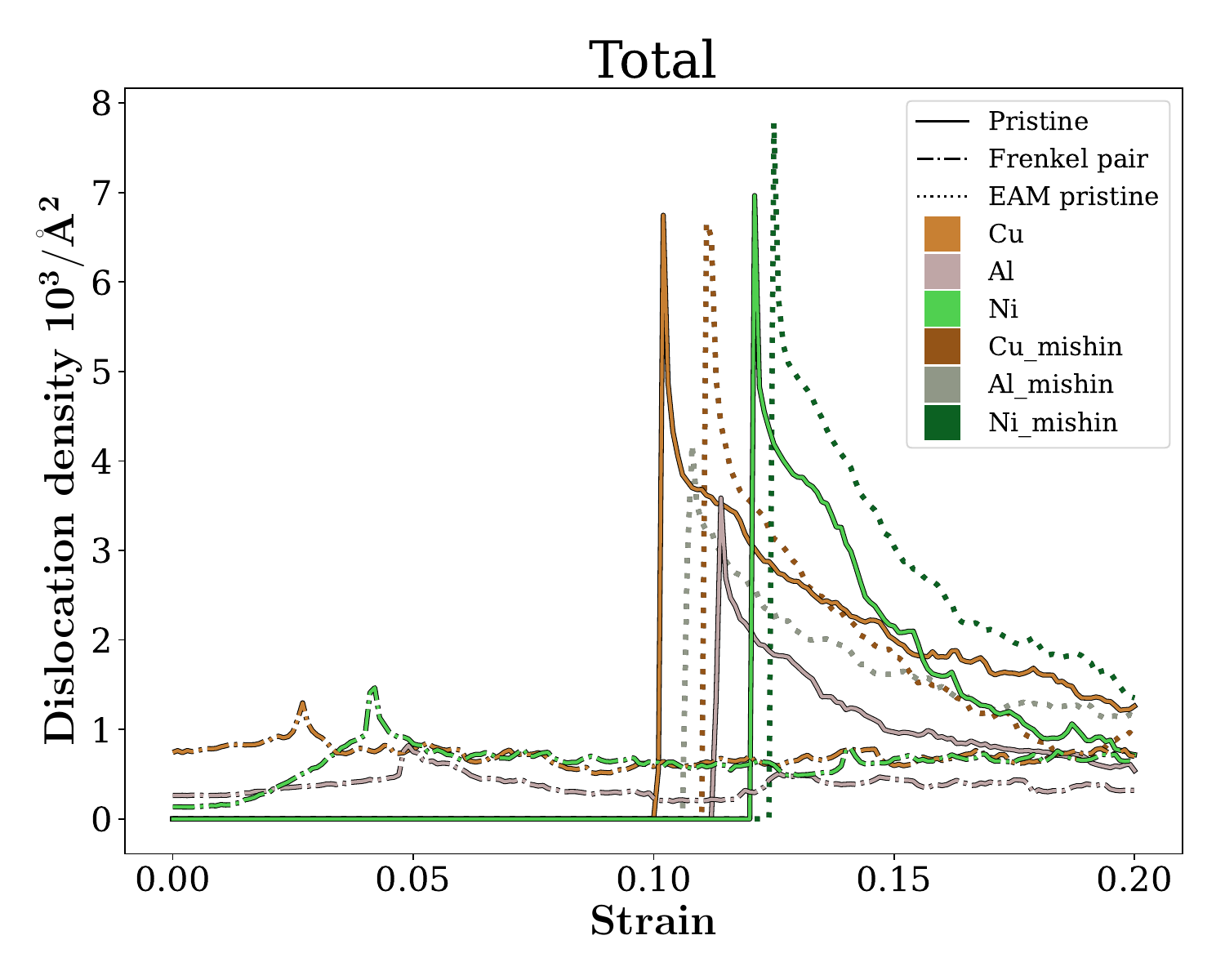}
      \caption{\hkl<111>}
\end{subfigure}
\begin{subfigure}[t]{0.24\textwidth}
    \includegraphics[width=\textwidth]{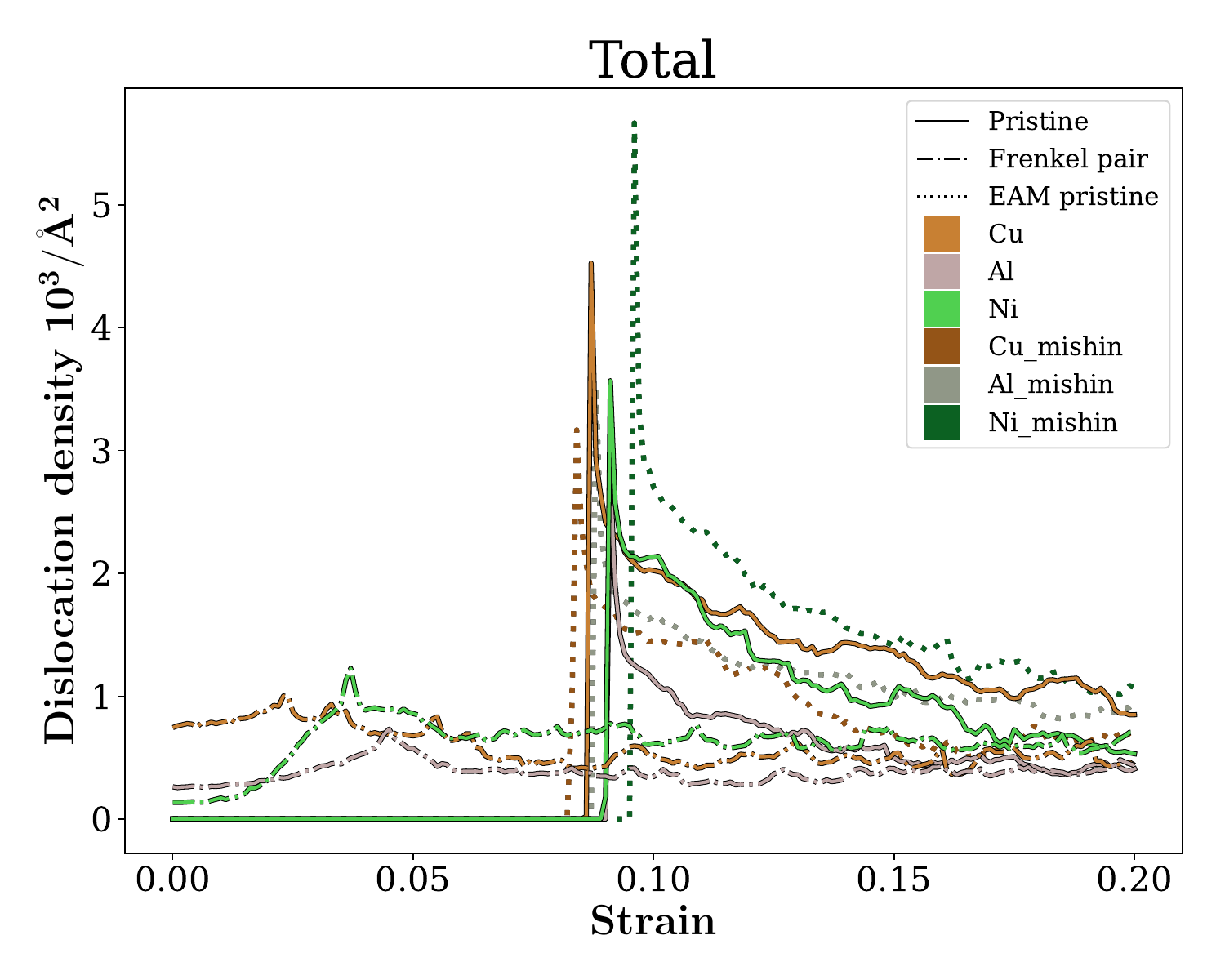}
    \caption{\hkl<112>}
\end{subfigure}
\begin{subfigure}[t]{0.24\textwidth}
    \includegraphics[width=\textwidth]{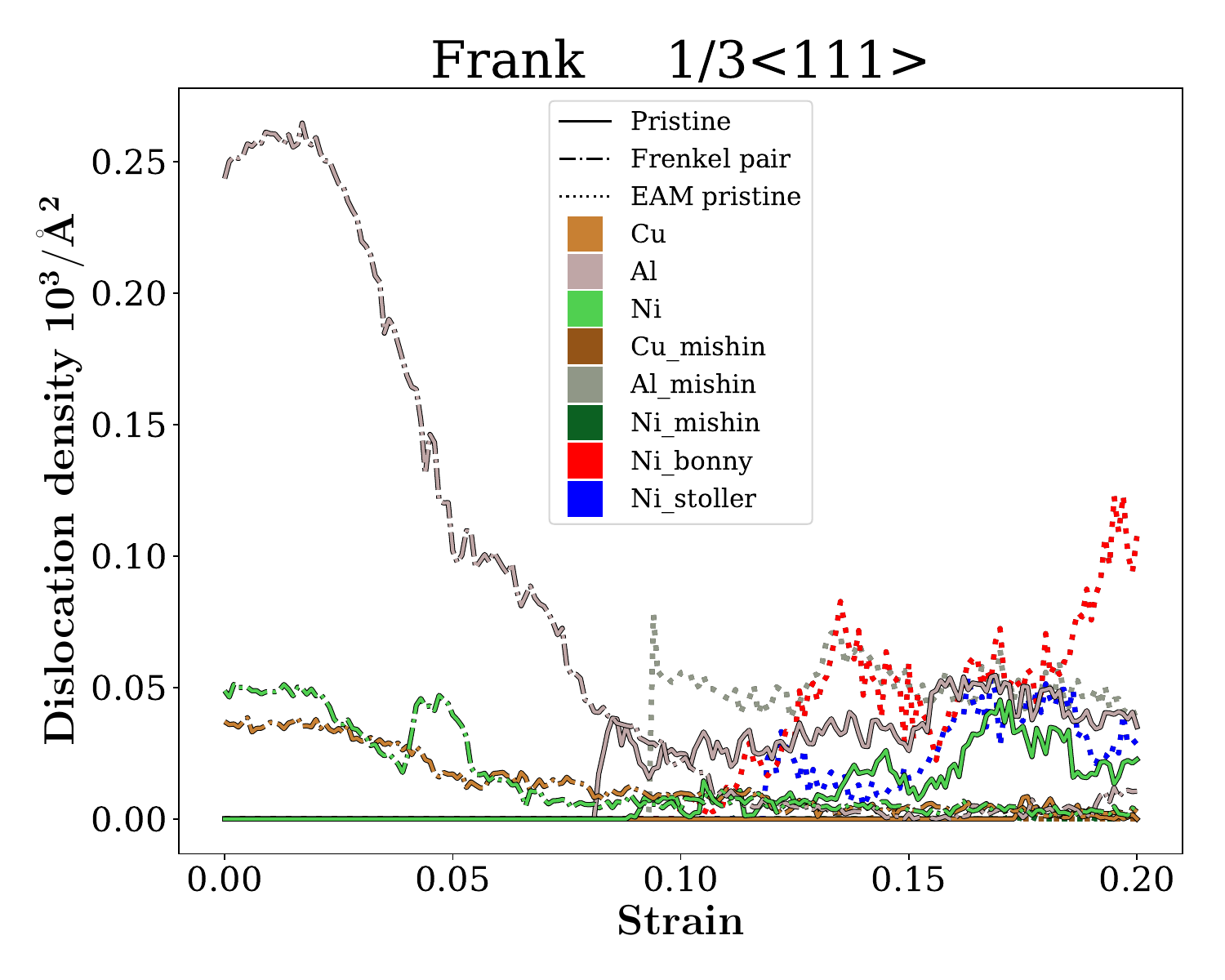}
      \caption{\hkl<100>  }
\end{subfigure}
\begin{subfigure}[t]{0.24\textwidth}
    \includegraphics[width=\textwidth]{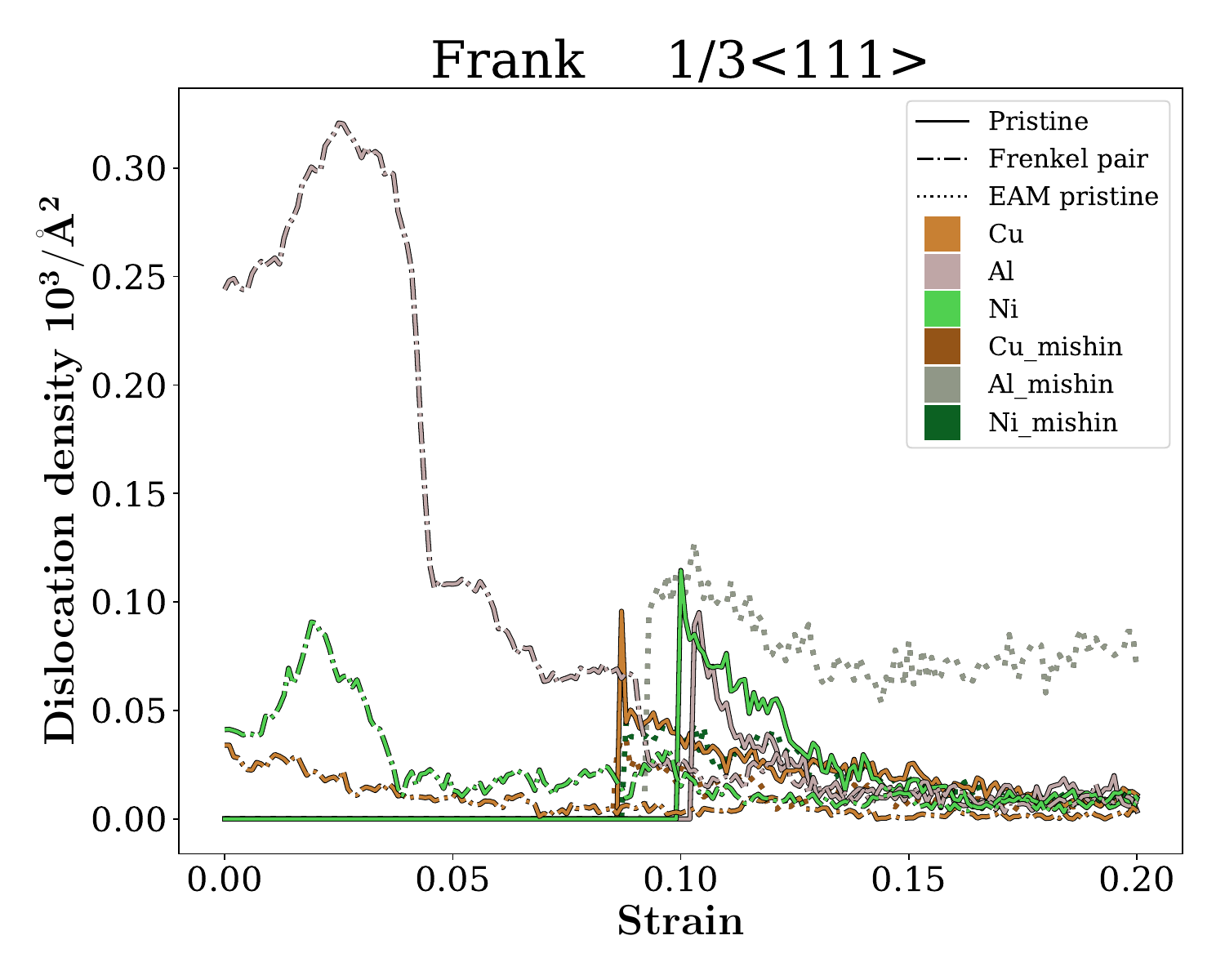}
    \caption{\hkl<110>}
\end{subfigure}
\begin{subfigure}[t]{0.24\textwidth}
    \includegraphics[width=\textwidth]{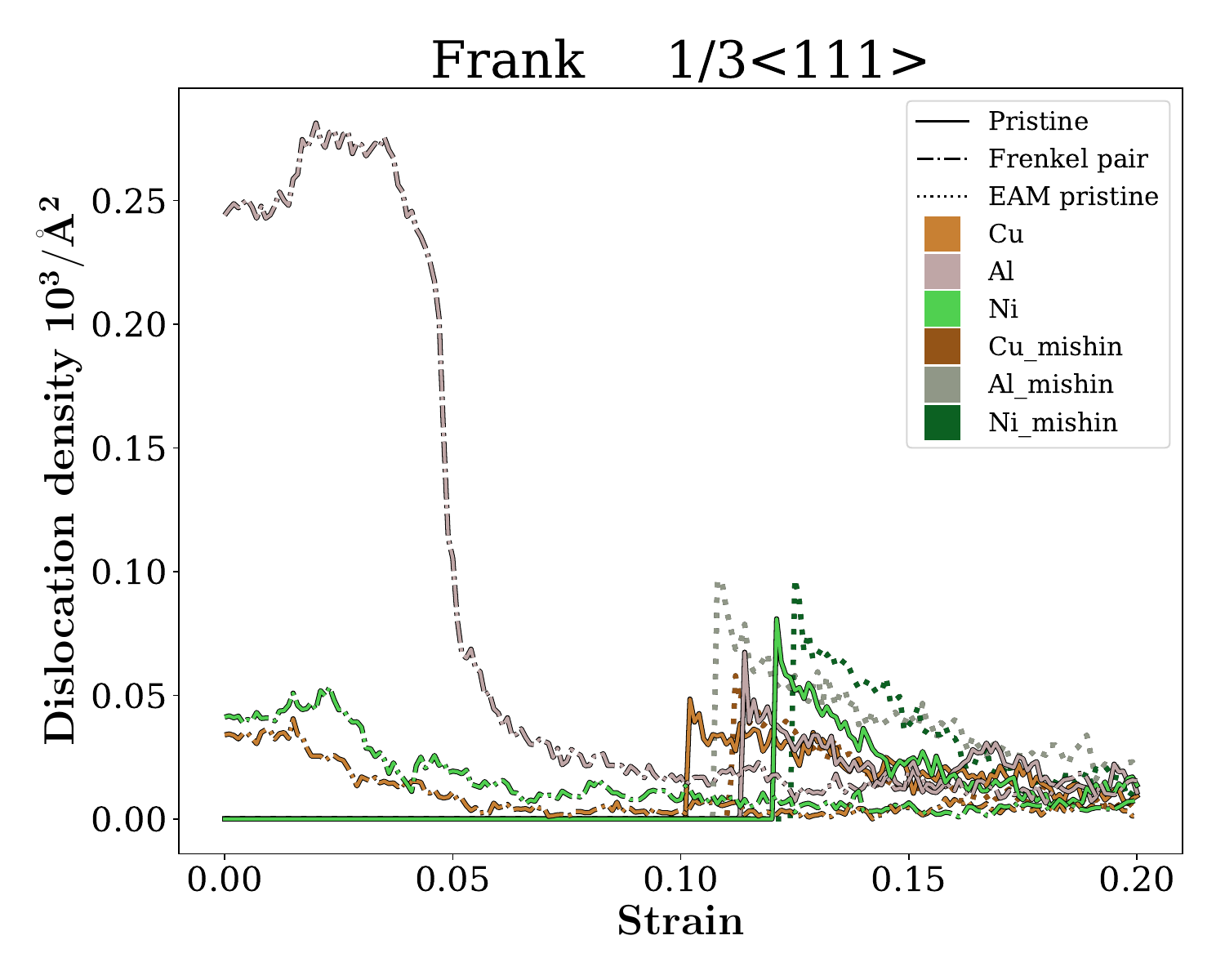}
      \caption{\hkl<111>}
\end{subfigure}
\begin{subfigure}[t]{0.24\textwidth}
    \includegraphics[width=\textwidth]{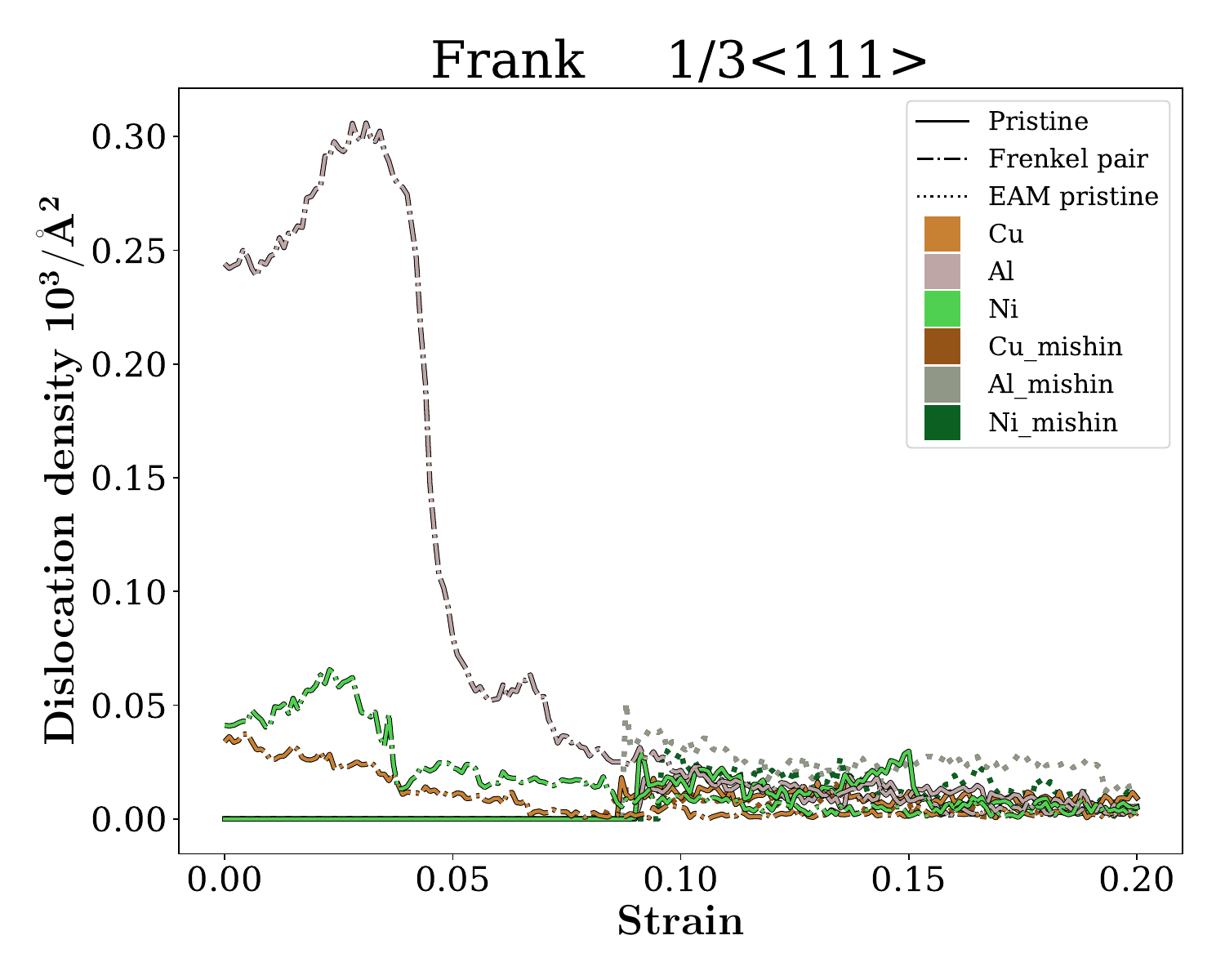}
    \caption{\hkl<112>}
\end{subfigure}
\begin{subfigure}[t]{0.24\textwidth}
    \includegraphics[width=\textwidth]{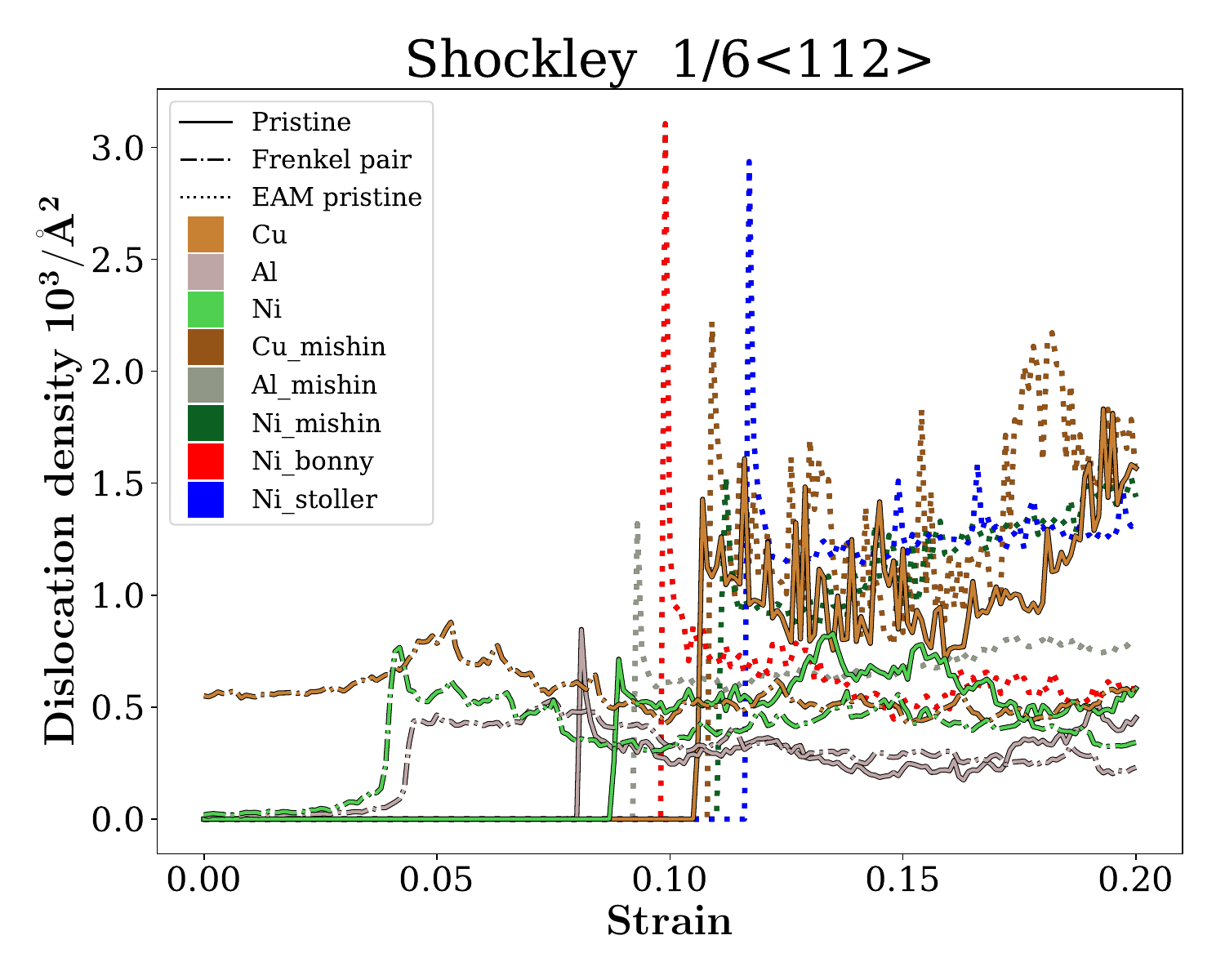}
      \caption{\hkl<100>  }
\end{subfigure}
\begin{subfigure}[t]{0.24\textwidth}
    \includegraphics[width=\textwidth]{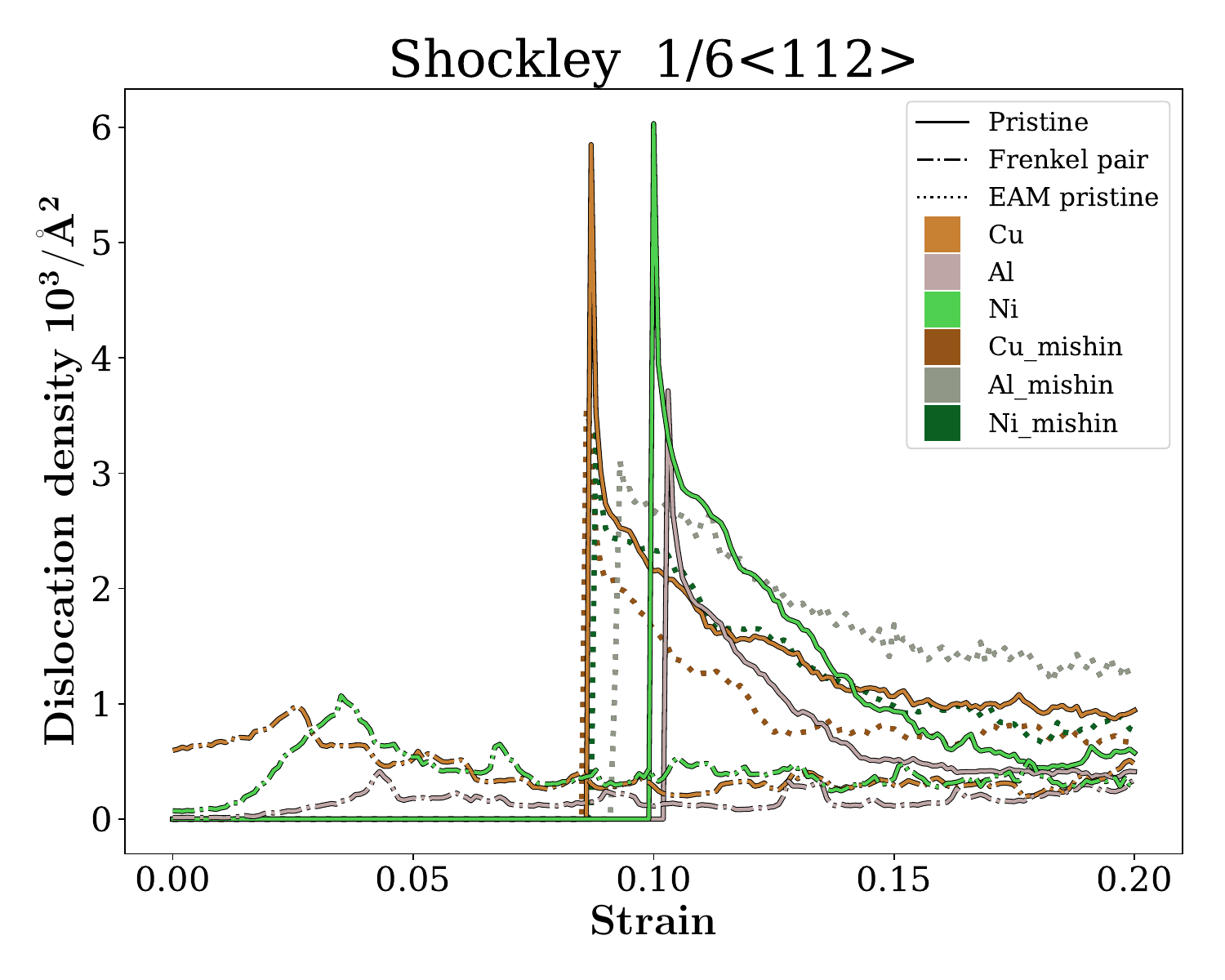}
    \caption{\hkl<110>}
\end{subfigure}
\begin{subfigure}[t]{0.24\textwidth}
    \includegraphics[width=\textwidth]{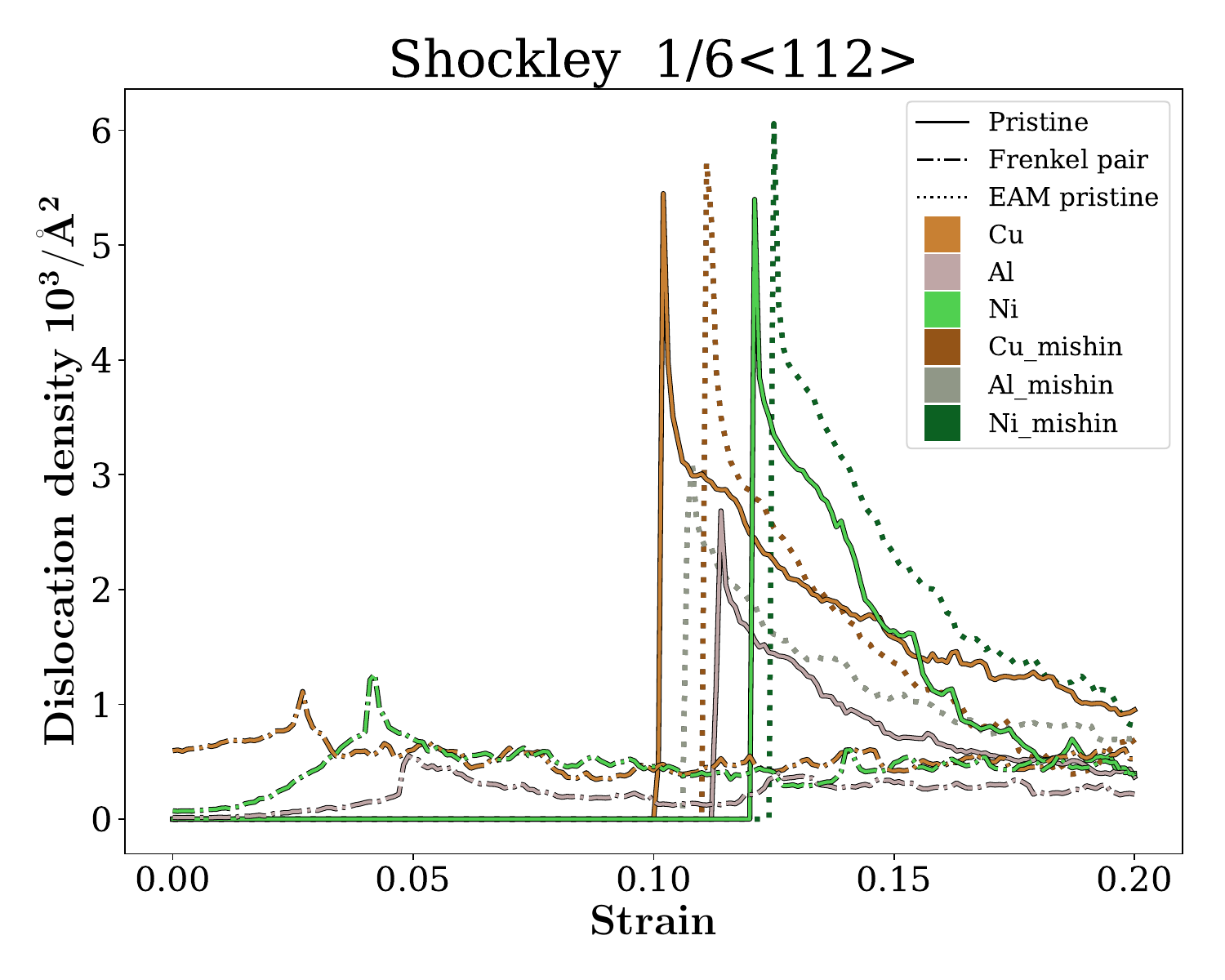}
      \caption{\hkl<111>}
\end{subfigure}
\begin{subfigure}[t]{0.24\textwidth}
    \includegraphics[width=\textwidth]{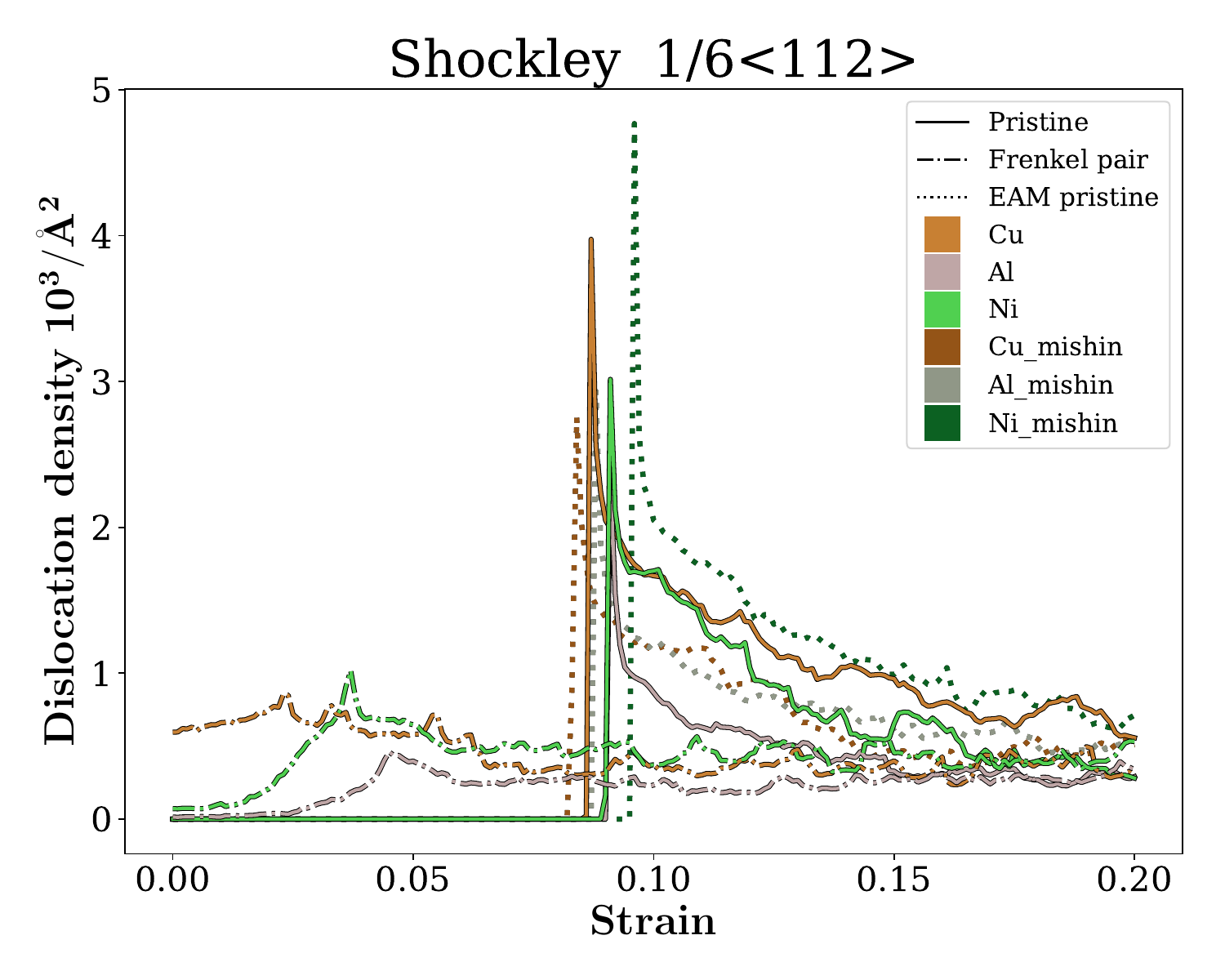}
    \caption{\hkl<112>}
\end{subfigure}
\begin{subfigure}[t]{0.24\textwidth}
    \includegraphics[width=\textwidth]{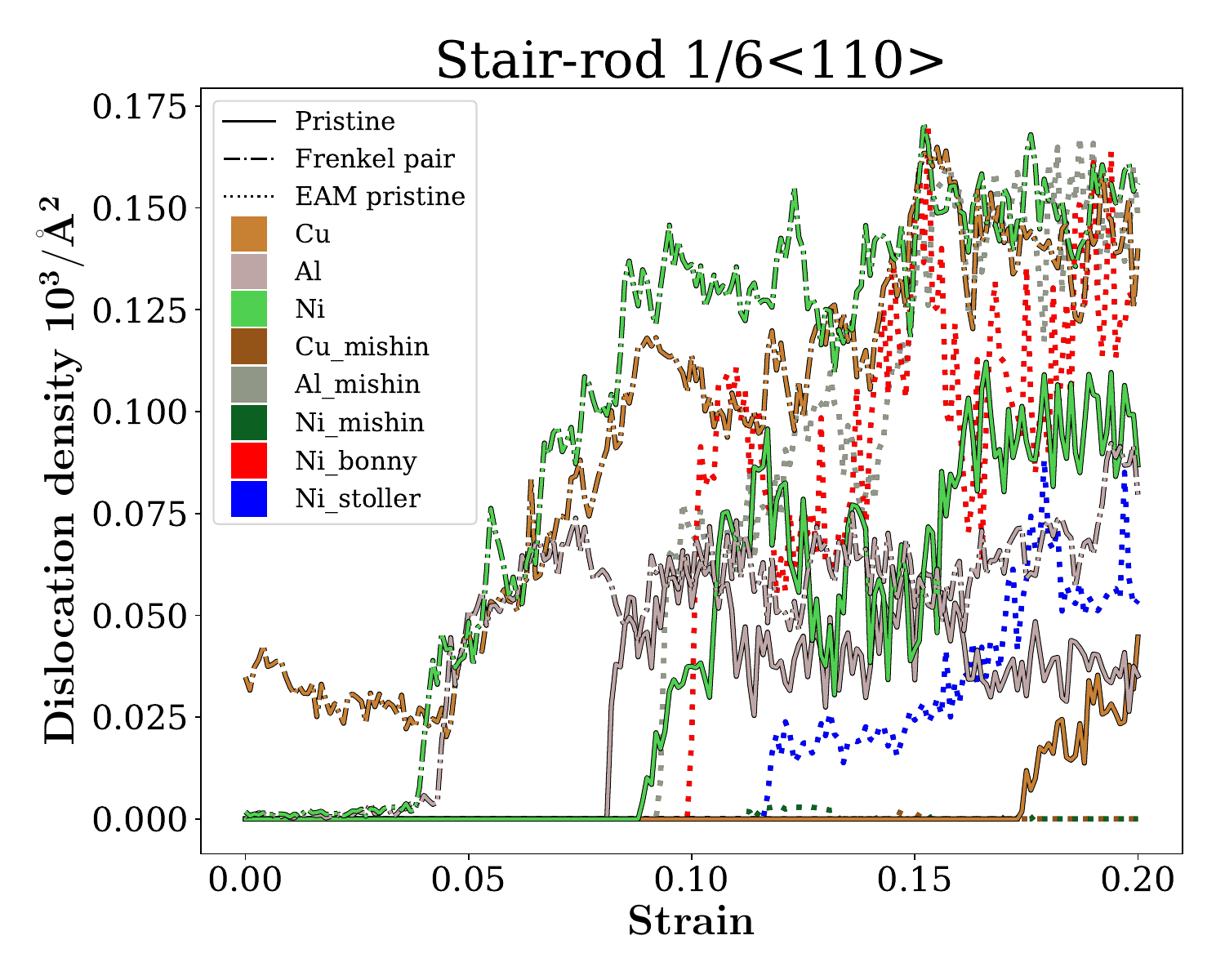}
      \caption{\hkl<100>  }
\end{subfigure}
\begin{subfigure}[t]{0.24\textwidth}
    \includegraphics[width=\textwidth]{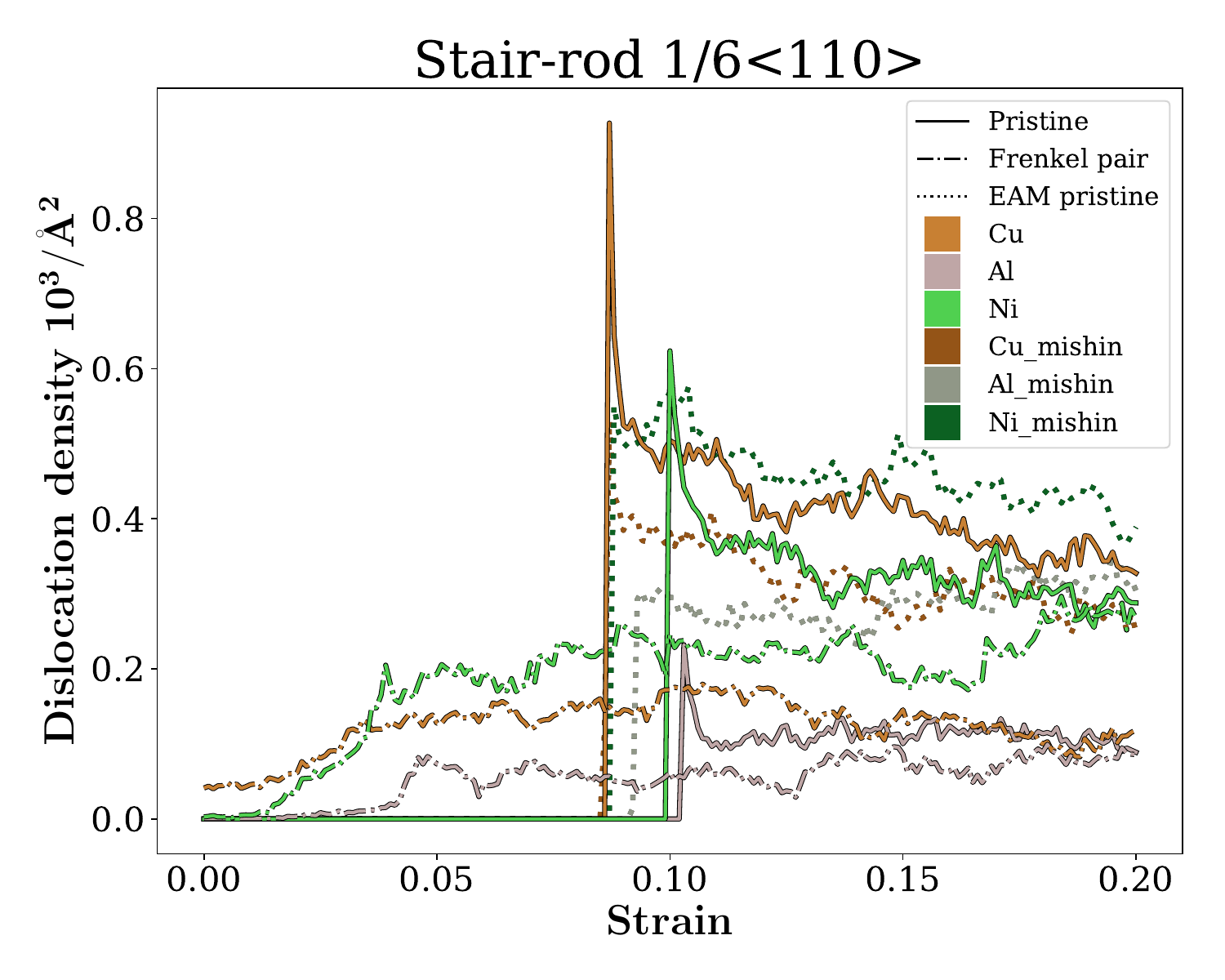}
    \caption{\hkl<110>}
\end{subfigure}
\begin{subfigure}[t]{0.24\textwidth}
    \includegraphics[width=\textwidth]{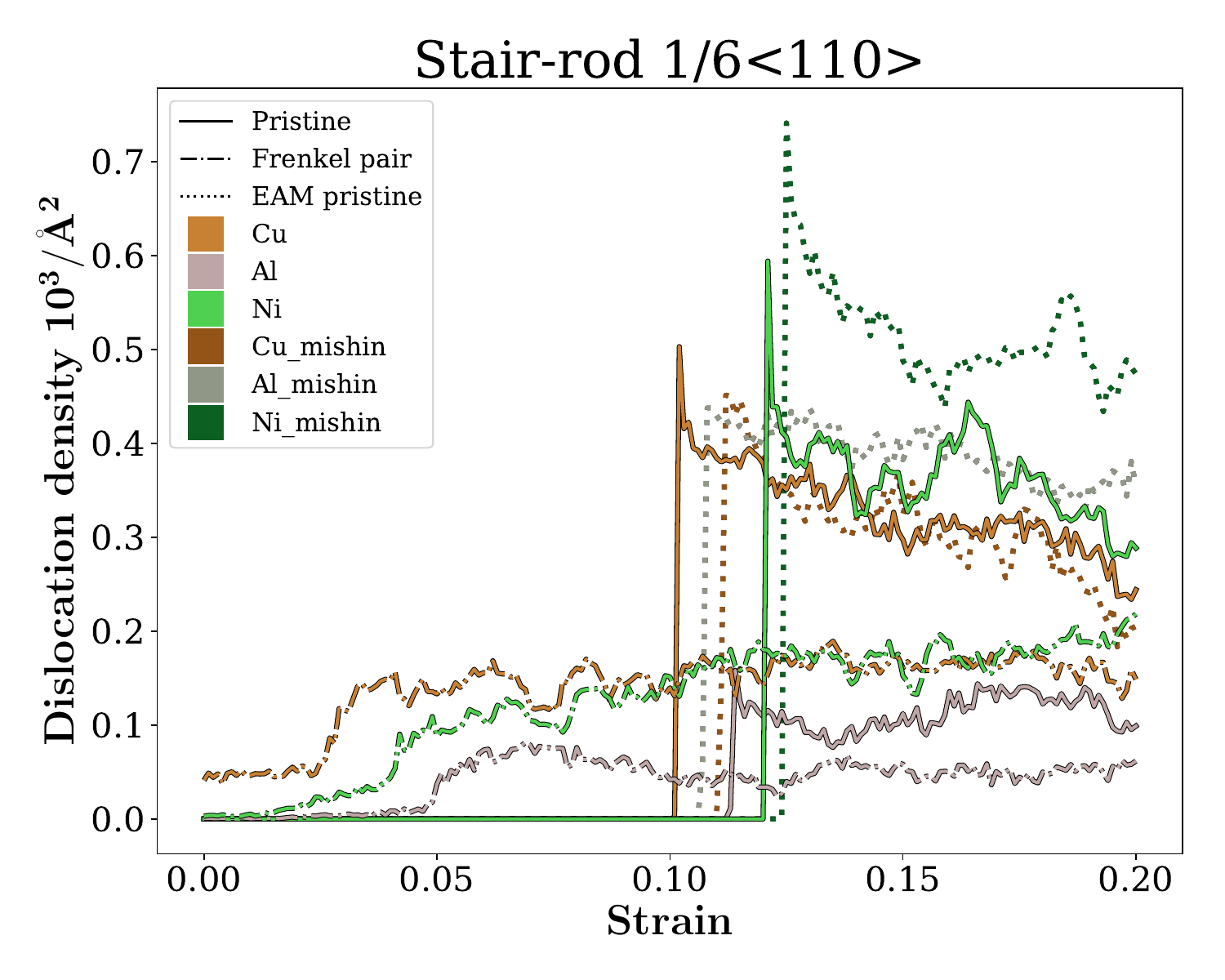}
      \caption{\hkl<111>}
\end{subfigure}
\begin{subfigure}[t]{0.24\textwidth}
    \includegraphics[width=\textwidth]{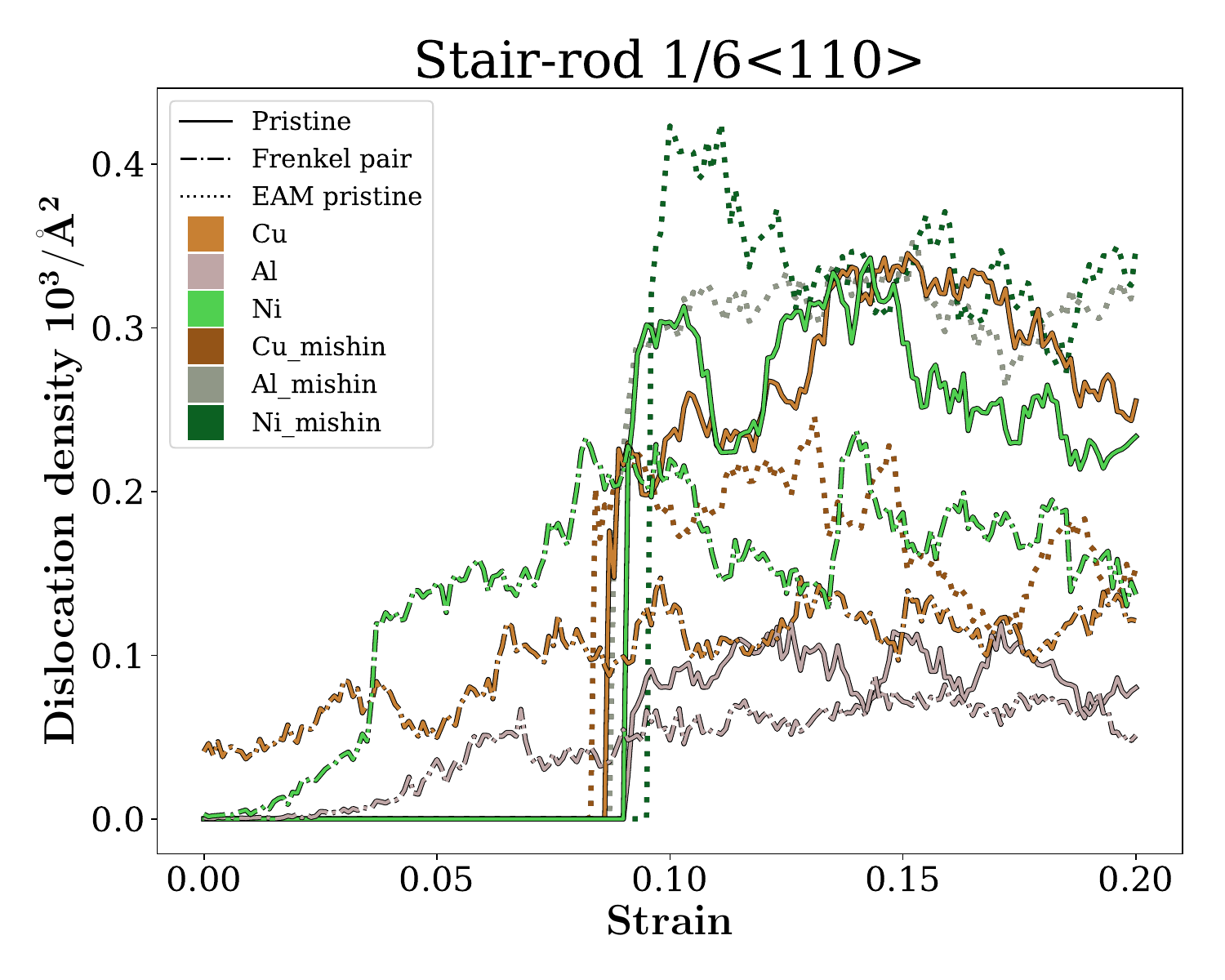}
    \caption{\hkl<112>}
\end{subfigure}
    \caption{Dislocation density as a function of strain for various types of dislocations (rows) and different loading directions (columns). The dislocation types are shown in the caption of the individual panels: "Total" includes all types of dislocations found in the structures; "Frank 1/3\hkl<111>", "Shockley 1/6\hkl<112>" and "Stair-rod 1/6\hkl<112>" refer to the corresponding dislocation types. The loading directions \hkl<100>, \hkl<110>, \hkl<111> and \hkl<112> for each case are given under the corresponding panels.  
    Pristine refers to the cells with no pre-existing defects and Frenkel pair refers to the cells relaxed after insertion of Frenkel pairs. Results from simulations with the developed tabGAP potentials and EAM potentials in the literature~\cite{mishin2001structural,PhysRevB.59.3393,stoller2016impact,bonny2013interatomic}.}
    \label{fig:dislocations}
\end{figure*}

In Fig.\ref{fig:dislocations} we analyse the dislocation densities in all three materials in different loading directions. In most cases, the total dislocation density at the end of the compression is higher for the pristine structures. Furthermore, in pristine materials the total dislocation density rises rapidly reaching the peak at the yielding point. After that the total density first drops significantly, but then decrease of the density slows down reaching a steady state value. Although there is a large difference in the total dislocation densities after yielding between the cells with the initially pristine and damaged structures, under the continued compression the dislocation density converges to the same steady-state values in all loading directions, except for the \hkl<100> one. The path-independence and loss of microstructure history has been also demonstrated during the simulation of compression in BCC materials (tantalum)~\cite{zepeda2017probing}. This suggests that in most cases one could not determine from the final structure whether the initial one contained defects. Neither it is possible to do when analyzing the type of dislocations present in the metal after yielding. We note that the main difference appears in the density of stair-rod dislocations, which is significantly higher in the initially pristine cell. 

In the \hkl<100> direction, on the other hand, there is a clear difference between the damaged and pristine structures, especially for Cu. This is due to the different dislocation types as is shown in Fig.\ref{fig:CNA} with different loading directions in Cu. Looking at the Frank dislocation densities we can observe that in all cases they are very low. A curious detail is that in some cases the EAM potentials show somewhat higher Frank type dislocation densities in Al and Ni. Additionally, comparing the tabGAP to EAM potentials we can generally conclude that EAM potentials have higher dislocation densities during compression, with the results of Cu being closest to its tabGAP counterpart.

\clearpage

\section{Conclusions}

We have developed machine-learned interatomic potentials for Al, Cu and Ni using the Gaussian approximation potential method. Fast and accurate tabGAP versions were created and extensively tested. The tabGAPs show similar accuracy as their GAP counterparts, while being orders of magnitude more computationally efficient, enabling large scale simulations containing millions of atoms. The potentials were validated and are in good agreement with DFT calculations and experimental data for various material properties. Using the developed potentials, we simulate the threshold displacement energies as a function of lattice direction for each of the elements. In addition to this, uniaxial compression simulations were performed on both pristine simulation cells and cells containing pre-existing defects. The stress strain curves for the pristine cases were in agreement with what has been reported previously. However, in Ni we could see large differences between different interatomic potentials. The compression simulations demonstrate the feasibility to simulate large-scale systems on the order of millions of atoms with the machine-learned tabGAPs. Furthermore, the diverse DFT databases for the training of the potentials can be augmented and used for future work on alloys.

\section*{Acknowledgments}

This work has received funding from the Academy of Finland through the HEADFORE project (grant number 333225).
The authors wish to thank the Finnish Computing Competence Infrastructure (FCCI) and CSC -- IT Center for Science for supporting this project with computational and data storage resources.
This work has been partially carried out within the framework of the EUROfusion Consortium, funded by the European Union via the Euratom Research and Training Programme (Grant Agreement No 101052200 — EUROfusion). Views and opinions expressed are however those of the author(s) only and do not necessarily reflect those of the European Union or the European Commission. Neither the European Union nor the European Commission can be held responsible for them.

\bibliography{mybib}

\end{document}